\documentclass[a4paper,11pt]{article}
\pdfoutput=1 
\usepackage{jcappub} 
\usepackage[T1]{fontenc} 

\usepackage{xcolor}
\usepackage{mathtools} 
\usepackage{bm}
\usepackage{mathrsfs}
\usepackage{amsmath}	
\usepackage{amssymb}	
\usepackage{subcaption}

\newcommand{\rd}{\mathrm{d}}

\graphicspath{%
	{./fig/} %
}	

\title{The $n$-point streaming model: how velocities shape correlation functions in redshift space}

\author[a,b]{Joseph Kuruvilla}
\author[b]{and Cristiano Porciani}
\affiliation[a]{Universit\'e Paris-Saclay, CNRS,  Institut d'astrophysique spatiale, 91405, Orsay, France}
\affiliation[b]{Argelander-Institut f\"ur Astronomie, Universit{\"a}t Bonn, Auf dem H\"ugel 71, D-53121 Bonn, Germany}

\emailAdd{joseph.kuruvilla@universite-paris-saclay.fr}
\emailAdd{porciani@astro.uni-bonn.de}

\abstract{
Starting from first principles,
we derive the fundamental equations that relate the $n$-point correlation functions in real and redshift space.  Our result generalises the so-called `streaming model' to higher-order statistics:
the full $n$-point correlation in redshift-space is
obtained as an integral of its real-space counterpart times the joint probability density of $n-1$ relative line-of-sight peculiar velocities.
Equations for the connected $n$-point correlation functions are obtained by recursively applying the generalised streaming model for decreasing $n$.
Our results are exact within the distant-observer approximation and completely independent of the nature of the tracers for which the correlations are evaluated. 
Focusing on 3-point statistics, we use an $N$-body simulation to study the joint probability density function
of the relative line-of-sight velocities of pairs of particles in a triplet.
On large scales,
we find that this distribution is approximately Gaussian and that its moments can be accurately computed with standard perturbation theory.
We use 
this information to
formulate a phenomenological 3-point Gaussian streaming model. A practical implementation is obtained by
using perturbation theory at leading order to
approximate several statistics in real space. 
In spite of this simplification, the resulting 
predictions for the matter 3-point correlation function in redshift space are in rather good agreement with
measurements performed in the simulation. 
We discuss the limitations of the simplified model and suggest a number of possible improvements.
Our results find direct applications in the analysis
of galaxy clustering but also set the basis
for studying 
3-point statistics 
with future peculiar-velocity surveys and experiments based on the kinetic Sunyaev-Zel'dovich effect.
}


\begin{document}
\maketitle
\flushbottom

\section{\label{sec:intro}Introduction}

Maps of the large-scale structure of the Universe obtained from galaxy redshift surveys suffer from the so-called redshift-space distortions (RSD) generated by galaxy peculiar velocities \cite{Jackson72,SargentTurner77}.
RSD break the isotropy of galaxy $N$-point statistics by introducing an angular dependence with respect to the direction of the line of sight (los)
\cite{Kaiser87,hamilton98}.
The degree of anisotropy depends on the growth rate of cosmic structure
and can thus be used to probe dark energy and test gravity theories. 
Achieving this goal, however, requires modelling daunting non-linear and non-perturbative physics as motions within virialised galaxy clusters alter galaxy statistics on significantly large scales.

The introduction of the streaming model for the 2-point correlation function \cite{Peebles80} represents a key milestone in this development. The basic idea is to compute the distorted anisotropic two-point  correlation function (in `redshift space') by an integral transformation of  the underlying isotropic correlation function (in `real space') combined with the distribution function of the relative los velocities
 of galaxy pairs.
 However,
 since the moments of this `pairwise velocity distribution function' (PVD) are strongly scale dependent and difficult to predict from first principles, the streaming model has been often considered as a rather impractical tool to use for cosmological inferences (although it is exact in the distant-observer approximation).
 Assuming that the PVD is Gaussian for large spatial separations and that
 its mean and variance can be evaluated using perturbation theory
 formed a successful step forward in this direction \cite{Fisher95,ReidWhite11,Carlson+13,White+15,Vlah+16}. 
 This `Gaussian streaming model' has been successfully applied to galaxy redshift surveys \cite{reid+12,Samushia+14,Alam+15,Chuang+17,Satpathy+17}.
 In a parallel line of research, several authors have discussed how to go beyond the Gaussian approximation by incorporating higher-order cumulants of the PVD \cite{Bianchi+15,Uhlemann+15,Bianchi+16,KuruvillaPorciani18,Cuesta+20}.

In this paper, we derive an exact streaming model for generic $n$-point correlation functions ($n$PCFs) with $n\geq 2$. In full analogy with the 2-point case, we find that the $n$-point correlation in redshift space is given by
an integral transformation of its real-space counterpart multiplied by the multivariate distribution of the relative los velocities between $n-1$ galaxy pairs in a $n$-tuple.
After studying the properties of this distribution for triplets of dark-matter particles in a large $N$-body simulation, we formulate a Gaussian streaming model for the 3PCF and test its performance against the simulation.

Measurements of the 3PCF have a long history that reflects the development of galaxy surveys. Pioneering studies, dating back to the 1970s, were based on a few thousand galaxy positions on the sky 
\cite{PeeblesGroth75,GrothPeebles77,Peebles81}. Early redshift surveys provided samples containing a few hundred objects  
\cite{Bean+83,Efstathiou+84,Hale-Sutton+89}.
A measurement with much larger signal-to-noise ratio was performed using
nearly 20,000 galaxies from the Las Campanas Redshift Survey \cite{JingBorner98}. 
Eventually, in the early 2000s, the advent of multi-fiber spectrographs provided homogeneous samples with $10^{5-6}$ galaxies at low redshift.
The 3PCF was measured from the Two-Degree Field Galaxy Redshift Survey \cite{JingBorner04,Wang+04,Gaztanaga+05_2dfgrs}, different generations of 
the Sloan Digital Sky Survey \cite{Kayo+04,Nicol+06,Ross+06,Kulkarni+07,Gaztanaga+09,McBride+11a,Marin11,Guo+14,Guo+15,Slepian+17}, and the WiggleZ Dark Energy Survey \cite{Marin+13}. 
Recently, it was also possible to extend the analysis at redshifts $0.5<z<1$ by using nearly 50,000 galaxies from the
VIMOS Public Extragalactic Redshift Survey \cite{Moresco+17}.

In spite of this impressive progress, estimates of 3-point statistics
on large scales still suffer from systematic shifts generated by rare statistical fluctuations, meaning that substantially larger volumes need to be covered in order to obtain unbiased measurements, e.g. \cite{Nicol+06}.
Fortunately, dark-energy science is providing a strong motivation for
building such unprecedentedly large samples. This led the community to
develop and build dedicated facilities 
like the Dark  Energy Spectroscopic Instrument (\textsc{Desi}, \cite{desi13}), the \textsc{Euclid} mission \cite{Euclid13}, the Wide-Field Infrared Survey Telescope (\textsc{Wfirst}, \cite{WFIRST}), the Prime Focus Spectrograph (\textsc{Pfs}, \cite{PFS}), the Large Synoptic Survey Telescope (\textsc{Lsst}, \cite{LSST}) and the  Spectro-Photometer  for  the  History  of  the  Universe,  Epoch  of  Reionization,  and Ices Explorer (\textsc{SPHEREx}, \cite{spherex}).

Several authors have recently highlighted that combining two- and three-point clustering statistics with data of this calibre
will ultimately lead to a sizeable information gain about the cosmological parameters \cite{Sefusatti+06,Gil+15,Gil+17,Karagiannis+18,YankelevichPorciani18,ChudaykinIvanov19,GualdiVerde20}.
In particular, 3-point clustering statistics (either in configuration or Fourier space) are expected to: i) remove the degeneracy between the
amplitude of dark-matter perturbations and the
galaxy linear bias coefficient that plagues 2-point statistics  \cite{Fry94,FriemanGaztanaga94,Matarrese+97}
and  constrain the linear growth rate of matter fluctuations \cite{Hoffman+15};
ii) provide an accurate determination of galaxy biasing \cite{Gil+15, YankelevichPorciani18}; 
iii) constrain the level of primordial non Gaussianity \cite{Scoccimarro+04_PNG,SefusattiKomatsu07,TommasoPorciani10,Tellarini+2016,Karagiannis+18};
iv) help distinguish between alternative models like coupled dark-energy cosmologies \cite{Moresco+14}; v) constrain neutrino masses \cite{ChudaykinIvanov19, Hahn+20} .

In order to keep these promises and fully exploit the forthcoming data,
it is essential to make fast progress from the theoretical point of view as well. Historically, most models of the 3PCF were based on the 
basic `hierarchical clustering' ansatz \cite{Peebles80} 
or on the phenomenological halo model \cite{Wang+04, Guo+15}. It is only recently that more quantitative techniques have received increased attention. For instance, perturbation theory has been used to compute a model for the 3PCF in redshift space \cite{SlepianEisenstein17_modelling3PCF} in analogy with previous results
obtained in Fourier space \cite{SCF99}.
Our work provides a framework for further developing this line of research
along a path that was already very successful for 2-point statistics.

The paper is organised as follows.
In section~\ref{sec:streaming-model}, we review the basic
concepts of RSD and derive the fundamental equations
of the generalised streaming model for the $n$PCF. 
This first part is very general and technical.
We then focus on applications of the theory to the 3PCF.
With this goal in mind, in section~\ref{sec:triplewise}, 
we use an $N$-body simulation
and perturbation theory to study the properties
of the bivariate distribution of the relative los velocities between particle pairs in a triplet.
Motivated by the resuls, in section~\ref{sec:3ptGSM}, 
we formulate the 3-point Gaussian
streaming model and test it against the simulation.
Finally, we summarise our results in section~\ref{sec:conclusions}.

\section{\label{sec:streaming-model}The streaming model}

We start with a note.
Busy readers who want to focus on applications of the theory to the 3PCF may want to skip large parts of 
this section on first reading but
will want to read sections~\ref{sec:triplewise}, \ref{sec:3ptGSM} and
\ref{sec:conclusions} in their entirety. 
To help them scan for desired information and
skip those parts that are more conceptual,
we recommend familiarising themselves with section~\ref{sec:rsd},
equation~(\ref{eq:2ptstreamfinal}),
the short sentence following equation~(\ref{eq:multipdfdef}) that provides a definition in words of the functions we denote by ${\mathcal P}^{(n)}_{\boldsymbol{w}_\parallel}$,
and the beginning of section~\ref{sec:app3pt} until
equation~(\ref{eq:3ptstreamrversion}).

\subsection{\label{sec:rsd}Redshift-space distortions}
The distance to a galaxy, quasar or galaxy cluster is generally estimated starting from the observed redshift of spectral lines in its electromagnetic spectrum. 
This conversion assumes an unperturbed Friedmann model of the Universe with instantaneous expansion factor $a$ and thus
a perfect Hubble flow with instantaneous Hubble parameter $H$. Therefore, this distance estimate is never exact with actual data due to the
presence of peculiar velocities.
In the distant-observer (or plane-parallel) approximation \citep{hamilton98}, 
a single los direction  $\hat{\boldsymbol{s}}$ can be defined for all objects. Hence,
the actual comoving distance $\bm{x}$ and the redshift-based estimate
$\bm{x}_{\mathrm{s}}$ satisfy the relation
\begin{equation}
\bm{x}_{\mathrm{s}} = \bm{x} + (\bm{v}\cdot\hat{\bm{s}}) \,\hat{\bm{s}} \, .
\label{eq:rsd-displacement}
\end{equation}
where $\boldsymbol{v}$ denotes the peculiar velocity $\boldsymbol{u}$ divided by the factor $aH$.
The locations described by the coordinates $\bm{x}_{\mathrm{s}}$
and $\bm{x}$ are commonly referred to as the `redshift space'
and the `real space' position, respectively.
Consider two tracers of the large-scale structure with real-space separation $\boldsymbol{r}_{12}=\boldsymbol{x}_2-\boldsymbol{x}_1$. Their redshift-space separation along the los is then
\begin{equation}
s_{12\parallel} = (\bm{x}_{\mathrm{s}_2} - \bm{x}_{\mathrm{s}_1}) \cdot \hat{\bm{s}} = r_{12\parallel} + w_{12\parallel} \, ,
\label{eq:rsd_sep}
\end{equation}
where $r_{12\parallel} = \bm{r}_{12}\cdot \hat{\bm{s}}$ and  $w_{12\parallel} = (\bm{v}_2 - \bm{v}_1) \cdot \hat{\bm{s}}$.
On the other hand, in the perpendicular plane, the real- and redshift-space separations coincide, i.e. $\bm{s}_{12\perp} = \bm{r}_{12\perp}$. 

\subsection{Phase-space densities and correlation functions}
\label{sec:fnintro}
Let us consider a system consisting of $N$ particles in 3-dimensional space.
At any instant of time, each particle is characterised by its comoving position $\bm{x}_i$ and the (rescaled) peculiar velocity $\bm{v}_i$ (with $1\leq i\leq N)$.
We introduce the $n$-particle phase-space densities \cite{Yvon35,Huang87,Kardar07}
\begin{align}
&f_n(\boldsymbol{x}_{\mathrm{A}_1},\dots,\boldsymbol{x}_{\mathrm{A}_n},\bm{v}_{\mathrm{A}_1},\dots,\bm{v}_{\mathrm{A}_n})=
  \sum_{i_1=1}^{N} \sum_{i_2 \neq i_1}\ldots\sum_{i_n\neq i_1,\ldots,i_{n-1}} \nonumber\\
& \left\langle  \delta_{\rm D}^{(3)}( \boldsymbol{x}_{\mathrm{A}_1}-\boldsymbol{x}_{i_1}) \dots \delta_{\rm D}^{(3)}( \boldsymbol{x}_{\mathrm{A}_n}-\boldsymbol{x}_{i_n})\,\delta_{\rm D}^{(3)}( \boldsymbol{v}_{\mathrm{A}_1}-\boldsymbol{v}_{i_1}) \dots \delta_{\rm D}^{(3)}( \boldsymbol{v}_{\mathrm{A}_n}-\boldsymbol{v}_{i_n}) \right\rangle\;,
\end{align}
where $\delta_{\rm D}^{(n)}$ is the Dirac delta distribution in $\mathbb{R}^{n}$ and the brackets denote averaging over an ensemble of realisations. 
Before we proceed, let us clarify our notation.
The symbols $\bm{x}_{\mathrm{A}_i}\in \mathbb{R}^3$ and $\bm{v}_{\mathrm{A}_i}\in\mathbb{R}^3$ denote the 
independent variables of the $f_n$ functions.
On the other hand, as we have already mentioned, $\bm{x}_{i_j}$ and
$\bm{v}_{i_j}$ indicate the position and velocity
of the $i_j^{\mathrm{th}}$ particle.
The indices $\{i_1,\dots,i_n\}$ specify a set of $n$ different particles and the sums
run over all possible $n$-tuples that can be formed with $N$ particles.
Note that $f_n$ is normalised to the total number of ordered $n$-tuples
of particles:
$\displaystyle \int f_n\, \rd \bm{x}_{\mathrm{A}_1} \dots \rd \bm{x}_{\mathrm{A}_n}\, \rd \bm{v}_{\mathrm{A}_1} \dots \rd \bm{v}_{\mathrm{A}_n} = N!/(N-n)!$.
Assuming statistical isotropy and homogeneity as well as that $N\rightarrow\infty$, it follows that $f_1=\bar{n}\,{\mathcal P}^{(1)}_{\bm{v}}$
where $\bar{n}$ denotes the mean particle number density per unit volume
and ${\mathcal P}^{(1)}_{\bm{v}}$ is the probability density function (PDF) of peculiar velocities that can only depend on $v^2$ and is normalised
such that $4\pi\,\int {\mathcal P}^{(1)}_{\bm{v}}\,v^2\,\mathrm{d}v=1$ \cite{KuruvillaPorciani18}.
Under the same assumptions, 
the $n$-point spatial correlation function of the particles in configuration space $(n\geq 2$) can be expressed as
\begin{equation}
\mathcal{F}_n = \frac{\displaystyle \int f_n
\, \rd \bm{v}_{\mathrm{A}_1}\dots \rd\bm{v}_{\mathrm{A}_n}}
{\left(\displaystyle \int f_1
\,\rd\bm{v}\right)^n} = \frac{1}{\bar{n}^n}\,\int f_n
\, \rd \bm{v}_{\mathrm{A}_1}\dots \rd\bm{v}_{\mathrm{A}_n}
\,,
\end{equation}
where we did not write explicitly the arguments of the correlation functions to simplify notation.
The irreducible (or connected) spatial $n$-point correlation functions can be expressed in terms of the ${\mathcal F}_n$.
For instance, ${\mathcal F}_2$ and the 2-point connected function $\xi$ satisfy the relation
\begin{equation}
\mathcal{F}_2(r)=1+\xi(r)\;,    
\end{equation}    
where 
$r=|\bm{x}_{{\mathrm A}_2}-\bm{x}_{{\mathrm A}_1}|$ denotes the comoving separation between the points at which the functions are evaluated.
Similarly, $\mathcal{F}_3$ is related to the 3-point connected function $\zeta$ by
\begin{equation}
\mathcal{F}_3(r_{12}, r_{23}, r_{31})=1+\xi(r_{12})+\xi(r_{23})+\xi(r_{31})+ \zeta(r_{12}, r_{23}, r_{31})\;,    
\end{equation}  
where the different $r_{ij}=|\bm{x}_{{\mathrm A}j} -\bm{x}_{{\mathrm A}i}|$ indicate the comoving separations between pairs of points in a triplet.

Analogous considerations apply in redshift space, where we can introduce
the  $n$-particle phase-space densities $g_n$ and the $n$-point spatial correlation functions

\begin{equation}
\mathcal{G}_n  = \frac{\displaystyle \int g_n \, \rd \bm{v}_{\mathrm{A}_1}\dots \rd\bm{v}_{\mathrm{A}_n}}
{\left(\displaystyle \int g_1\,\rd\bm{v}\right)^n}  = \frac{1}{\bar{n}^n}\,\int g_n
\, \rd \bm{v}_{\mathrm{A}_1}\dots \rd\bm{v}_{\mathrm{A}_n}
\,.
\label{eq:npoint-red}
\end{equation}
Since redshift-space distortions appear along the line of sight, these functions are not isotropic. However,
due to the invariance under rotations along the los, $\mathcal{G}_2$ and $\xi_\mathrm{s}$ only depend on the modulus of $\bm{s}_\perp$:
\begin{equation}
\mathcal{G}_2(s_\parallel,s_\perp)=1+\xi_\mathrm{s}(s_\parallel, s_\perp)\;.
\label{eq:g2xis}
\end{equation}    
Similarly, we can write
\begin{equation}
\mathcal{G}_3(\triangle_{123}) = 1+\xi_{\mathrm{s}}(s_{12\parallel},s_{12\perp})+\xi_{\mathrm{s}}(s_{23\parallel},s_{23\perp})+\xi_{\mathrm{s}}(s_{31\parallel}, s_{31\perp})+ \zeta_{\mathrm{s}}(\triangle_{123})\;,  
\label{eq:G3expl}
\end{equation}  
although the compact notation above needs further explanation. 
First of all, there are multiple ways to parameterize the triangle $\triangle_{123}\equiv\{\bm{s}_{12}, \bm{s}_{23}, \bm{s}_{31}\}$. Since, by definition, $\bm{s}_{12}+ \bm{s}_{23}+\bm{s}_{31}=0$, picking two of the legs automatically determines the third one. For instance, we could write
\begin{equation}
\mathcal{G}_3(\bm{s}_{12}, \bm{s}_{23}) = 1+\xi_{\mathrm{s}}(s_{12\parallel},s_{12\perp})+\xi_{\mathrm{s}}(s_{23\parallel},s_{23\perp})+\xi_{\mathrm{s}}(s_{31\parallel}, s_{31\perp})+ \zeta_{\mathrm{s}}(\bm{s}_{12}, \bm{s}_{23})\;, 
\label{eq:G3expl2}
\end{equation}
even though also this notation does not reflect the full picture.
In fact,
$\mathcal{G}_3$ and
$\zeta_\mathrm{s}$ only depend on $ s_{12\parallel},s_{12\perp}, s_{23\parallel}, s_{23\perp}$ and  $\cos \theta_\perp=\hat{\bm{s}}_{12\perp}\cdot \hat{\bm{s}}_{23\perp}$.
Since, $s_{31\perp}^2=s_{12\perp}^2+s_{23\perp}^2+2s_{12\perp}s_{23\perp}\cos \theta_\perp$ 
and $s_{31\parallel}=-(s_{12\parallel}+s_{23\parallel})$, we can equivalently express the functional dependence of
$\zeta_\mathrm{s}$
in terms of five separations: $s_{12\perp}, s_{12\parallel}, s_{23\perp}, s_{23\parallel}$ and $s_{31\perp}$ (as we will do in sections~\ref{sec:losproj} and \ref{sec:3ptGSM}).
However, the 3PCFs $\mathcal{G}_3$ and $\zeta_{\mathrm{s}}$ do not depend on the labelling of the vertices of $\triangle_{123}$, e.g. $\zeta_\mathrm{s}(\bm{s}_{12}, \bm{s}_{23})=
\zeta_\mathrm{s}(\bm{s}_{13}, \bm{s}_{32})=
\zeta_\mathrm{s}(\bm{s}_{21}, \bm{s}_{13})=
\zeta_\mathrm{s}(\bm{s}_{23}, \bm{s}_{31})=
\zeta_\mathrm{s}(\bm{s}_{31}, \bm{s}_{12})=
\zeta_\mathrm{s}(\bm{s}_{32}, \bm{s}_{21})$,
whereas using  $s_{12\perp}, s_{12\parallel}, s_{23\perp}, s_{23\parallel}$ and $s_{31\perp}$ associates different  parameter sets to different labellings. For instance, in a measurement, a single triplet of points would contribute to six different triangular configurations thus introducing unnecessary covariances and repetitions.
Fixing the labelling so that  $s_{12}\geq s_{23}\geq s_{31}$ provides a simple solution to this issue 
\cite{YankelevichPorciani18} but we will not
adopt this convention in this work.

\subsection{The streaming model for the 2-point correlation function}
In this section, we outline the original derivation of the streaming model for the 2PCF presented in \cite{KuruvillaPorciani18}.
By definition, the phase-space distributions $f_2$ and $g_2$ differ only by the coordinate change in equation~(\ref{eq:rsd_sep}). We can thus combine
equations~(\ref{eq:npoint-red}) and (\ref{eq:g2xis}) and write\footnote{To avoid the proliferation of subscripts, whenever possible (i.e. when we discuss
explicit examples for the 2 and 3PCFs
instead of the generic $n$-point case), we use the indices A, B, $\dots$ instead
of $\mathrm{A}_1, \mathrm{A}_2, \dots$.}
\begin{equation}
1+\xi_{\mathrm{s}}(s_\parallel,s_\perp)=
\frac{1}{\bar{n}^2}\,\int  f_2(s_\parallel-w_{\parallel},s_\perp,\bm{v}_\mathrm{A},\bm{v}_\mathrm{B})\,\delta_\mathrm{D}^{(1)}(w_\parallel-v_{\mathrm{B}\parallel}+v_{\mathrm{A}\parallel})\,\mathrm{d}w_\parallel\,\mathrm{d}\bm{v}_\mathrm{A} \,\mathrm{d}\bm{v}_\mathrm{B}\;.
\label{eq:prestreaming2}
\end{equation}
We now multiply the integrand in the right-hand side (rhs) of the last equation by the quantity
\begin{equation}
\frac{\bar{n}^2\,\left[ 1+\xi\left(\sqrt{(s_\parallel-w_\parallel)^2+s_\perp^2} \right)\right]}{\displaystyle \int f_2(s_\parallel-w_\parallel,s_\perp,\bm{v}_\mathrm{A},\bm{v}_\mathrm{B})\,\mathrm{d}\bm{v}_\mathrm{A} \,\mathrm{d}\bm{v}_\mathrm{B}}
=\frac{\bar{n}^2\,\left\{1+\xi[r(s_\parallel, s_\perp, w_\parallel )]\right\}}{\displaystyle \int f_2[r_\parallel(s_\parallel, w_\parallel),r_\perp(s_\perp),\bm{v}_\mathrm{A},\bm{v}_\mathrm{B}]\,\mathrm{d}\bm{v}_\mathrm{A} \,\mathrm{d}\bm{v}_\mathrm{B}}\;,
\end{equation}
(which is identically one) and define the pairwise-velocity PDF at fixed real-space separations $r_\parallel=s_\parallel-w_\parallel$
and $r_\perp=s_\perp$
as
\begin{align}
{\mathcal P}^{(2)}_{w_\parallel}\left[w_\parallel|\bm{r}\left(s_\parallel,s_\perp,w_\parallel\right)\right]& = \frac{\displaystyle \int  f_2(s_\parallel-w_\parallel,s_\perp,\bm{v}_\mathrm{A},\bm{v}_\mathrm{B})\,\delta_\mathrm{D}^{(1)}(w_\parallel-v_{\mathrm{B}\parallel}+v_{\mathrm{A}\parallel})\,\mathrm{d}\bm{v}_\mathrm{A} \,\mathrm{d}\bm{v}_\mathrm{B}}{\displaystyle \int  f_2(s_\parallel-w_\parallel,s_\perp,\bm{v}_\mathrm{A},\bm{v}_\mathrm{B})\,\mathrm{d}\bm{v}_\mathrm{A} \,\mathrm{d}\bm{v}_\mathrm{B}} \nonumber \\
& = \frac{\displaystyle \int  f_2(s_\parallel-w_\parallel,s_\perp,\bm{v}_\mathrm{A},\bm{v}_\mathrm{B})\,\delta_\mathrm{D}^{(1)}(w_\parallel-v_{\mathrm{B}\parallel}+v_{\mathrm{A}\parallel})\,\mathrm{d}\bm{v}_\mathrm{A} \,\mathrm{d}\bm{v}_\mathrm{B}}
{\bar{n}^2\,\left[ 1+\xi\left(\sqrt{(s_\parallel-w_\parallel)^2+s_\perp^2}\right)\right]}\;.
\end{align}
Equation~(\ref{eq:prestreaming2}) thus reduces to the fundamental equation of the streaming model
\begin{align}
1+\xi_{\mathrm{s}}(s_\parallel,s_\perp) &= \int \left[1+\xi\left(\sqrt{(s_\parallel-w_\parallel)^2+s_\perp^2}\,\right)\right]\,{\mathcal P}^{(2)}_{w_\parallel}\left[w_\parallel| \bm{r}\left(s_\parallel,s_\perp,w_\parallel\right)\right]\,\mathrm{d}w_\parallel\; \nonumber \\
& = \int \left[1+\xi(\check{r})\right]\,{\mathcal P}^{(2)}_{w_\parallel}(s_{\parallel} - r_\parallel| \check{\bm{r}})\,\mathrm{d}r_\parallel\;.
\label{eq:2ptstreamfinal}
\end{align}
where a descending wedge symbol highlights variables that are derived and not independent.

\subsection{The streaming model for the $n$-point correlation function \label{sec:n-point}}

The reasoning above can be generalised to derive a streaming model for the $n$PCF.
An ordered $n$-tuple of points is fully described by the position of one of them together with $n-1$ independent separation vectors.\footnote{\label{footone}Convenient choices could be either the `star rays' $\bm{r}_{12}, \bm{r}_{13},\dots,\bm{r}_{1n}$ computed with respect to one of the points or the `polygon sides' $\bm{r}_{12}, \bm{r}_{23},\dots,\bm{r}_{(n-1)n}$ computed between points with consecutive labels. We adopt this second option.}
Then, the $n$-point analogue of equation~(\ref{eq:prestreaming2}) is
\begin{align}
&\mathcal{G}_n =  \frac{1}{\bar{n}^n} \int  f_{n}( s_{12\parallel}-w_{12\parallel},\ldots,s_{mn\parallel} -w_{mn\parallel},\bm{s}_{12\perp},\ldots,\bm{s}_{mn\perp}, \bm{v}_{\mathrm{A}_1},\ldots,\bm{v}_{\mathrm{A}_n}) 
\label{eq:gn-red}\\  &\,\delta_{\rm D}^{(1)}(w_{12\parallel} - v_{\mathrm{A}_2\parallel}+v_{\mathrm{A}_1\parallel})\ldots\delta_{\rm D}^{(1)}(w_{mn\parallel} - v_{\mathrm{A}_n\parallel}+v_{\mathrm{A}_m\parallel}) \, \rd w_{12\parallel} \ldots \rd w_{mn\parallel} \, \, \rd \bm{v}_{\mathrm{A}_1}\ldots \rd\bm{v}_{\mathrm{A}_n} \;,\nonumber
\end{align}
where 
the subscript $m$ is a short for the index $n-1$. 
We now multiply and divide the integrand in the rhs of equation~(\ref{eq:gn-red}) by $\bar{n}^n\,{\mathcal F}_n/\int f_n \, \mathrm{d}\bm{v}_{\mathrm{A}_1}\ldots\mathrm{d}\bm{v}_{\mathrm{A}_n}$ 
(which is identically one)
and define 
\begin{align}
    \mathcal{P}^{(n)}_{\bm{w}_\parallel}
    =  \frac{
    \displaystyle \int  f_{n} \,
    \displaystyle \delta_{\rm D}^{(1)}(w_{12\parallel} - v_{\mathrm{A}_2\parallel}+v_{\mathrm{A}_1\parallel})
\ldots\delta_{\rm D}^{(1)}(w_{mn\parallel} - v_{\mathrm{A}_n\parallel}+v_{\mathrm{A}_m\parallel}) \, \rd \bm{v}_{\mathrm{A}_1}\ldots\rd\bm{v}_{\mathrm{A}_n}}{\displaystyle \bar{n}^n\,\mathcal{F}_n}\;,
\label{eq:multipdfdef}
\end{align}
where $f_n$ has the same functional dependencies as in equation~(\ref{eq:gn-red}).
This is the joint PDF of the $n-1$ relative pairwise (i.e. for unordered 2-subsets of points) los velocities that fully determine the redshift-space distortions for a fixed $n$-tuple configuration in real space (bear in mind that $w_{n1\parallel}=-w_{12\parallel}-\dots-w_{mn\parallel}$).
It follows immediately from the definition above
that ${\mathcal P}^{(n)}_{\boldsymbol{w}\parallel}$
is symmetric under particle exchange and parity transformations.
By combining equations~(\ref{eq:gn-red}) and
(\ref{eq:multipdfdef}) we obtain the streaming model
for $n$-point statistics
\begin{equation}
    \mathcal{G}_n  = \int \mathcal{F}_n \, \mathcal{P}^{(n)}_{\bm{w}_\parallel} \, \rd w_{12\parallel} \ldots \rd w_{mn\parallel} \, ,
    \label{eq:general-streaming-model-npoint}
\end{equation}
which is one of the central results of this paper.
Note that equation~(\ref{eq:general-streaming-model-npoint}) is exact under the distant-observer approximation and the assumption of statistical homogeneity and isotropy in real space. 
For dark matter,
our particle-based approach holds true even in multi-stream regions and fully accounts for density-velocity correlations. At the same time, the $n$-point streaming model obtained above applies to any population of tracers of the large-scale structure (e.g. galaxies or their host dark-matter halos) without making any assumptions regarding their interactions.

\subsubsection{Application to the 3-point correlation function }
\label{sec:app3pt}
The main focus of this paper is 3-point statistics. We therefore
give a closer look at the streaming model for the 3PCF.
After setting $n=3$, equation~(\ref{eq:general-streaming-model-npoint}) gives 
\begin{align}
	1+\xi_{\mathrm{s}}&(s_{12\parallel},s_{12\perp}) +\xi_{\mathrm{s}}(s_{23\parallel}, s_{23\perp})+\xi_{\mathrm{s}}(\check{s}_{31\parallel},s_{31\perp})+ \zeta_{\mathrm{s}}(\bm{s}_{12}, \bm{s}_{23})  \nonumber \\ 
	= &  \int \left[1+\xi(\check{r}_{12})+\xi(\check{r}_{23})+\xi(\check{r}_{31})+ \zeta(\check{r}_{12}, \check{r}_{23}, \check{r}_{31})\right] \nonumber \\
&\ \ \ \ \ \ \ \ \ \ \ \ \ \ \ \ \ \ \ \ \ \  \mathcal{P}^{(3)}_{\bm{w}_{\parallel}}\left[w_{12\parallel},  w_{23\parallel} |\, \check{\bm{r}}_{12}(\bm{s}_{12},w_{12\parallel}), \check{\bm{r}}_{23}(\bm{s}_{23},w_{23\parallel}) \right] \, \rd w_{12\parallel}\, \rd w_{23\parallel}\label{eq:3ptstreaming_w}  \\
	= &  \int \left[1+\xi(\check{r}_{12})+\xi(\check{r}_{23})+\xi(\check{r}_{31})+ \zeta(\check{r}_{12}, \check{r}_{23}, \check{r}_{31})\right] \nonumber \\
&\ \ \ \ \ \ \ \ \ \ \ \ \ \ \ \ \ \ \ \ \ \  \mathcal{P}^{(3)}_{\bm{w}_\parallel}\left(s_{12\parallel}-r_{12\parallel},  s_{23\parallel}-r_{23\parallel} |\, \check{\bm{r}}_{12}, \check{\bm{r}}_{23}\right) \, \rd r_{12\parallel}\, \rd r_{23\parallel} \;.
\label{eq:3ptstreaming_r}
\end{align}
where $s_{31\parallel}$ and $s_{31\perp}$ have been defined in the text following equation~(\ref{eq:G3expl}) and for the derived variables we have
$\check{r}_{12}=[(s_{12\parallel}-w_{12\parallel})^2+s^2_{12\perp}]^{1/2}=(r_{12\parallel}^2+s^2_{12\perp})^{1/2}$,
 $\check{r}_{23}=[(s_{23\parallel}-w_{23\parallel})^2+s^2_{23\perp}]^{1/2}=(r_{23\parallel}^2+s^2_{23\perp})^{1/2}$
and $\check{r}_{31} = [ (-s_{12\parallel} - s_{23\parallel}+w_{12\parallel}+w_{23\parallel})^2 + s^2_{31\perp}]^{1/2}=[(-r_{12\parallel}-r_{23\parallel})^2+ s^2_{31\perp}]^{1/2}$.

We can now use the streaming model for the 2PCF
to replace all appearances of $\xi_\mathrm{s}$ and write an equation for the connected 3PCF in redshift space: 
\begin{align}
-2&+\zeta_{\mathrm s}(\boldsymbol{s}_{12},\boldsymbol{s}_{23})=
-\int \left[1+\xi(\check{r}_{12})\right]\,{\mathcal P}^{(2)}_{w_{12\parallel}}(w_{12\parallel}| \check{\boldsymbol{r}}_{12})\,\mathrm{d}w_{12\parallel}\nonumber\\&-
\int \left[1+\xi(\check{r}_{23})\right]\,{\mathcal P}^{(2)}_{w_{23\parallel}}(w_{23\parallel}| \check{\boldsymbol{r}}_{23})\,\mathrm{d}w_{23\parallel}-\int \left[1+\xi(\check{r}_{31})\right]\,{\mathcal P}^{(2)}_{w_{31\parallel}}(w_{31\parallel}| \check{\boldsymbol{r}}_{31})\,\mathrm{d}w_{31\parallel}
\label{eq:connectedzetas}\\
&+
\int \left[1 +\xi(\check{r}_{12})+\xi(\check{r}_{23})+\xi(\check{r}_{31})
+\zeta(\check{r}_{12},\check{r}_{23},\check{r}_{31})\right]\,
{\mathcal P}^{(3)}_{\bm{w}_\parallel}(w_{12\parallel}, w_{23\parallel} | \check{\boldsymbol{r}}_{12},\check{\boldsymbol{r}}_{23})\,\mathrm{d}w_{12\parallel}\,\mathrm{d}w_{23\parallel}\; \nonumber \\
&= - \int [1+\xi(\check{r}_{12})] \mathcal{P}^{(2)}_{w_{12\parallel}}(s_{12\parallel}-r_{12\parallel}|\, \check{\bm{r}}_{12})\,\rd r_{12\parallel} - \int [1+\xi(\check{r}_{23})] \mathcal{P}^{(2)}_{w_{23\parallel}}(s_{23\parallel}-r_{23\parallel}|\, \check{\bm{r}}_{23})\, \rd r_{23\parallel} \nonumber \\
& \,\,\,\,\,\,- \int [1+\xi(\check{r}_{31})] \mathcal{P}^{(2)}_{w_{31\parallel}}(s_{31\parallel}-r_{31\parallel}|\,
\check{\bm{r}}_{31}
)\, \rd r_{31\parallel} 
+ \int \left[1+\xi(\check{r}_{12})+\xi(\check{r}_{23})+\xi(\check{r}_{31})\right. \nonumber \\
&\quad\quad\quad\quad\quad\quad\left.+ \zeta(\check{r}_{12}, \check{r}_{23}, \check{r}_{31})\right] \mathcal{P}^{(3)}_{\bm{w}_\parallel}\left(s_{12\parallel}-r_{12\parallel},  s_{23\parallel}-r_{23\parallel} |\, \check{\bm{r}}_{12}, \check{\bm{r}}_{23}\right) \, \rd r_{12\parallel}\, \rd r_{23\parallel} \label{eq:3ptstreamrversion}
     \;.
\end{align}
Since $\xi(\check{r}_{12})$ does not depend on $w_{23\parallel}$,
$\xi(\check{r}_{23})$ does not depend on $w_{12\parallel}$, and
the term
$\xi(\check{r}_{31})$ in the last row is a function of $t_\parallel=-w_{12}-w_{23}$ but
does not depend on $p_\parallel=w_{12\parallel}-w_{23\parallel}$,
we can write
\begin{align}
-2&+\zeta_{\mathrm s}(\boldsymbol{s}_{12},\boldsymbol{s}_{23})=\nonumber\\
&\int \left[-2 +\zeta(\check{r}_{12},\check{r}_{23},\check{r}_{31})\right]\,
{\mathcal P}^{(3)}_{\bm{w}_{\parallel}}(w_{12\parallel}, w_{23\parallel} | \check{\boldsymbol{r}}_{12},\check{\boldsymbol{r}}_{23})\,\mathrm{d}w_{12\parallel}\,\mathrm{d}w_{23\parallel}\nonumber\\
&+\int \left[1+\xi(\check{r}_{12})\right]\,\left[
\int {\mathcal P}^{(3)}_{\bm{w}_\parallel}(w_{12\parallel}, w_{23\parallel} | \check{\boldsymbol{r}}_{12},\check{\boldsymbol{r}}_{23})\,\mathrm{d}w_{23\parallel}-
{\mathcal P}^{(2)}_{w_{12\parallel}}(w_{12\parallel}| \check{\boldsymbol{r}}_{12})\right]\,\mathrm{d}w_{12\parallel}\nonumber\\
&+\int \left[1+\xi(\check{r}_{23})\right]\,\left[\int {\mathcal P}^{(3)}_{\bm{w}_\parallel}(w_{12\parallel}, w_{23\parallel} | \check{\boldsymbol{r}}_{12},\check{\boldsymbol{r}}_{23})\,\mathrm{d}w_{12\parallel}-{\mathcal P}^{(2)}_{w_{23\parallel}}(w_{23\parallel}| \check{\boldsymbol{r}}_{23})\right]\,\mathrm{d}w_{23\parallel}\label{eq:3ptdfullexpanded}\\
&+\int \left[1+\xi(\check{r}_{31})\right]\,\left[ \frac{1}{2}\,\int {\mathcal P}^{(3)}_{\bm{w}_\parallel}\left(\frac{-t_\parallel+p_\parallel}{2}, \frac{-t_\parallel-p_\parallel}{2} | \check{\boldsymbol{r}}_{12},\check{\boldsymbol{r}}_{23}\right)\,\mathrm{d}p_\parallel
-{\mathcal P}^{(2)}_{t_\parallel}(t_\parallel| \check{\boldsymbol{r}}_{31})\right]\,\mathrm{d}t_\parallel\;,\nonumber
\end{align}
where we have changed the integration variables from $w_{12\parallel}$ and
$w_{23\parallel}$ to $t_\parallel$ and $p_\parallel$ in the last line. The asymmetry of this term reflects the fact that
we have picked $w_{12\parallel}$ and $w_{23\parallel}$ as
the independent variables for ${\mathcal P}^{(3)}_{\boldsymbol{w}_\parallel}$.

\subsection{The streaming model for the connected correlation functions}

The procedure discussed above can be iterated to write down the streaming model for the connected $n$PCFs. The course of action consists of three basic steps: (i) start by writing down
equation~(\ref{eq:general-streaming-model-npoint});
(ii) express ${\mathcal G}_n$ and ${\mathcal F}_n$ in terms of the connected functions of order 2 to $n$; (iii) recursively apply the streaming model for the connected functions of order $n-1$ to 2.

We now derive an alternative formulation of the streaming model that
only involves connected correlation functions.
In order to facilitate understanding, we first discuss 2-point statistics
and then generalise the derivation to $n$-point correlations.
\subsubsection{2-point statistics}
Our starting point is the introduction of the connected  2-point phase-space density $f_2^{(c)}=f_2-f_1f_1$. The corresponding
quantity in redshift-space is
\begin{equation}
g_2^{(c)}=\int f_2^{(c)}\,\delta_\mathrm{D}^{(1)}(w_\parallel-v_{\mathrm{B}\parallel}+v_{\mathrm{A}\parallel})\,\mathrm{d}w_\parallel\;,
\end{equation}
so that the 2PCF
\begin{equation}
   \xi_{\mathrm{s}}(s_\parallel,s_\perp)=\frac{1}{\bar{n}^2}\,\int g_2^{(c)}\, \mathrm{d}\bm{v}_\mathrm{A} \,\mathrm{d}\bm{v}_\mathrm{B}=\frac{1}{\bar{n}^2}\,
   \int f_2^{(c)}\,\,\delta_\mathrm{D}^{(1)}(w_\parallel-v_{\mathrm{B}\parallel}+v_{\mathrm{A}\parallel})\, \mathrm{d}\bm{v}_\mathrm{A} \,\mathrm{d}\bm{v}_\mathrm{B}\;.
\end{equation}
We then multiply the integrand on the rhs by $\bar{n}^2\xi/\int f_2^{(c)} \,\mathrm{d}\bm{v}_\mathrm{A} \,\mathrm{d}\bm{v}_\mathrm{B}$ which is always identical to one.
By rearranging the terms, we obtain
\begin{equation}
\xi_{\mathrm{s}}(s_\parallel,s_\perp)=\int \xi\left[\sqrt{(s_\parallel-w_\parallel)^2+s_\perp^2}\,\right]\,{\mathcal C}^{(2)}_{w_\parallel}\left[w_\parallel| \check{\bm{r}}\left(s_\parallel,s_\perp,w_\parallel\right)\right]\,\mathrm{d}w_\parallel\;,
\label{connstream}
\end{equation}
with
\begin{equation}
{\mathcal C}^{(2)}=\frac{\displaystyle \int f_2^{(c)}\,\delta_\mathrm{D}^{(1)}(w_\parallel-v_{\mathrm{B}\parallel}+v_{\mathrm{A}\parallel})\,\mathrm{d}\bm{v}_\mathrm{A} \,\mathrm{d}\bm{v}_\mathrm{B}}{\displaystyle \int f_2^{(c)} \,\mathrm{d}\bm{v}_\mathrm{A} \,\mathrm{d}\bm{v}_\mathrm{B}}=
\frac{\displaystyle \int f_2^{(c)}\,\delta_\mathrm{D}^{(1)}(w_\parallel-v_{\mathrm{B}\parallel}+v_{\mathrm{A}\parallel})\,\mathrm{d}\bm{v}_\mathrm{A} \,\mathrm{d}\bm{v}_\mathrm{B}}{\bar{n}^2\,\xi}\;.
\label{eq:C2def}
\end{equation}
Equation~(\ref{connstream}) embodies the streaming model for the connected part of the 2PCF. Here ${\mathcal C}^{(2)}$
accounts for the relative los velocity between particles forming  `correlated pairs'.
In order to better grasp its meaning,
we replace $f_2^{(c)}=f_2-f_1\,f_1$ in equation~(\ref{eq:C2def})
and express ${\mathcal C^{(2)}}$ in terms of 
${\mathcal P}^{(2)}_{w_\parallel}$ to obtain
\begin{align}
{\mathcal C}^{(2)}=\frac{(1+\xi)\,{\mathcal P}^{(2)}_{w_\parallel}-{\mathcal R}_{w_\parallel}^{(2)}}{\xi}\;,
\label{newpdf}
\end{align}
with
\begin{align}
{\mathcal R}_{w_\parallel}^{(2)}&=
    \frac{\displaystyle \int f_1\,f_1\,\delta_\mathrm{D}^{(1)}(w_\parallel-v_{\mathrm{B}\parallel}+v_{\mathrm{A}\parallel})\,\mathrm{d}\bm{v}_\mathrm{A} \,\mathrm{d}\bm{v}_\mathrm{B}}{\bar{n}^2}\nonumber\\
    &=\int {\mathcal P}_{\bm{v}_\mathrm{A}}^{(1)}\,{\mathcal P}_{\bm{v}_\mathrm{B}}^{(1)}\,\delta_\mathrm{D}^{(1)}(w_\parallel-v_{\mathrm{B}\parallel}+v_{\mathrm{A}\parallel})\,\mathrm{d}\bm{v}_\mathrm{A} \,\mathrm{d}\bm{v}_\mathrm{B}
    \;.
    \label{eq:R2def}
\end{align}
As discussed in section~\ref{sec:fnintro},
due to statistical homogeneity,
${\mathcal P}_{\bm{v}_\mathrm{A}}^{(1)}$ and ${\mathcal P}_{\bm{v}_\mathrm{B}}^{(1)}$ assume the same
functional form.
Let us denote by ${\mathcal P}_{v_{\parallel}}^{(1)}$ the PDF of the los velocity obtained marginalising ${\mathcal P}_{\bm{v}}^{(1)}$ over the two perpendicular directions.\footnote{Because of statistical isotropy, the PDF of the velocity components parallel to any axis must assume the same form.} Then, equation~(\ref{eq:R2def}) reduces to
\begin{align}
{\mathcal R}_{w_\parallel}^{(2)}(w_\parallel)=
\int {\mathcal P}_{v_{\parallel}}^{(1)}({v}_{\mathrm{A}\parallel})\,{\mathcal P}_{v_{\parallel}}^{(1)}(w_\parallel+v_{\mathrm{A}\parallel})\,\mathrm{d}{v}_{\mathrm{A}\parallel}\;.
    \label{eq:R2def1d}
\end{align}
While ${\mathcal P}^{(2)}_{w_\parallel}$ gives the PDF of
the relative los velocity between particles in all pairs with a given real-space separation, 
 ${\mathcal R}_{w_\parallel}^{(2)}$ is the
 distribution of $w_\parallel$ generated by sampling (allowing repetitions) two particles at random irrespective of their separation.
This provides an operational way to compute ${\mathcal C}^{(2)}$ from simulations.
Note that, although $\int {\mathcal C}^{(2)}\,\mathrm{d}w_{\parallel}=1$, ${\mathcal C}^{(2)}$ is not a PDF (this is why we do not write the subscript $w_\parallel$ for it) and can assume negative values.  
In brief, this function quantifies the excess (or defect) probability to get  pairs with a given $w_\parallel$ with respect to random.

By substituting equation~(\ref{newpdf}) into equation~(\ref{connstream}) we get
\begin{equation}
\xi_{\mathrm{s}}(s_\parallel,s_\perp)=\int [1+\xi(\check{r})]\,{\mathcal P}^{(2)}_{w_\parallel}(w_\parallel|
\check{\boldsymbol{r}}
)\,\mathrm{d}w_\parallel- \int {\mathcal R}_{w_\parallel}^{(2)}(w_\parallel)\,\rd w_\parallel\;,
\end{equation}
and, after integrating over $w_\parallel$, it is obvious that the second term is identically equal to one and that equation~(\ref{connstream}) is equivalent to the classic streaming model.

\subsubsection{3-point statistics}
The reasoning above can be generalised to $n$-point statistics.
After repeating the same basic steps, we obtain
\begin{equation}
    \mathcal{G}^{(c)}_n  = \int \mathcal{F}^{(c)}_n \, \mathcal{C}^{(n)}(w_{12\parallel},\dots,w_{mn\parallel})
    \, \rd w_{12\parallel} \ldots \rd w_{mn\parallel} \, ,
    \label{eq:general-streaming-model-connected}
\end{equation}
where 
\begin{align}
    \mathcal{C}^{(n)}
    =  \frac{
    \displaystyle \int  f_{n}^{(c)} \,
    \displaystyle \delta_{\rm D}^{(1)}(w_{12\parallel} - v_{\mathrm{A}_2\parallel}+v_{\mathrm{A}_1\parallel})
\ldots\delta_{\rm D}^{(1)}(w_{mn\parallel} - v_{\mathrm{A}_n\parallel}+v_{\mathrm{A}_m\parallel}) \, \rd \bm{v}_{\mathrm{A}_1}\ldots\rd\bm{v}_{\mathrm{A}_n}}{\displaystyle \bar{n}^n\,\mathcal{F}_n^{(c)}}\;.
\label{eq:cndef}
\end{align}
In particular, for $n=3$, we have
\begin{equation}
\zeta_{\mathrm s}(\boldsymbol{s}_{12},\boldsymbol{s}_{23})= \int
\zeta(\check{r}_{12},\check{r}_{23},\check{r}_{31})\,
{\mathcal C}^{(3)}(w_{12\parallel}, w_{23\parallel} | \check{\boldsymbol{r}}_{12},\check{\boldsymbol{r}}_{23})\,\mathrm{d}w_{12\parallel}\,\mathrm{d}w_{23\parallel} \, ,
\label{eq:3ptconnectedstream}
\end{equation}
where
\begin{align}
    {\mathcal C}^{(3)}=\frac{
    {\displaystyle \int f_3^{(c)}\,\delta_{\rm D}^{(1)}(w_{12\parallel} - v_{\mathrm{B}\parallel}+v_{\mathrm{A}\parallel})\,
\delta_{\rm D}^{(1)}(w_{23\parallel} - v_{\mathrm{C}\parallel}+v_{\mathrm{B}\parallel}) \, \rd \bm{v}_{\mathrm{A}}\,\rd\bm{v}_{\mathrm{B}}\,\rd\bm{v}_{\mathrm{C}}}}{\displaystyle \bar{n}^3\,\zeta}\;.
\label{eq:c3def}
\end{align}
Since $f_3^{(c)}=f_3-(f_2^{(c)}\,f_1+{\mathrm{symm.}})-f_1\,f_1\,f_1$ and
$f_2^{(c)}=f_2-f_1\,f_1$, it follows that
$f_3^{(c)}=f_3-(f_2\,f_1+{\mathrm{symm.}})+2\,f_1\,f_1\,f_1$.
It is thus convenient to re-write ${\mathcal C}^{(3)}$ as
\begin{align}
{\mathcal C}^{(3)}=
\frac{\left[1+\xi_{12}+\xi_{23}+\xi_{31}+\zeta\right]\,{\mathcal P}^{(3)}_{\bm{w}_\parallel}
-\left[(1+\xi_{12})
\,{\mathcal Q}^{(3{\mathrm{AB}})}_{{\boldsymbol w}_\parallel}
+{\mathrm{symm.}}\right]
+2\,{\mathcal R}^{(3)}_{\bm{w}_\parallel}}
{\zeta}\;.
\label{eq:c3defexp}
\end{align}
where we have used $\xi_{ij}$ as a short for $\xi(r_{ij})$ and
the PDFs ${\mathcal Q}^{(3{\mathrm{AB}})}_{{\boldsymbol w}_\parallel}$ and ${\mathcal R}_{{\boldsymbol w}_\parallel}$ are defined as:
\begin{align}
{\mathcal Q}^{(3{\mathrm{AB}})}_{{\boldsymbol w}_\parallel}&=
    \frac{\displaystyle\int f_2(\mathrm{A},\mathrm{B})\,
f_1(\mathrm{C})\,\delta_{\rm D}^{(1)}(w_{12\parallel} - v_{\mathrm{B}\parallel}+v_{\mathrm{A}\parallel})\,\delta_{\rm D}^{(1)}(w_{23\parallel}- v_{\mathrm{C}\parallel}+v_{\mathrm{B}\parallel})\, \rd \bm{v}_{\mathrm{A}}\,\rd\bm{v}_{\mathrm{B}}\,\rd\bm{v}_{\mathrm{C}}}{\bar{n}^3\,\xi_{12}} \nonumber\\
&=\displaystyle\int G(v_{\mathrm{A}\parallel},v_{\mathrm{A}\parallel}+w_{12}|\bm{r}_{\mathrm{12}})\,
{\mathcal P}_{v_{\parallel}}^{(1)}(v_{\mathrm{A}\parallel}+w_{12\parallel}+w_{23\parallel})\, \rd v_{\mathrm{A}\parallel}\;,
\end{align}
(the function $G$ is defined such that
${\mathcal P}_{w_\parallel}^{(2)}(w_{12}|\bm{r}_{\mathrm{12}})=\int G(v_{\mathrm{A}\parallel},v_{\mathrm{A}\parallel}+w_{12}|\bm{r}_{\mathrm{12}})\,\rd v_{\mathrm{A}\parallel}$)
and
\begin{align}
{\mathcal R}^{(3)}_{\boldsymbol{w}_\parallel}&=
 \frac{\displaystyle \int f_1(\mathrm{A})\,f_1(\mathrm{B})\,f_1(\mathrm{C})\,\delta_\mathrm{D}^{(1)}(w_{12\parallel}-v_{\mathrm{B}\parallel}+v_{\mathrm{A}\parallel})\,\delta_{\rm D}^{(1)}(w_{23\parallel} - v_{\mathrm{C}\parallel}+v_{\mathrm{B}\parallel})\,\mathrm{d}\bm{v}_\mathrm{A} \,\mathrm{d}\bm{v}_\mathrm{B}\,\rd\bm{v}_{\mathrm{C}}}{\bar{n}^3}\nonumber\\
 &= \int {\mathcal P}^{(1)}_{\bm{v}_\mathrm{A}}
\,{\mathcal P}^{(1)}_{\bm{v}_\mathrm{B}}
\,{\mathcal P}^{(1)}_{\bm{v}_\mathrm{C}}
\,\delta_\mathrm{D}^{(1)}(w_{12\parallel}-v_{\mathrm{B}\parallel}+v_{\mathrm{A}\parallel})\,\delta_{\rm D}^{(1)}(w_{23\parallel} - v_{\mathrm{C}\parallel}+v_{\mathrm{B}\parallel})\,\mathrm{d}\bm{v}_\mathrm{A} \,\mathrm{d}\bm{v}_\mathrm{B}\,\rd\bm{v}_{\mathrm{C}}\nonumber\\
&=
\int
{\mathcal P}_{v_{\parallel}}^{(1)}(v_{\mathrm{A}\parallel})\,
{\mathcal P}_{v_{\parallel}}^{(1)}(v_{\mathrm{A}\parallel}+w_{12\parallel})
\,{\mathcal P}_{v_{\parallel}}^{(1)}(v_{\mathrm{A}\parallel}+w_{12\parallel}+w_{23\parallel})\,\mathrm{d}v_{\mathrm{A}\parallel} \;.
\label{eq:r3_pdf}
\end{align}
In full analogy with the 2-point case, equation~(\ref{eq:c3def}) provides
an operational way to compute $\mathcal{C}^{(3)}$ in practice.
The first term on the rhs is proportional to $\mathcal{P}^{(3)}_{\bm{w}\parallel}$ and
thus represents the (rescaled) bivariate distribution of the relative los velocities in actual triplets of particles.
The next three terms are proportional to ${\mathcal Q}^{(3ij)}_{\boldsymbol{w}_\parallel}$ i.e. to 
the bivariate distribution of the relative los velocities in triplets that are formed by an actual pair of particles with a fixed separation $\bm{r}_{ij}$ and a third particle which is randomly selected (irrespective from its actual position). Finally,
the last term accounts for the contribution of fully random triplets.
Note that, by definition, $\int \mathcal{C}^{(3)}(w_{12\parallel}, w_{23\parallel} | \boldsymbol{r}_{12},\boldsymbol{r}_{23})\,\mathrm{d}w_{12\parallel}\,\mathrm{d}w_{23\parallel}=1$.
Substituting equation~(\ref{eq:c3defexp}) into (\ref{eq:3ptconnectedstream}) and taking into account that $\int {\mathcal R}^{(3)}_{\bm{w}_\parallel}\,\rd w_{12\parallel}\,\rd w_{23\parallel}=1$ gives back equation~(\ref{eq:connectedzetas})

\subsection{Collisionless systems}
\label{sec:collisionless}

So far we have considered the most general and complete description of an $N$-body system and our equations are exact. 
However, great simplifications are possible in particular cases. For instance, systems composed by very many particles interacting exclusively through long-range forces are conveniently
described by kinetic equations of the Jeans-Vlasov type.
This corresponds to neglecting two-body and higher-order velocity correlations, i.e. to assuming that
\begin{equation}
f_n(\boldsymbol{x}_{\mathrm{A}_1},\dots,\boldsymbol{x}_{\mathrm{A}_n},\bm{v}_{\mathrm{A}_1},\dots,\bm{v}_{\mathrm{A}_n})\propto
\langle \prod_{j=1}^n \bar{f}(\boldsymbol{x}_{\mathrm{A}_j},\bm{v}_{\mathrm{A}_j})\rangle\;,
\label{eq:vlasovclosure}
\end{equation}
where $\bar{f}$ denotes the macroscopic coarse-grained phase-space
density that satisfies Vlasov equation.
The approximation holds true for time scales comparable to the collision time.
Since dark-matter particles form a collisionless system for the entire life of the Universe,
equation~(\ref{eq:vlasovclosure}) is often implicitly assumed in the cosmological literature. 
In order to compare our results with previous work, we recast our equations in terms of $\bar{f}$. 
After introducing the mass density contrast $\delta(\bm{x})$,
in the single-stream regime, we can write
$\bar{f}(\bm{x},\bm{v})=\bar{n}\,[1+\delta(\bm{x})]\,\delta_{\mathrm D}^{(3)}[\bm{v}-\mathsf{v}(\bm{x})]$ (where $\mathsf{v}(\bm{x})$ denotes the continuous velocity field), while
$\bar{f}(\bm{x},\bm{v})=\bar{n}\,[1+\delta(\bm{x})]\,F_{\bm{v}}(\bm{x},\bm{v})$ with $\int F_{\bm{v}}(\bm{x},\bm{v})\,\rd\bm{v}=1$
in the multi-stream case. Therefore,\footnote{To simplify the notation, from now on we use the symbols $\bm{x}_i$ and $\bm{v}_i$ to indicate generic
positions and velocities. This differs from 
section~\ref{sec:fnintro} where we used the same symbols
to indicate the location and velocity of the $i^{\rm th}$ particle.}
\begin{align}
    \mathcal{P}^{(2)}_{w_\parallel}(w_{12\parallel}| \bm{x}_2-\bm{x}_1)
    =  \frac{
    \langle [1+\delta(\bm{x}_1)] \,[1+\delta(\bm{x}_2)]\,\mathcal{K}^{(2)}(w_{12\parallel},\bm{x}_1,\bm{x}_2)\rangle} 
    {1+\xi(r)}\;,
\label{eq:p2collisionless}
\end{align}
where\footnote{In the full solution, the term $F_{\bm{v}_1}F_{\bm{v}_2}\,$ should be replaced with
$F_{\bm{v}_1}F_{\bm{v}_2}+G_{\bm{v}_1,\bm{v}_2}$ where the function $G_{\bm{v}_1,\bm{v}_2}$ accounts for velocity correlations. }
\begin{equation}
\mathcal{K}^{(2)}(w_{12\parallel},\bm{x}_1,\bm{x}_2)=\int F_{\bm{v}_1}(\bm{x}_1,\bm{v}_1)
\,F_{\bm{v}_2}(\bm{x}_2,\bm{v}_2)\,
\delta_{\rm D}^{(1)}(w_{12\parallel} - v_{2\parallel}+v_{1\parallel}) \, 
\rd \bm{v}_{1}\,\rd\bm{v}_{2}\;.    
\end{equation}
which, in the single-stream regime, reduces to
$ \mathcal{K}^{(2)}(w_{12\parallel},\bm{x}_1,\bm{x}_2)= 
\delta_{\rm D}^{(1)}[w_{12\parallel} - v_{\parallel}(\bm{x}_2)+v_{\parallel}(\bm{x}_1)]$.
By Fourier transforming $\mathcal{K}^{(2)}$, we obtain the characteristic function
\begin{align}
    \widetilde{\mathcal{P}}^{(2)}_{{w}_\parallel}(k| \bm{x}_2-\bm{x}_1)
    = 
     \frac{
    \langle [1+\delta(\bm{x}_1)] \,[1+\delta(\bm{x}_2)]\,\widetilde{\mathcal{K}}^{(2)}(k,\bm{x}_1,\bm{x}_2)\rangle} 
    {1+\xi(r)}\;,
\end{align}
with
\begin{equation}
\widetilde{\mathcal{K}}^{(2)}(k,\bm{x}_1,\bm{x}_2)=\int F_{\bm{v}_1}(\bm{x}_1,\bm{v}_1)
\,F_{\bm{v}_2}(\bm{x}_2,\bm{v}_2)\,
e^{ik(v_{\mathrm{A}_2\parallel}+v_{\mathrm{A}_1\parallel})} \,
\rd \bm{v}_{1}\,\rd\bm{v}_{2}\;.    
\end{equation}
The pairwise-velocity distribution is therefore fully determined
by the so-called\footnote{With an abuse of notation due to the fact that it was originally derived assuming a single-stream fluid \cite{Scoccimarro04}, equation~(\ref{eq:charfun}) is usually written as $1+{\mathcal M}(J,\bm{x}_2-\bm{x}_1)=\langle [1+\delta(\bm{x}_1)]\,[1+\delta(\bm{x}_2)]\, e^{iJv_{21\parallel}}\rangle$ \cite[e.g.][]{VlahWhite18}.} `moment generating function'
(which is actually a characteristic function)
\begin{equation}
1+{\mathcal M}(J,\bm{x}_2-\bm{x}_1)=\langle [1+\delta(\bm{x}_1)]\,[1+\delta(\bm{x}_2)]\,\widetilde{\mathcal{K}}^{(2)}(J,\bm{x}_1,\bm{x}_2)\rangle \, .
\label{eq:charfun}
\end{equation}
The streaming model for the 2PCF can be derived by applying a cumulant expansion to it, i.e. by expanding
$\ln [1+{\mathcal M}(J,\bm{x}_2-\bm{x}_1)]$ in $J$
\cite{Scoccimarro04, VlahWhite18}.
This approach has been generalised to 3-point statistics
in \cite{VlahWhite18}. 
Their equation~7.7 provides the Fourier-space version of the streaming model. Compared to our equation~(\ref{eq:connectedzetas}) in real space, their expression is missing several terms. This difference stems from the incorrect assumption that
the Fourier transform of $\langle[1+\delta(\bm{x}_1)]\,
[1+\delta(\bm{x}_2)]\,[1+\delta(\bm{x}_3)] \rangle$ gives the bispectrum, i.e. the full 3PCF has been replaced with its connected part in \cite{VlahWhite18}.
Note that,
\begin{align}
    \mathcal{P}^{(3)}_{\bm{w}_\parallel}(w_{12\parallel},w_{23\parallel}| \bm{r}_{12},\bm{r}_{23})
    =  \frac{
    \langle [1+\delta(\bm{x}_1)] \,[1+\delta(\bm{x}_2)]\,[1+\delta(\bm{x}_3)]\,\mathcal{K}^{(3)}(w_{12\parallel},w_{23\parallel},\bm{x}_1,\bm{x}_2,\bm{x}_3)\rangle} 
    {1+\xi(r_{12})+\xi(r_{23})+\xi(r_{31})+\zeta(r_{12},r_{23},r_{31})}\;,
\end{align}
where
\begin{align}
\mathcal{K}^{(3)}(w_{12\parallel},w_{23\parallel},\bm{x}_1,\bm{x}_2,\bm{x}_3)=\int F_{\bm{v}_1}(\bm{x}_1,\bm{v}_1)
\,&F_{\bm{v}_2}(\bm{x}_2,\bm{v}_2)\,F_{\bm{v}_3}(\bm{x}_3,\bm{v}_3)\,
\delta_{\rm D}^{(1)}(w_{12\parallel} - v_{2\parallel}+v_{1\parallel}) \nonumber\\
&\delta_{\rm D}^{(1)}(w_{23\parallel} - v_{3\parallel}+v_{2\parallel}) 
\, 
\rd \bm{v}_{1}\,\rd\bm{v}_{2}\,\rd\bm{v}_{3}\;.    
\end{align}
By direct integration, we find that 
\begin{equation}
\int \mathcal{K}^{(3)}(w_{12\parallel},w_{23\parallel},\bm{x}_1,\bm{x}_2,\bm{x}_3)\,\rd w_{12\parallel}\,\rd\bm{v}_3=
\mathcal{K}^{(2)}(w_{12\parallel},\bm{x}_1,\bm{x}_2)\;,
\end{equation}
and
\begin{align}
&\int {\mathcal P}^{(3)}_{\bm{w}_\parallel}(w_\parallel, q_\parallel | \boldsymbol{r}_{12},\boldsymbol{r}_{23})\,\mathrm{d}q_\parallel-{\mathcal P}^{(2)}_{w_\parallel}(w_\parallel| \boldsymbol{r}_{12})= \\
&\left\langle\left\{
\frac{
     [1+\delta(\bm{x}_1)] \,[1+\delta(\bm{x}_2)]\,[1+\delta(\bm{x}_3)]}
    {1+\xi(r_{12})+\xi(r_{23})+\xi(r_{31})+\zeta(r_{12},r_{23},r_{31})}
    -\frac{
 [1+\delta(\bm{x}_1)] \,[1+\delta(\bm{x}_2)]} 
    {1+\xi(r_{12})}\right\}\,
    \mathcal{K}^{(2)}(w_{12\parallel},\bm{x}_1,\bm{x}_2)\right\rangle
    \nonumber
\;,
\end{align}
which gives the difference between the triplet weighted and the pair weighted averages
of the function $\mathcal{K}^{(2)}(w_{12\parallel},\bm{x}_1,\bm{x}_2)$
and does not necessarily vanish.

\section{\label{sec:triplewise}The joint distribution of pairwise velocities in a triplet}
The joint distribution of pairwise los velocities for a given triangle in configuration space, ${\mathcal P}_{\boldsymbol{w}_\parallel}^{(3)}(w_{12\parallel}, w_{23\parallel} | \boldsymbol{r}_{12},\boldsymbol{r}_{23})\equiv {\mathcal P}^{(3)}_{\boldsymbol{w}_\parallel}(w_{12\parallel}, w_{23\parallel} | \triangle_{123})$,
is a central quantity in the streaming model for the 3PCF. In this section, we use a large $N$-body simulation and perturbative techniques to study its properties.

\subsection{\label{sec:sim}$N$-body simulation}
We use the public code \textsc{Gadget-2} \cite{Springel05} to simulate
the formation of the large-scale structure of the Universe within a periodic cubic box with a side of 1.2 $h^{-1} \mathrm{Gpc}$. 
We assume the base $\Lambda$CDM model that provides the best fit to the
2015 power spectra determined by the  \textsc{Planck} satellite in combination with lensing reconstruction and external data \cite{Planck15I}.
In brief, the flat background is characterised by the density parameters $\Omega_{\mathrm{m}} = 0.3089$ (total matter), $\Omega_{\Lambda} = 0.6911$ (cosmological constant), $\Omega_{\mathrm{b}} = 0.0486$ (baryonic matter) and by the present-day value of the Hubble parameter of $H_{0} \equiv H(z=0) = 100\,h \,\mathrm{km\,s}^{-1}\mathrm{Mpc}^{-1}$ with $h=0.6774$.
The primordial spectral index of the density perturbations
is $n_s = 0.9667$ and the
linear rms fluctuation measured in spheres of 8 $h^{-1} \mathrm{Mpc}$ is $\sigma_8 = 0.8159$.
The matter content of the simulation box is discretised into $1024^3$ identical particles, each with a mass of $M_{\mathrm{part}} = 1.379 \times 10^{11}\, h^{-1} \mathrm{M}_\odot$. 
The input linear power spectrum of the matter perturbations is obtained using the Code for Anisotropies in the Microwave Background (\textsc{camb}\footnote{camb.info}, \cite{CAMB}).
Gaussian initial conditions are generated at redshift $z=50$ according to second-order Lagrangian perturbation theory using the \textsc{Music} code \citep{MUSIC11}. 

\subsection{Basic properties of 
${\mathcal P}^{(3)} \texorpdfstring{_{\boldsymbol{w}_\parallel}}{_w}(w_{12\parallel}, w_{23\parallel} |\triangle_{123})$}

\begin{figure}[tbp]
    \centering
 	\includegraphics[scale=0.75]{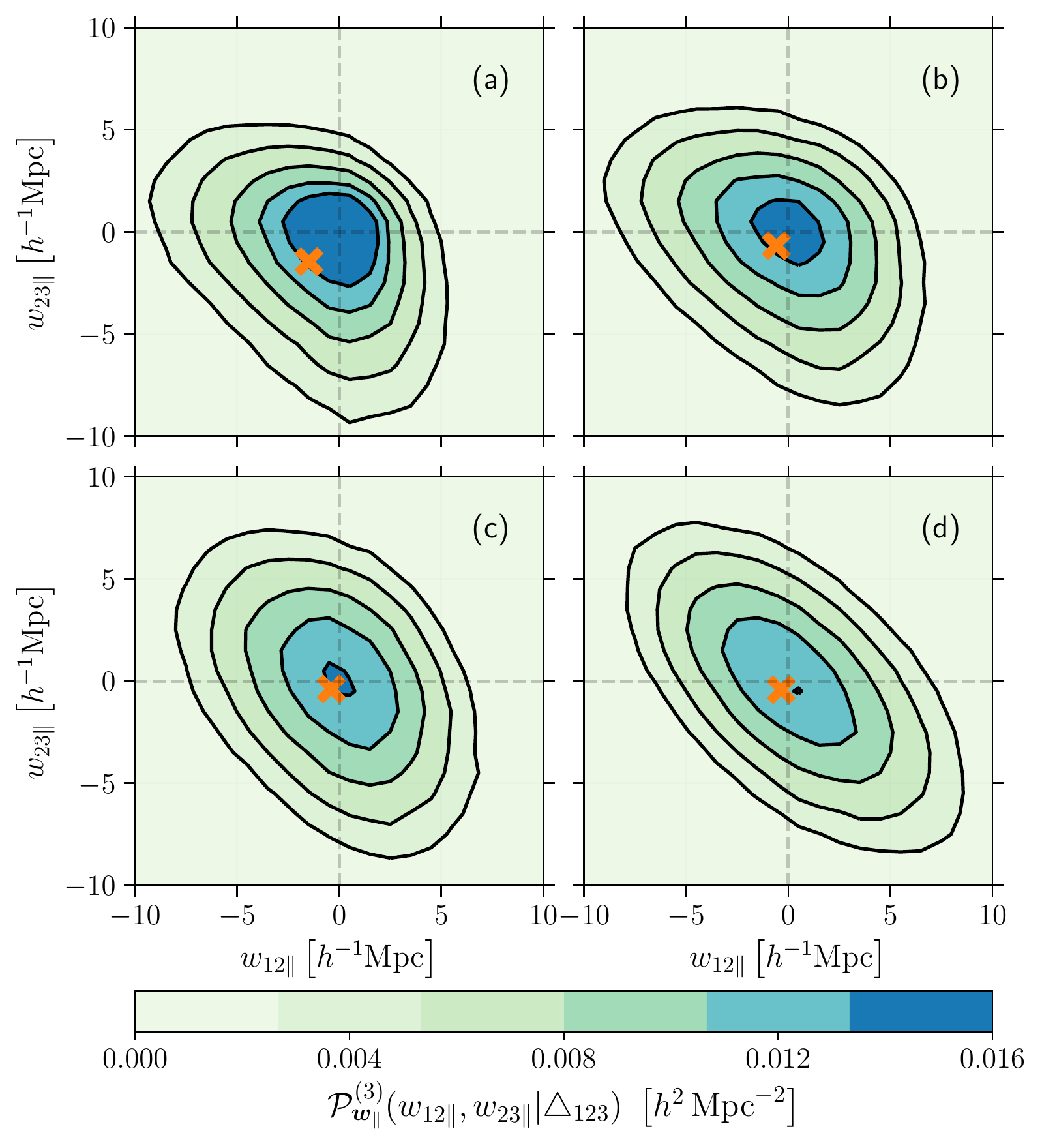}
    \caption{Contour levels for the joint probability distribution of the relative los velocities $w_{12\parallel}$ and $w_{23\parallel}$ extracted from our $N$-body simulation.
    The mean is indicated with a cross.
    The four panels correspond to different triangular configurations with  $\{r_{12\parallel},r_{12\perp},r_{23\parallel}, r_{23\perp}, r_{31\perp} \}$ lying
    within ($5 \,h^{-1}$Mpc wide) bins centred at
    $\{7.5, 7.5, 7.5, 7.5, 7.5\}_{(\mathrm{a})}$,
    $\{27.5, 17.5, 17.5, 17.5, 27.5\}_{(\mathrm{b})}$,
    $\{22.5, 32.5, 42.5, 32.5, 27.5\}_{(\mathrm{c})}$,
    $\{52.5, 47.5, 57.5, 42.5, 62.5\}_{(\mathrm{d})}$ in units of $h^{-1}$ Mpc.}
    \label{fig:equi}
\end{figure}

We measure 
${\mathcal P}^{(3)}_{\boldsymbol{w}_\parallel}(w_{12\parallel}, w_{23\parallel} |\triangle_{123})$ 
from the final output of our $N$-body
simulation at $z=0$.
This is a demanding task as it requires identifying all particle triplets
with a given $r_{12\parallel}, r_{12\perp}, r_{23\parallel}, r_{23\perp}$ and $r_{31\perp}$.
herefore, we consider a subsample of
$100^3$ randomly selected simulation particles.
Four examples are shown in figure~\ref{fig:equi}.
Note that the distribution is always unimodal with a mode which is close
to $(w_{12\parallel},w_{23\parallel})=(0,0)$. On the other hand, the mean los pairwise velocities (indicated with a cross in the plot) are negative.
In general, contour levels are not symmetric but tend to become elliptical for large separations.
The pairwise velocities $w_{12\parallel}$ and
$w_{23\parallel}$ anti-correlate on these scales due to the opposite sign of $v_{2\parallel}$ in their definition.

\begin{figure}
    \centering
 	\includegraphics[scale=0.75]{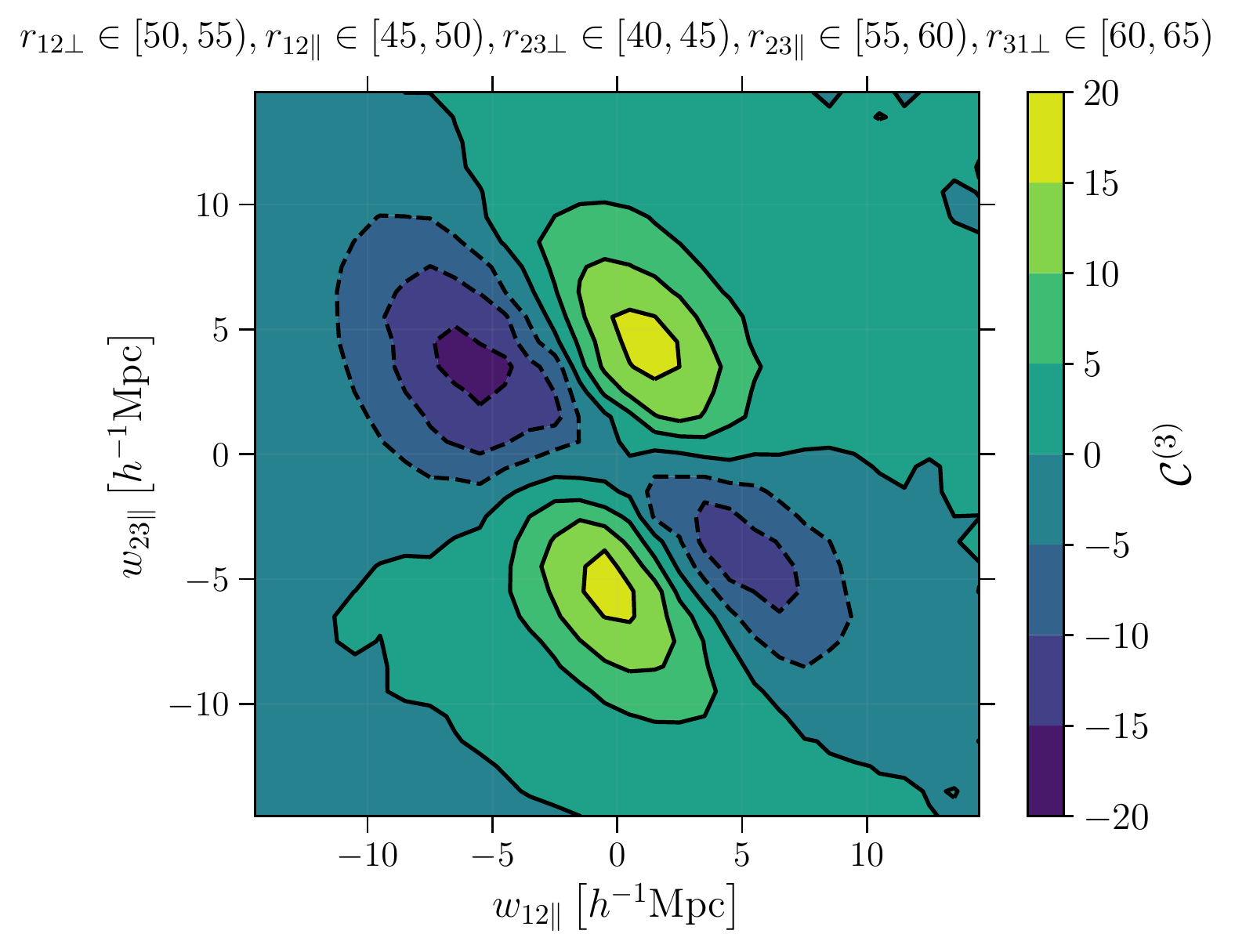}
    \caption{Contour levels for the function 
    ${\mathcal C}^{(3)}(w_{12\parallel},w_{23\parallel}|\triangle_{123})$ extracted from our $N$-body simulation by combining several PDFs as in equation~(\ref{eq:c3defexp}). The side lengths that define the specific triangular configuration we consider are listed on top of the figure in units of $h^{-1}\mathrm{Mpc}$.}
    \label{fig:connected_pdf}
\end{figure}

In figure~\ref{fig:connected_pdf}, we show one example of the function ${\mathcal C}^{(3)}(w_{12\parallel},w_{23\parallel}|\triangle_{123})$ for 
the same triangular configuration considered in the bottom-right panel of figure~\ref{fig:equi}. As expected,  ${\mathcal C}^{(3)}(w_{12\parallel},w_{23\parallel}|\triangle_{123})$ is more complex than the corresponding ${\mathcal P}^{(3)}_{\boldsymbol{w}_{\parallel}}(w_{12\parallel}, w_{23\parallel} |\triangle_{123})$.
The function ${\mathcal C}^{(3)}$ shows a typical quadrupolar structure with 
correlated steps in $w_{12\parallel}$ and $w_{23\parallel}$ giving a positive signal and anti-correlated ones producing a negative output value.
Figures~\ref{fig:equi} and \ref{fig:connected_pdf} suggest that
 ${\mathcal P}^{(3)}_{\boldsymbol{w}_{\parallel}}(w_{12\parallel}, w_{23\parallel} |\triangle_{123})$ is best suited for simple approximations in terms of analytical PDFs.
 We will pursue this phenomenological approach in section~\ref{sec:3ptGSM}.

\subsection{\label{sec:linear-theory}Moments of 
${\mathcal P}^{(3)}\texorpdfstring{_{\boldsymbol{w}_\parallel}}{_w}(w_{12\parallel}, w_{23\parallel} |\triangle_{123})$:
perturbative predictions at leading order}
In this section, we compute the first two moments of the joint distribution of $w_{12\parallel}$ and $w_{23\parallel}$
using standard perturbation theory at leading order (LO) and compare the results against our simulation. 

\subsubsection{Mean relative velocities between particle pairs in a triplet}
\label{sec:meanvel}
Standard perturbation theory assumes that
the matter content of the Universe is in the single-stream regime
and, at any given time, describes it in terms of two continuous fields:
the mass density contrast $\delta(\bm{x})$ and the peculiar velocity $\bm{u}(\bm{x})$. Linear perturbations in $\delta$ grow proportionally
to the growth factor $D$ while those in $\bm{u}$ grow proportionally to
$aHfD$ with $f=\rd \log D/\rd \log a$. The Fourier transforms of the linear terms are related as
\begin{align}
    \tilde{\bm{u}}(\bm{k})=aHf\,\frac{i\bm{k}}{k^2}\,\tilde{\delta}(\bm{k})\;.
    \label{eq:vft}
\end{align}
To make equations shorter, we follow the notation introduced
in sections~\ref{sec:rsd} and \ref{sec:collisionless}
and describe peculiar
velocities in terms of the vector field $\mathsf{v}(\bm{x})=\bm{u}(\bm{x})/(aH)$, i.e. in terms of the comoving separation vector that gives rise to a Hubble velocity $\bm{u}$. However, we continue referring to $\mathsf{v}$ as a velocity.

Let us consider the mean pairwise (relative) velocity
\begin{equation}
 \langle \bm{w}_{12}|\bm{r}_{12}\rangle_{\mathrm p}=\int \bm{w}_{12}\,\mathcal{P}^{(2)}_{\bm{w}_{12}}(\bm{w}_{12}| \bm{r}_{12})\,\rd\bm{w}_{12}  \, ,
\end{equation}
where the subscript p indicates a pair-weighted average, i.e. an average taken over all particle pairs with separation $\bm{r}_{12}$, and $\mathcal{P}^{(2)}_{\bm{w}_{12}}$ generalises
equation~(\ref{eq:p2collisionless}) to the full vector $\bm{w}_{12}$.
In the single-stream regime,
since the number of particles at one location
is proportional to $1+\delta$ (see section~\ref{sec:collisionless}), we can write
\begin{align}
\langle \bm{w}_{12}|\bm{r}_{12} \rangle_{\mathrm p}  =\displaystyle \frac{\langle(1+\delta_{1})(1+\delta_{2})(\bm{v}_{2}-\bm{v}_{1})\rangle}{\langle(1+\delta_{1})(1+\delta_{2})\rangle}\;, 
\end{align}
where $\delta_i$ and $\bm{v}_i$ are short for $\delta(\bm{x}_i)$ and
$\mathsf{v}(\bm{x}_i)$.
At LO in the perturbations, 
$\langle \bm{w}_{12}|\bm{r}_{12} \rangle_{\mathrm p}  
 \simeq \langle \delta_{1}\bm{v}_{2} \rangle - \langle \delta_{2} \bm{v}_{1} \rangle $ and, making use of equation~(\ref{eq:vft}), it is 
 straightforward to show that
 \begin{equation}
    \langle \delta_{1}\bm{v}_{2} \rangle= -\frac{f}{2\pi^2} \, \hat{\bm{r}}_{12} \int_0^{\infty}\!\!\!\!  k \,j_1(k\,r_{12})\,P(k)\, \rd k\, ,
 \end{equation}
where $j_1(x) = \sin (x)/x^2- \cos (x)/x$, and $P(k)$ denotes the linear matter power spectrum.
Putting everything together, one obtains \cite{Fisher95}
\begin{equation}
\langle \bm{w}_{12}|\bm{r}_{12} \rangle_{\mathrm p}  
 \simeq \displaystyle - \frac{f}{\pi^2} \, \hat{\bm{r}}_{12} \int_0^{\infty}\!\!\!\!  k \,j_1(k\,r_{12})\,P(k)\, \rd k=\bar{w}(r_{12})\,\hat{\bm{r}}_{12}\, ,
 \label{eq:mean-radial-velocity}
\end{equation}
 where the symbol
$\simeq$ indicates that the expression has been truncated to LO.
Note that, because of gravity, the particles in a pair approach each other on average, i.e. $\bar{w}(r_{12})<0$.

We now want to generalise this calculation to particle triplets
with separations $\triangle_{123}=(\bm{r}_{12},\bm{r}_{23},\bm{r}_{31})$.
In this case, there are three mean relative velocities to consider:
$\langle \bm{w}_{12}|\triangle_{123} \rangle_\mathrm{t}$,
$\langle \bm{w}_{23}|\triangle_{123} \rangle_\mathrm{t}$,
and $\langle \bm{w}_{31}|\triangle_{123} \rangle_\mathrm{t}$
(the subscript t, here, denotes that averages are taken over all particle triplets with separations $\triangle_{123}$). For instance, to LO in the perturbations,
\begin{eqnarray}
\langle \bm{w}_{12}|\triangle_{123} \rangle_\mathrm{t}   = & \displaystyle \frac{\langle(1+\delta_{1})(1+\delta_{2}) (1+\delta_{3})(\bm{v}_{2}-\bm{v}_{1})\rangle}{\langle (1+\delta_{1})(1+\delta_{2}) (1+\delta_{3})\rangle} \nonumber \\
 \simeq & \langle \delta_{1}\bm{v}_{2} \rangle - \langle \delta_{2} \bm{v}_{1} \rangle + \langle \delta_{3}\bm{v}_{2} \rangle - \langle \delta_{3} \bm{v}_{1} \rangle\nonumber \\
 =&\bar{w}(r_{12})\,\hat{\bm{r}}_{12}-\frac{1}{2}\left[
 \bar{w}(r_{23})\,\hat{\bm{r}}_{23}+\bar{w}(r_{31})\,\hat{\bm{r}}_{31}\right]
 \, .
 \label{eq:mean-radial-velocity-three}
 \end{eqnarray}
Note that the mean relative velocity between a particle pair in a triplet is not purely radial but
 has also a transverse component in the plane of the triangle defined by the particles. This is generated by the gravitational influence of the third particle on the pair.
 In order to separate the radial and transverse components,
 let us first denote by $\chi= \arccos(\hat{\bm r}_{12} \cdot \hat{\bm r}_{23})$ 
 the (shortest) rotation angle from $\hat{\bm r}_{12}$ to $\hat{\bm r}_{23}$ 
 around the normal vector $\bm{n}=\hat{\bm{r}}_{12}\times\hat{\bm{r}}_{23}=\hat{\bm{n}}\,\sin \chi$ (with $0\leq \chi < \pi$ and $\sin \chi\geq 0$).
 We then build a right-handed Cartesian coordinate system with unit axes $\{\hat{\bm{r}}_{12},  \hat{\bm{t}},  \hat{\bm{n}}\}$ such that 
 $\hat{\bm{t}}=\hat{\bm{n}}\times \hat{\bm{r}}_{12}=(\hat{\bm{r}}_{23}-\cos\chi\,\hat{\bm{r}}_{12})/\sin\chi$
 (see also appendix A in \cite{YankelevichPorciani18}).
 By construction, $\hat{\bm{t}}$ lies in the plane of $\triangle_{123}$,
 is orthogonal to $\hat{\bm{r}}_{12}$, and always points towards
the half-plane that contains point 3 with respect to the $\hat{\bm{r}}_{12}$ direction.
 Since $\bm{r}_{12}+\bm{r}_{23}+\bm{r}_{31}=0$, it follows that
$\bm{r}_{31}\cdot \hat{\bm{r}}_{12}=-(r_{12}+r_{23}\,\cos \chi)$
and $\bm{r}_{31}\cdot \hat{\bm{t}}=-r_{23}\,\sin \chi$.
We can thus decompose the mean relative velocity between a particle pair in a triplet into its
radial and transverse components (by symmetry, there cannot be any component along $\hat{\bm{n}}$ as motions in the two vertical directions are equally likely)
\begin{align}
   \langle \bm{w}_{12}|\triangle_{123} \rangle_\mathrm{t}  & = \langle \bm{w}_{12}\cdot \hat{\bm{r}}_{12} |\triangle_{123} \rangle_{\mathrm t}\,\hat{\bm{r}}_{12} +
   \langle \bm{w}_{12}\cdot \hat{\bm{t}}|\triangle_{123} \rangle_{\mathrm t}\, \hat{\bm{t}}\nonumber\\
&   =R_{12}(\triangle_{123})\,\hat{\bm{r}}_{12}+T_{12}(\triangle_{123})\,\hat{\bm{t}}\;,
   \label{eq:decomposition}
\end{align}
obtaining
 \begin{equation}
 R_{12}(\triangle_{123})=
 \bar{w}(r_{12})-\frac{1}{2}\left[
 \bar{w}(r_{23})\,\cos \chi-\bar{w}(r_{31})\,\frac{r_{12}+r_{23}\cos\chi}{\sqrt{r_{12}^2+r_{23}^2+2r_{12}r_{23}\cos\chi}}\right]\;,
 \label{eq:R(triangle)}
 \end{equation}
 \begin{equation}
 T_{12}(\triangle_{123})=
 -\frac{1}{2}\left[
 \bar{w}(r_{23})-\bar{w}(r_{31})\,\frac{r_{23}}{\sqrt{r_{12}^2+r_{23}^2+2r_{12}r_{23}\cos\chi}}\right]\,\sin \chi\;,
 \label{eq:Ttriangle}
 \end{equation}
where we have parameterized the shape of $\triangle_{123}$ in terms of $r_{12}, r_{23}$ and
$\chi$ (since $r_{31}^2=r_{12}^2+r_{23}^2+2r_{12}r_{23}\cos \chi$ and $\sin \chi=\sqrt{1-\cos^2 \chi}$ it is straightforward to use the three side lengths instead).
Equations~(\ref{eq:R(triangle)}) and (\ref{eq:Ttriangle})  describe how the presence of the third particle influences the mean radial velocity in a pair and gives rise to a transverse component.
Depending on the exact geometrical configuration,  $R_{12}(\triangle_{123})$ can be larger or smaller than $\bar{w}(r_{12})$ and $T_{12}(\triangle_{123})$ positive
or negative. If $\bm{r}_{12}$ is the base of an isosceles triangle, for instance, then $R_{12}=\bar{w}(r_{12})+\bar{w}(r_{23})(r_{12}/r_{31})/2$ and $T_{12}=0$. This reflects the fact that the `gravitational pulls' due to the third particle add up to generate a larger relative velocity in the radial direction but exactly cancel out in the transverse one.
For equilateral triangles, this reduces to
$R_{12}=3\bar{w}(r_{12})/2$ and $T_{12}=0$.
Considering a degenerate triangle with $\chi=0$ 
gives $R_{12}=\bar{w}(r_{12})-[\bar{w}(r_{23})-\bar{w}(r_{31})]/2$ and $T_{12}=0$.
A note is in order here. Triangles with the same shape can have opposite orientations (intended as winding orders, i.e. signed areas of opposite signs) and both $\hat{\bm{t}}$ and $\hat{\bm{n}}$ flip sign if the winding order of $\triangle_{123}$ is switched (e.g. by reflecting the triangle with respect to $\bm{r}_{12}$). It follows that, if one disregards orientation and takes the average among all triangles with given side lenghts, then 
the transverse part of the mean relative velocity between a particle pair in a triplet is a null vector as triangles with opposite winding orders give identical contributions in opposite directions. As stated in equation~(\ref{eq:decomposition}),
with the term `transverse component'
we always refer to the projection along $\hat{\bm{t}}$ 
which does not vanish even when the average is taken irrespective of orientation.

The steps above can be repeated to decompose $\langle \bm{w}_{23}|\triangle_{123}\rangle_\mathrm{t}$ in its radial
and transverse parts.
In this case, we use a right-handed coordinate system with
unit axes $\{\hat{\bm{r}}_{23}, \hat{\bm{t}}', \hat{\bm{n}} \}$ 
where $\hat{\bm{t}}'=\hat{\bm{n}}\times \hat{\bm{r}}_{23}=(-\hat{\bm{r}}_{12}+\cos\chi\, \hat{\bm{r}}_{23})/\sin\chi$)
and write
$\langle \bm{w}_{23}|\triangle_{123} \rangle_\mathrm{t}=R_{23}(\triangle_{123})\,\hat{\bm{r}}_{23}+T_{23}(\triangle_{123})\,\hat{\bm{t}}'$.
The resulting radial and transverse components are, respectively,
\begin{equation}
 R_{23}(\triangle_{123})=
 \bar{w}(r_{23})-\frac{1}{2}\left[
 \bar{w}(r_{12})\,\cos \chi-\bar{w}(r_{31})\,\frac{r_{23}+r_{12}\cos\chi}{\sqrt{r_{12}^2+r_{23}^2+2r_{12}r_{23}\cos\chi}}\right]\;,
 \label{eq:R23(triangle)}
 \end{equation}
 \begin{equation}
 T_{23}(\triangle_{123})=
 \frac{1}{2}\left[
 \bar{w}(r_{12})-\bar{w}(r_{31})\,\frac{r_{12}}{\sqrt{r_{12}^2+r_{23}^2+2r_{12}r_{23}\cos\chi}}\right]\,\sin \chi\;.
 \label{eq:T23triangle}
 \end{equation}
\begin{figure}[tbp]
    \centering
 	\includegraphics[scale=0.7]{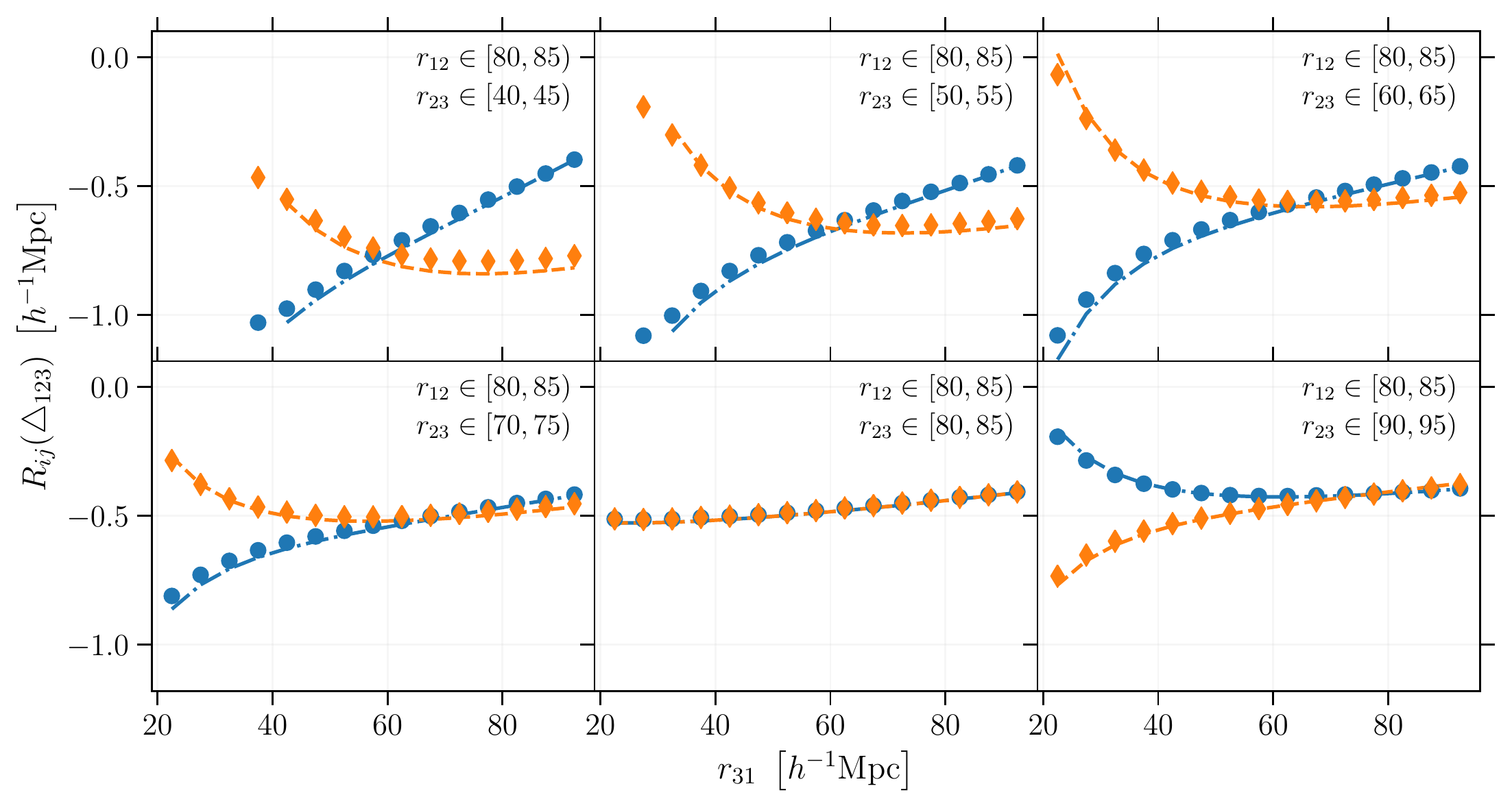}
 	\includegraphics[scale=0.7]{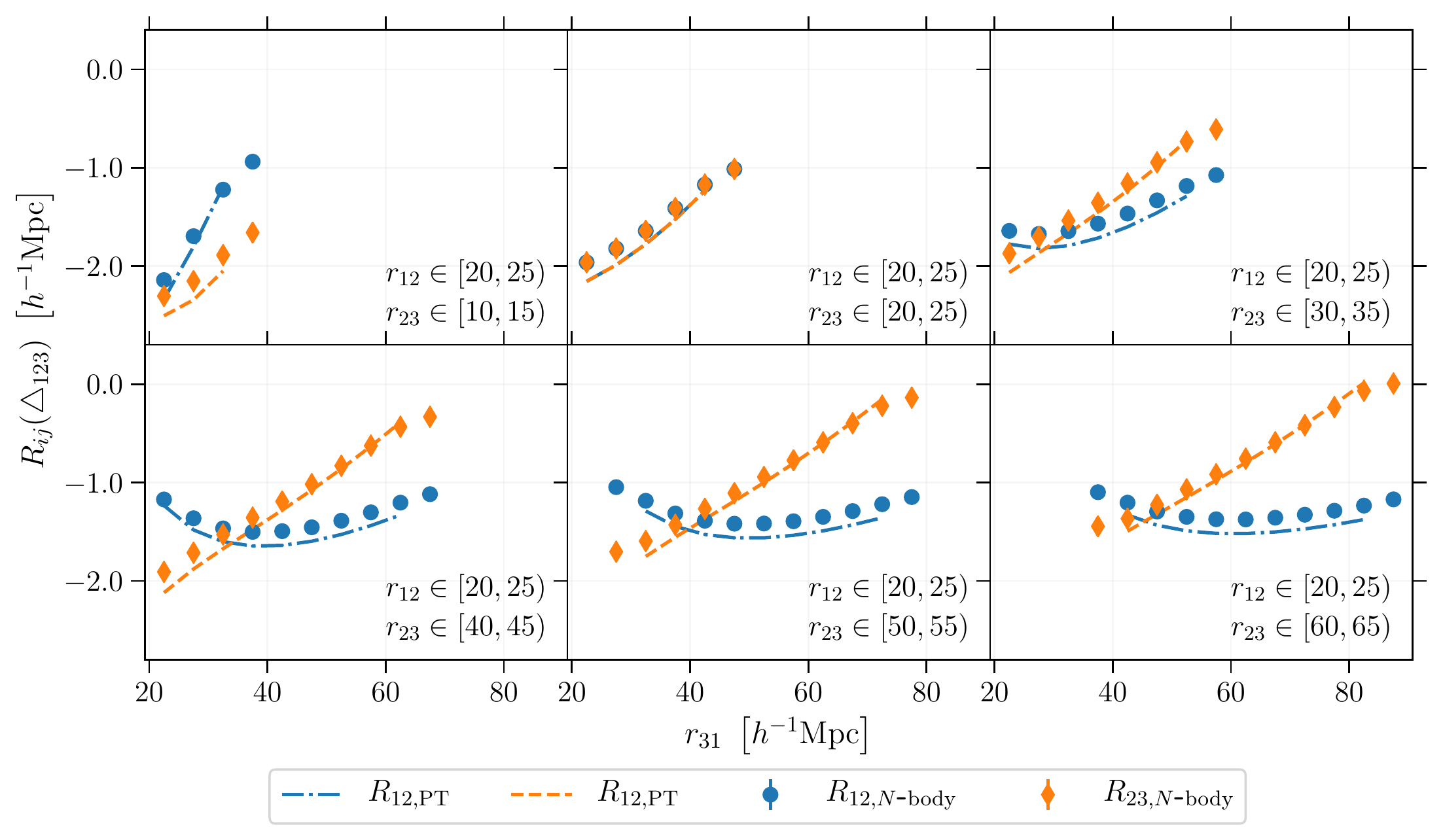}
    \caption{The radial component of the mean relative velocity between particle pairs in a triplet for different triangular configurations. Symbols with error bars denote measurements from our $N$-body simulation while the smooth curves show the predictions from the perturbative calculations at LO derived in section~\ref{sec:meanvel}. The labels give the particle separations $r_{12}$ and $r_{23}$ in units of $h^{-1} \mathrm{Mpc}$.
    }   
\label{fig:grid_radial}
\end{figure}
\begin{figure}[tbp]
    \centering
 	\includegraphics[scale=0.7]{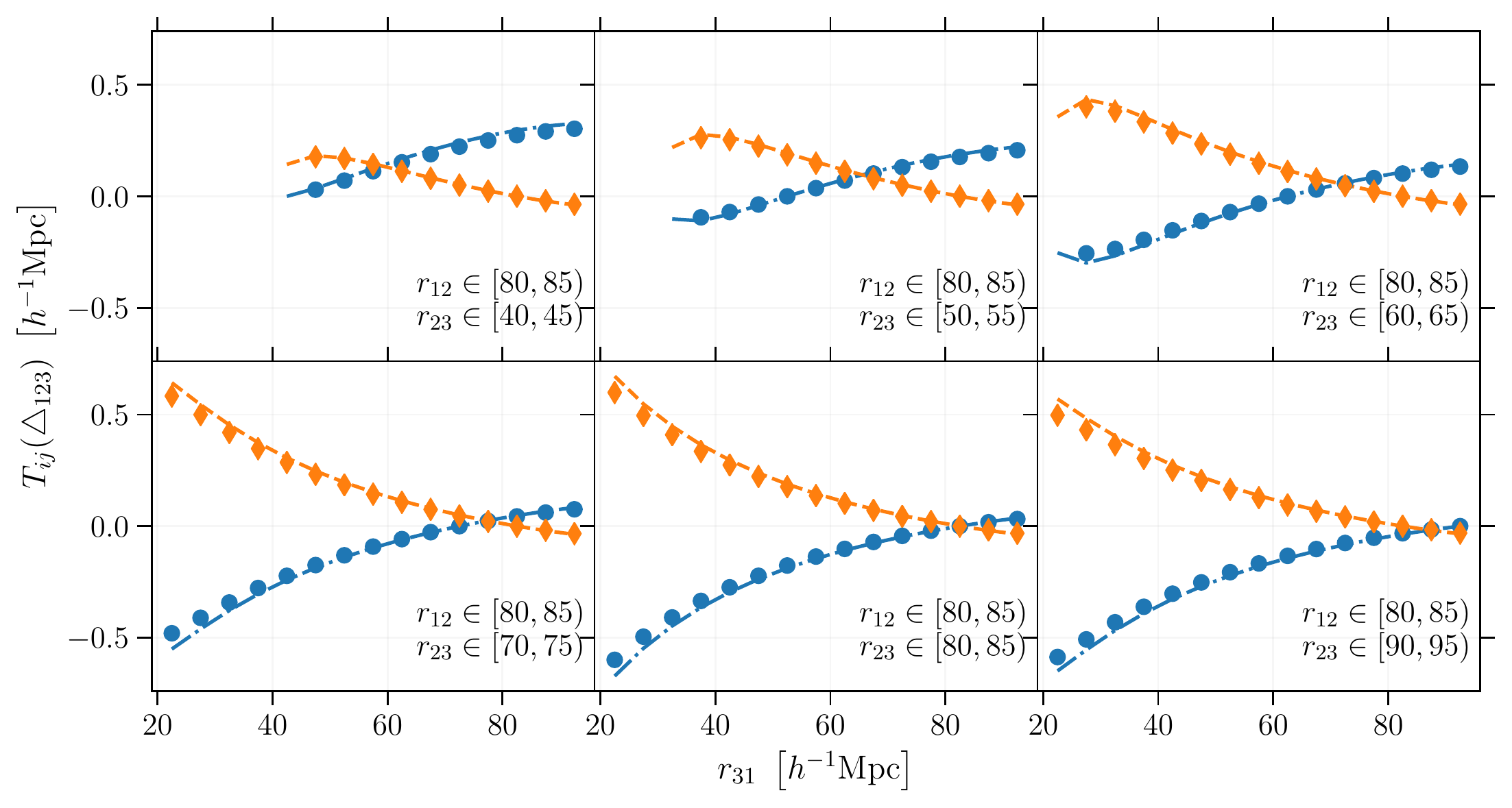}
 	\includegraphics[scale=0.7]{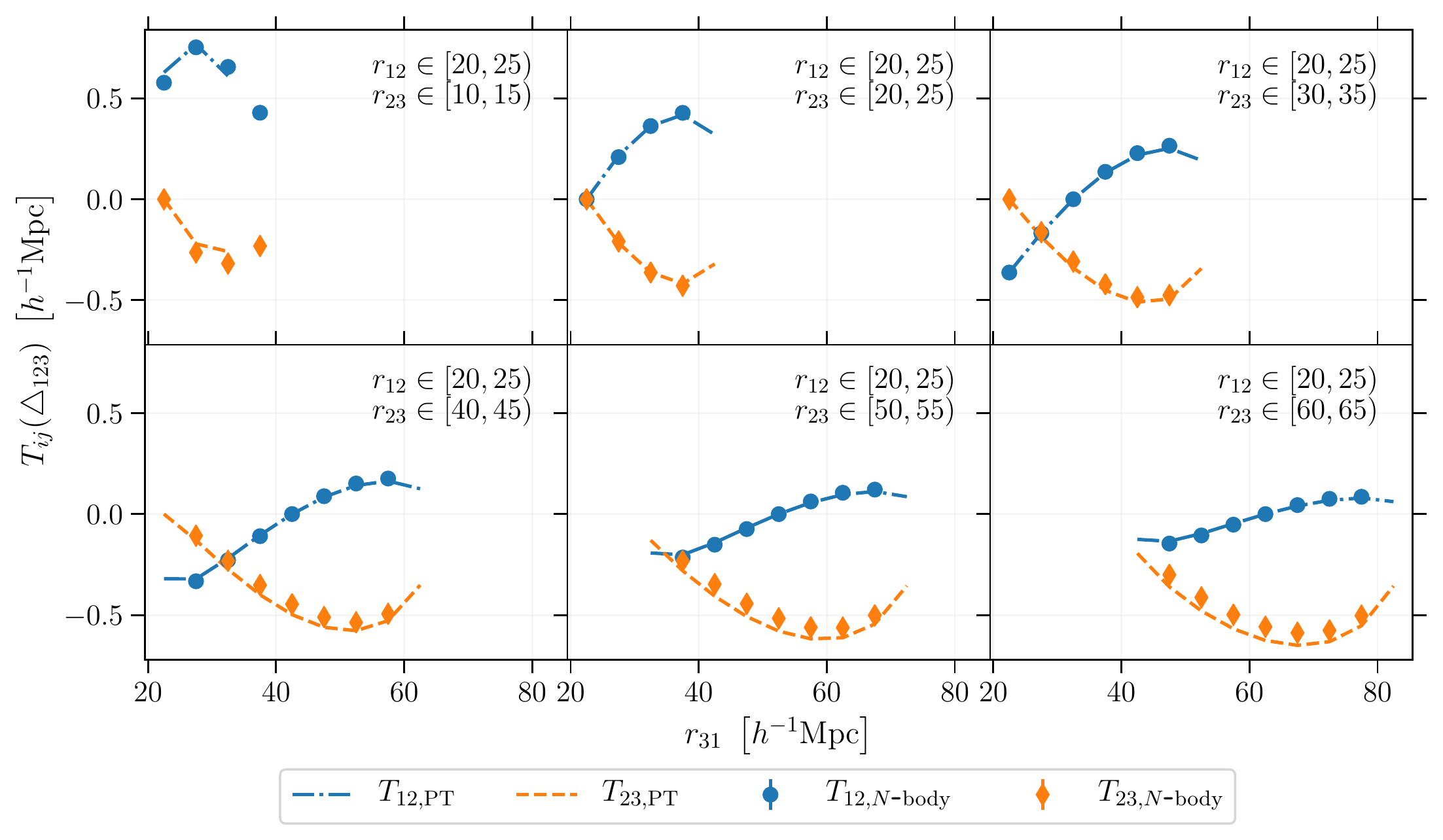}
    \caption{As in figure~\ref{fig:grid_radial} but for the transverse component.}
\label{fig:grid_transverse}
\end{figure}

In figures~\ref{fig:grid_radial} and \ref{fig:grid_transverse}, we compare the perturbative results at LO for $R_{12}$, $R_{23}$, $T_{12}$ and $T_{23}$
against measurements from the simulation introduced in section \ref{sec:sim}.
We consider triangular configurations $\triangle_{123}$ with different shapes and sizes (but we always average over winding order). 
In the top set of panels, we look at triangles with relatively large
values of $r_{12}$ and $r_{23}$. Here, $r_{12}\in [80,85) \, h^{-1} \mathrm{Mpc}$ and each sub panel corresponds to a different narrow range for $r_{23}$ as indicated by the labels.
Results are plotted as a function of $r_{31}$ (i.e. by varying $\chi$).
It is remarkable to see that the theoretical predictions match very well the measurements from the simulation for these large triangles.
In the bottom set of panels, we consider smaller triangles with
$r_{12}\in [20,25) \, h^{-1} \mathrm{Mpc}$ and also smaller values
for $r_{23}$. Also in this case, the LO predictions are 
quite accurate although to a lesser degree than in the top panel.
We conclude that the perturbative calculations are a reliable
tool to compute the mean relative velocity for triangular configurations
with scales $r \gtrsim 20\, h^{-1} \mathrm{Mpc}$).

\subsubsection{Dispersion of relative velocities between particle pairs in a triplet}
\label{sec:disptrip}
The second moment of the pairwise velocity
\begin{align}
\langle \bm{w}_{12}\,\bm{w}_{12}|\bm{r}_{12} \rangle_\mathrm{p} =& \displaystyle \frac{\langle(1+\delta_{1})(1+\delta_{2}) (\bm{v}_{2}-\bm{v}_{1}) (\bm{v}_{2}-\bm{v}_{1})\rangle}{\langle (1+\delta_{1})(1+\delta_{2})\rangle} 
\end{align}
is a dyadic tensor which,  to LO in the perturbations, reduces to
\begin{align}
\langle \bm{w}_{12}\,\bm{w}_{12}|\bm{r}_{12} \rangle_\mathrm{p} \simeq
\langle \bm{v}_{2}\bm{v}_{2} \rangle
-\langle \bm{v}_{2}\bm{v}_{1} \rangle
-\langle \bm{v}_{1}\bm{v}_{2} \rangle
+\langle \bm{v}_{1}\bm{v}_{1} \rangle\;.
\end{align}
Two-point correlations between linear velocity fields are conveniently written as \cite{Gorski88}
\begin{equation}
\langle v_{1i}\,v_{2j} \rangle \simeq \psi_{p}(r_{12})\, \delta_{ij} + [\psi_{r}(r_{12})-\psi_{p}(r_{12})] \, \hat{r}_{12i}\,\hat{r}_{12j} \; , 
\label{eq:sigmaexpansion}
\end{equation}
where the indices $i$ and $j$ denote the Cartesian components of the velocities (i.e. they run from 1 to 3), $\delta_{ij}$ is the Kronecker symbol, and
$\psi_{r}$ and $\psi_{p}$ are the radial and transverse correlation functions defined as
\begin{equation}
\psi_{r}(r_{12}) = \displaystyle \frac{f^2}{2\pi^2} \int_0^{\infty}  \Biggl[j_0(k\,r_{12}) - 2\, \frac{j_1(k\,r_{12})}{k\,r_{12}}\Biggr]\, P(k)\,\rd k\;  ,
\end{equation}
\begin{equation}
\psi_{p}(r_{12}) =  \displaystyle  \frac{f^2}{2\pi^2} \int_0^{\infty}   \frac{j_1(k\,r_{12})}{k\,r_{12}}\,P(k) \,\rd k\; ,
\end{equation}
with $j_0(x) = \sin(x)/x$. 
Note that, when $r_{12}\to 0$, $\psi_p\to \sigma^2_v$ and
$\psi_r\to \sigma^2_v$ where
\begin{equation}
\sigma^2_v = \frac{f^2}{6\pi^2} \int_0^{\infty} \!\!\!\!P(k) \,\rd k
\label{eq:sigmav}
\end{equation}
is the one-dimensional linear velocity dispersion, i.e. $\sigma_v^2=\langle v_i^2\rangle$.
Therefore, the velocity dispersion tensor at zero lag is isotropic
\begin{equation}
    \langle v_{1i}\,v_{1j}\rangle=\langle v_{2i}\,v_{2j}\rangle=\sigma^2_v \,\delta_{ij}\;.
\end{equation}
It follows that
\begin{align}
\langle w_{12i}\,w_{12j}|\bm{r}_{12} \rangle_\mathrm{p} \simeq
2\left[\sigma_v^2-\psi_p(r_{12}) \right]\,\delta_{ij}-
2\left[\psi_r(r_{12})-\psi_p(r_{12})\right] \, \hat{r}_{12i}\,\hat{r}_{12j}\;.
\end{align}
In other words, the second moments of the radial component is
\begin{equation}
\langle (\bm{w}_{12}\cdot \hat{\bm{r}}_{12})^2|\bm{r}_{12} \rangle_\mathrm{p} = 2\left[\sigma^2_v - \psi_{r}(r_{12})\right] \; ,
\label{eq:pairwise-radial-dispersion}
\end{equation}
while for each of the perpendicular components
(e.g. those along the unit vectors $\hat{\bm{n}}$ and $\hat{\bm{t}}$ introduced in section~\ref{sec:meanvel})
we have
\begin{equation}
\frac{1}{2}\langle[\bm{w}_{12}-(\bm{w}_{12}\cdot \hat{\bm{r}}_{12})\hat{\bm{r}}_{12}]^2|\bm{r}_{12}  \rangle_{\mathrm p} = 2\left[\sigma^2_v - \psi_{p}(r_{12})\right] \;.
\label{eq:pairwisetransvdisp}
\end{equation}
Moreover, the different Cartesian components are uncorrelated.

The calculations above can be easily extended to particle pairs in a triplet. In this case, we are interested in two types of combinations, e.g. 
\begin{align}
\langle \bm{w}_{12}\,\bm{w}_{12}|\triangle_{123} \rangle_\mathrm{t} =& \displaystyle \frac{\langle(1+\delta_{1})(1+\delta_{2}) (1+\delta_{3})(\bm{v}_{2}-\bm{v}_{1}) (\bm{v}_{2}-\bm{v}_{1})\rangle}{\langle (1+\delta_{1})(1+\delta_{2}) (1+\delta_{3})\rangle} \;,
\end{align}
and
\begin{align}
\langle \bm{w}_{12}\,\bm{w}_{23}|\triangle_{123} \rangle_\mathrm{t} =& \displaystyle \frac{\langle(1+\delta_{1})(1+\delta_{2}) (1+\delta_{3})(\bm{v}_{2}-\bm{v}_{1}) (\bm{v}_{3}-\bm{v}_{2})\rangle}{\langle (1+\delta_{1})(1+\delta_{2}) (1+\delta_{3})\rangle}\;.
 \label{eq:dispersion_12}
 \end{align}
To LO in the perturbations, they reduce to
\begin{align}
\langle \bm{w}_{12}\,\bm{w}_{12}|\triangle_{123} \rangle_\mathrm{t} \simeq &
\langle \bm{v}_{2}\bm{v}_{2} \rangle
-\langle \bm{v}_{2}\bm{v}_{1} \rangle
-\langle \bm{v}_{1}\bm{v}_{2} \rangle
+\langle \bm{v}_{1}\bm{v}_{1} \rangle\;,
\end{align}
and
\begin{align}
\langle \bm{w}_{12}\,\bm{w}_{23}|\triangle_{123} \rangle_\mathrm{t} \simeq &
\langle \bm{v}_{2}\bm{v}_{3} \rangle
-\langle \bm{v}_{2}\bm{v}_{2} \rangle
-\langle \bm{v}_{1}\bm{v}_{3} \rangle
+\langle \bm{v}_{1}\bm{v}_{2} \rangle\;.
\label{eq:crossexpanded}
\end{align}
that have exactly the same structure as equation~(\ref{eq:sigmaexpansion}).
Therefore, we conclude that
\begin{align}
\langle w_{12i}\,w_{12j}|\triangle_{123} \rangle_\mathrm{t} \simeq
2\left[\sigma_v^2-\psi_p(r_{12}) \right]\,\delta_{ij}-
2\left[\psi_r(r_{12})-\psi_p(r_{12})\right] \, \hat{r}_{12i}\,\hat{r}_{12j}\;,
\label{eq:matrix1212}
\end{align}
and
\begin{align}
\langle w_{12i}\,w_{23j}|\triangle_{123} \rangle_\mathrm{t} \simeq &
\left[\psi_p(r_{12})+\psi_p(r_{23})-\psi_p(r_{31})-\sigma^2_v\right]\,\delta_{ij} \nonumber\\
&+\left[\psi_r(r_{12})-\psi_p(r_{12}) \right]
 \, \hat{r}_{12i}\,\hat{r}_{12j} \nonumber\\
 & +\left[\psi_r(r_{23})-\psi_p(r_{23}) \right]
 \, \hat{r}_{23i}\,\hat{r}_{23j}\nonumber\\
&-\left[\psi_r(r_{31})-\psi_p(r_{31}) \right]
 \, \hat{r}_{31i}\,\hat{r}_{31j} \;.
 \label{eq:matrix1223}
\end{align}

\begin{figure}[tbp]
    \centering
 	\includegraphics[scale=0.7]{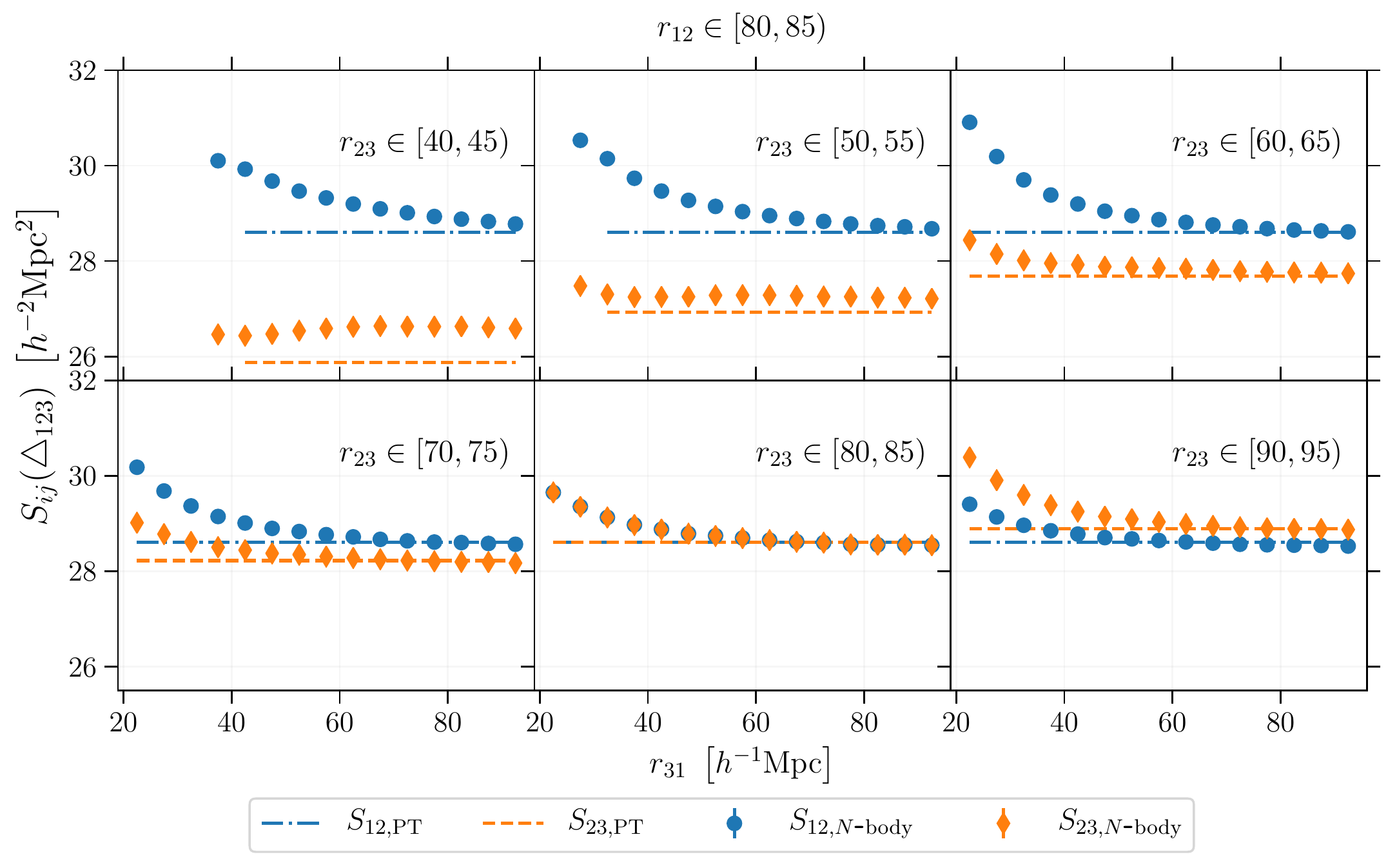}
 	\includegraphics[scale=0.7]{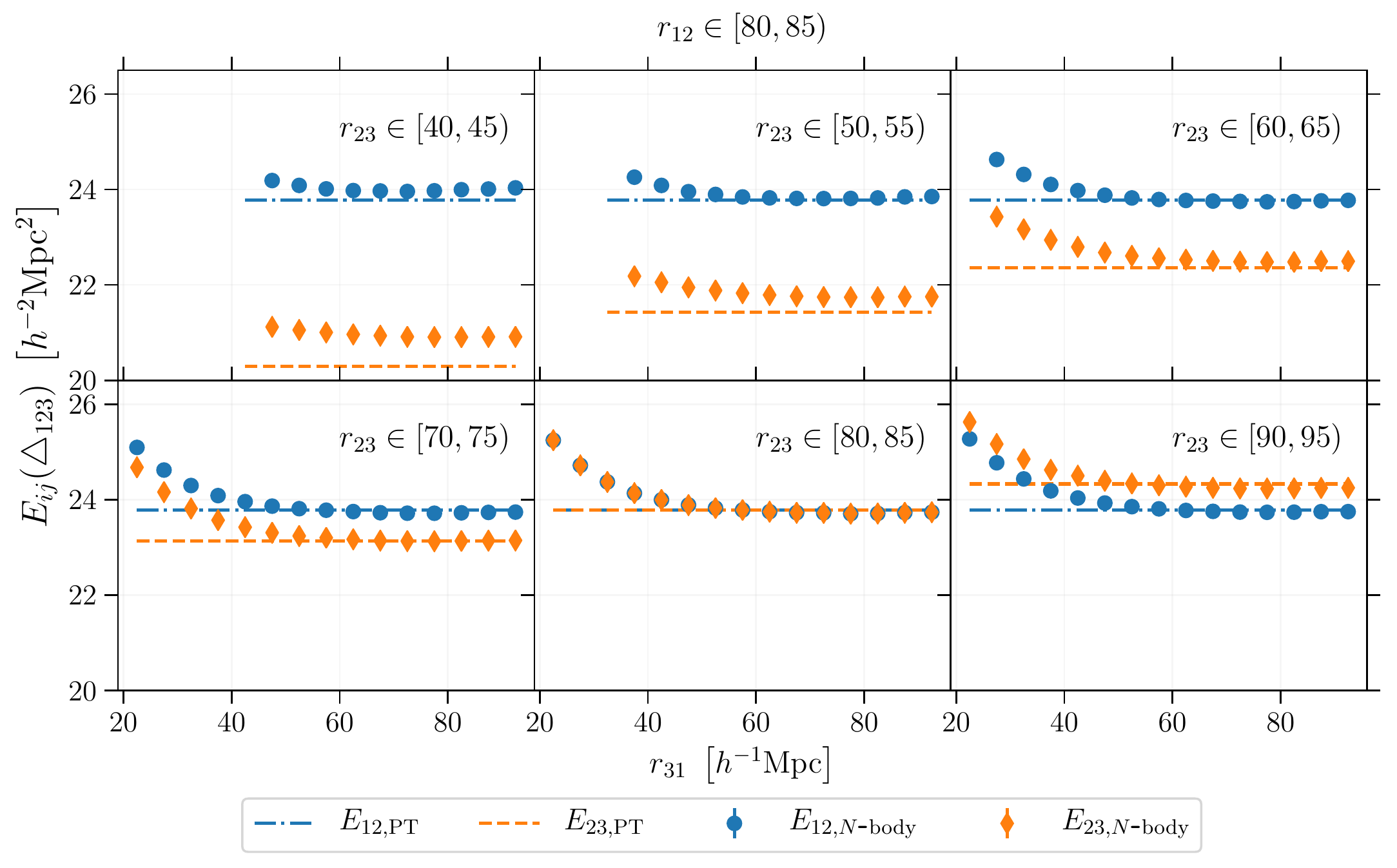}
    \caption{As in the top panel of figure~\ref{fig:grid_radial} but for the second moment of the radial (top) and transverse (bottom) components of the relative velocity between particle pairs in a triplet. Note that a constant offset has been added to the theoretical predctions as described at the end of  section~\ref{sec:disptrip}.}
\label{fig:grid_radial_std}
\end{figure}

In figure~\ref{fig:grid_radial_std}, we compare some of these perturbative results to measurements performed in our numerical simulation.
Shown are the second moments
of the radial (top panel) and transverse (bottom panel) components of the relative velocity between particle pairs in a triplet. Symbols with error bars display the $N$-body measurements while the constant lines
indicate the theoretical results to LO, i.e,
\begin{align}
S_{12}(\triangle_{123})=
\langle (\bm{w}_{12}\cdot \hat{\bm{r}}_{12})^2 |\triangle_{123} \rangle_\mathrm{t}&=2\left[\sigma^2_v-\psi_r(r_{12}) \right]\;,
\label{eq:sigmapar}\\
E_{12}(\triangle_{123})=\langle (\bm{w}_{12}\cdot \hat{\bm{t}})^2 |\triangle_{123} \rangle_\mathrm{t}&=2\left[\sigma^2_v-\psi_p(r_{12}) \right]\;,
\label{eq:sigmaperp}
\end{align}
and the corresponding results for $\bm{w}_{23}$. 
The first thing worth mentioning is that
the second moments are generally much larger
than the mean values shown in figures.~\ref{fig:grid_radial}
and \ref{fig:grid_transverse}.
The model, however, does not account for all the dispersion around the mean. In fact,
as previously noted in the literature \citep{ReidWhite11,Wang+14, Uhlemann+15}, the prediction for $\sigma^2_v$ given in equation~(\ref{eq:sigmav}) is not very accurate.
Being a zero-lag correlation, $\sigma^2_v$ is influenced by small-scale, non-perturbative physics. Adding a constant offset to 
equations~(\ref{eq:sigmapar}) and (\ref{eq:sigmaperp}) is a common fix that has been found to reproduce simulations well. We follow this approach and add a constant $C$ to $\sigma_v^2$ so that to match the measurements for the largest triangles we consider (i.e. the rightmost points in the bottom-right sub panels). 
This way, we find consistent offset values ($C\simeq 4.8 \,h^{-2} \mathrm{Mpc}^2$ within 1\%) for the dispersions in the radial and transverse components  
as well as in the pairwise velocity.
Keeping this shift fixed, we find that the theoretical predictions
are able to reproduce the measurements from the simulation quite well
for the largest triangles.
However, the level of agreement drops off rapidly when lower separation scales are considered.

\subsubsection{Projection along the line of sight}
\label{sec:losproj}
The los component of the relative velocities between particle pairs in a triplet
depends on the relative orientation of $\triangle_{123}$ with respect to the los (see Appendix A in \cite{YankelevichPorciani18} for a detailed discussion).
We set up a spherical coordinate system with $\hat{\bm{r}}_{12}$ as the polar axis and use $\theta = \arccos (\hat{\bm{r}}_{12}\cdot \hat{\bm{s}})$ as the polar angle ($0\leq \theta<\pi$). We also define the azimuthal angle $\phi$ ($0\leq \phi< 2\pi$) as the angle between $\hat{\bm{n}}$ and the projection of $\hat{\bm{s}}$ on to the plane perpendicular to $\hat{\bm{r}}_{12}$ so that
$\cos\phi=0$ whenever $\hat{\bm{s}}$ lies in the plane of the triangle.
It follows that
$\hat{\bm{t}}\cdot \hat{\bm{s}}=\sin \theta \sin \phi$,
$\hat{\bm{n}}\cdot \hat{\bm{s}}=\sin \theta \cos \phi$,
and $\hat{\bm{t}}'\cdot \hat{\bm{s}}=-\cos \theta \sin \chi+\sin \theta \sin \phi\cos\chi$.
For the scalar products between
the different pairwise separation vectors and the los direction, one thus finds \cite{SCF99, YankelevichPorciani18}, 
\begin{equation}
\mu_{12} = \hat{\bm{r}}_{12} \cdot \hat{\bm s} = \frac{r_{12\parallel}}{r_{12}} =  \cos\theta \; ,
\label{eq:mu12}
\end{equation}
\begin{equation}
\mu_{23} = \hat{\bm{r}}_{23} \cdot \hat{\bm  s} = \frac{r_{23\parallel}}{r_{23}} = \cos\theta\cos\chi + \sin\theta\sin\phi\sin\chi \; ,
\label{eq:mu23}
\end{equation}
\begin{equation}
\mu_{31} = \hat{\bm{r}}_{31} \cdot \hat{\bm s} = \frac{-(r_{12\parallel}+r_{23\parallel})}{r_{31}} = -\frac{r_{12}}{r_{31}}\mu_{12} - \frac{r_{23}}{r_{31}}\mu_{23}  \; .
\label{eq:mu31}
\end{equation}
Note that by flipping the winding order of $\triangle_{123}$ for a fixed los direction, $\cos\theta$ stays the same while both $\sin \phi$ and $\cos \phi$ change sign (i.e. $\phi \to \pi+\phi \mod 2\pi$) as $\hat{\bm{t}}$ and $\hat{\bm{n}}$ flip. 

Combining equations~(\ref{eq:mu12}), (\ref{eq:mu23}) and
(\ref{eq:mu31}) with the results obtained in section~\ref{sec:linear-theory}, we can eventually write the first and second moments for the projections of the relative velocities along the los, $w_{12\parallel}$ and $w_{23\parallel}$. In particular, 
equation~(\ref{eq:mean-radial-velocity-three}) gives
\begin{align}
\langle w_{12\parallel}|\triangle_{123}\rangle_\mathrm{t}\simeq
\bar{w}(r_{12})\,\mu_{12}-\frac{1}{2}\left[
 \bar{w}(r_{23})\,\mu_{23}+\bar{w}(r_{31})\,\mu_{31}\right]\;.
 \label{eq:meanw12final}
\end{align}
The very same expression can be derived from equation~(\ref{eq:decomposition})
and written as
\begin{align}
\langle w_{12\parallel}|\triangle_{123}\rangle_\mathrm{t}\simeq
R_{12}(\triangle_{123})\,\cos\theta+T_{12}(\triangle_{123})\,\sin\theta\sin\phi\;.
\end{align}
Similarly, we have
\begin{align}
\langle w_{23\parallel}|\triangle_{123}\rangle_\mathrm{t}\simeq
\bar{w}(r_{23})\,\mu_{23}-\frac{1}{2}\left[
 \bar{w}(r_{12})\,\mu_{12}+\bar{w}(r_{31})\,\mu_{31}\right]\;,
 \label{eq:meanw23final}
\end{align}
and
\begin{align}
\langle w_{23\parallel}|\triangle_{123}\rangle_\mathrm{t}\simeq
R_{23}(\triangle_{123})\,\mu_{23}+T_{23}(\triangle_{123})\,(-\cos\theta \sin \chi+\sin\theta\sin\phi\cos\chi)\;.
\end{align}
Moreover, from equation~(\ref{eq:matrix1212}) we derive
\begin{align}
\langle w_{12\parallel}^2|\triangle_{123}\rangle_\mathrm{t}&\simeq
2\left[\sigma_v^2-\psi_p(r_{12}) \right]-
2\left[\psi_r(r_{12})-\psi_p(r_{12})\right] \, \mu_{12}^2\nonumber\\
&=2\left[\sigma_v^2-\psi_\parallel(r_{12}) \right]\;,
\label{eq:dispw12final}
\end{align}
with $\psi_{\parallel}(r_{12}) = \mu^2_{12}\,\psi_{r}(r_{12}) + (1-\mu^2_{12})\,\psi_{p}(r_{12})$.
The corresponding expression for $\langle w_{23\parallel}^2|\triangle_{123}\rangle_\mathrm{t}$ is obtained
by replacing $r_{12}$ with $r_{23}$ in equation~(\ref{eq:dispw12final}).
Finally, equation~(\ref{eq:matrix1223}) implies
\begin{align}
\langle w_{12\parallel}\,w_{23\parallel}|\triangle_{123} \rangle_\mathrm{t} \simeq &
\left[\psi_p(r_{12})+\psi_p(r_{23})-\psi_p(r_{31})-\sigma^2_v\right]\,\nonumber\\
&+\left[\psi_r(r_{12})-\psi_p(r_{12}) \right]
 \, \mu_{12}^2 \nonumber\\
 & +\left[\psi_r(r_{23})-\psi_p(r_{23}) \right]
 \, \mu_{23}^2\nonumber\\
&-\left[\psi_r(r_{31})-\psi_p(r_{31}) \right]
 \, \mu_{31}^2 \nonumber\\
 =& \,\,\psi_{\parallel}(r_{12}) + \psi_{\parallel}(r_{23}) - \psi_{\parallel}(r_{31}) - \sigma^2_v 
 \;.
 \label{eq:w12w23final}
\end{align}
\begin{figure}
    \centering
    \includegraphics[scale=0.57]{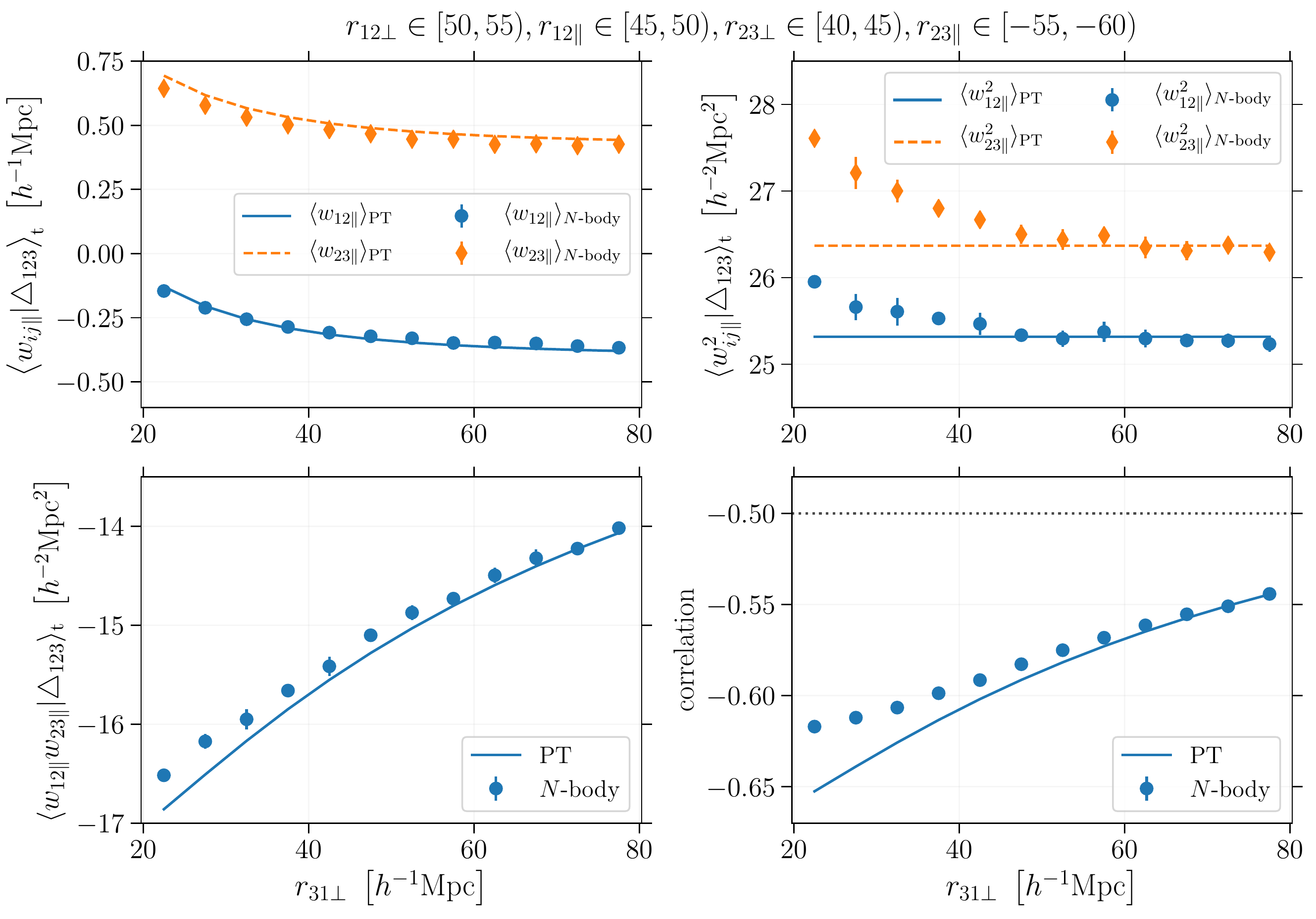}
    \caption{Moments of the relative los velocities between particle pairs in a triplet,
    $w_{12\parallel}$ and $w_{23\parallel}$.
    The mean values (top left),
    the second moments (top right), the second cross moment (bottom left), and the linear correlation coefficient (bottom right)
    are plotted for different triangular configurations $\triangle_{123}$. Symbols with error bars denote measurements from our $N$-body simulation while the smooth curves show the predictions from the perturbative calculations at LO derived in section~\ref{sec:losproj}. The labels give the particle separations in units of $h^{-1} \mathrm{Mpc}$.}
    \label{fig:cov-los}
\end{figure}
We compare these results with measurements from the simulation in figure~\ref{fig:cov-los}.
In this case, we bin our data based on the variables:
\begin{align}
r_{12\parallel}&=r_{12}\,\cos\theta\;,\nonumber\\
r_{12\perp}&=r_{12}\, |\sin\theta|\;, \nonumber \\
r_{23\parallel}&=r_{23}\,(\cos\theta\cos\chi + \sin\theta\sin\phi\sin\chi)\;, \label{eq:coordtransf}\\
r_{23\perp}&=r_{23}\,[1-(\cos\theta\cos\chi + \sin\theta\sin\phi\sin\chi)^2]^{1/2}\;, \nonumber \\
r_{31\perp}&= \left\{r_{12}^2+r_{23}^2+2r_{12}r_{23}\cos \chi- \left[r_{12}\cos\theta +r_{23}\,(\cos\theta\cos\chi + \sin\theta\sin\phi\sin\chi) \right]^2\right\}^{1/2}\;, \nonumber
\end{align}
(note that triangles with the same shape but opposite winding orders
correspond to different sets of these variables).
Results are plotted as a function of $r_{31\perp}$ by keeping the remaining four variables that define a triangular configuration fixed.
The top-left panel shows $\langle w_{12\parallel}|\triangle_{123}\rangle_\mathrm{t}$ and
$\langle w_{23\parallel}|\triangle_{123}\rangle_\mathrm{t}$.
Here, the theoretical predictions are in very good agreement with the numerical data confirming the results presented in 
figures~~\ref{fig:grid_radial} and \ref{fig:grid_transverse}. 
The top-right panel displays $\langle w_{12\parallel}^2|\triangle_{123}\rangle_\mathrm{t}$
 and $\langle w_{23\parallel}^2|\triangle_{123}\rangle_\mathrm{t}$ while the bottom-left panel shows $\langle w_{12\parallel}\,w_{23\parallel}|\triangle_{123} \rangle_\mathrm{t}$.
Once adjusted for the offset discussed in section~\ref{sec:disptrip},
the predictions for the second moments are excellent for
$r_{31\perp} \gtrsim 50\, h^{-1}\ \mathrm{Mpc}$ 
but tend to slightly underestimate the $N$-body results by a few percent at smaller separations. Likewise, the model for the cross second moment always agrees to better than 3\% with the simulation and gives better predictions when $r_{31\perp}$ is large.
Note that the linear correlation coefficient between
$w_{12\parallel}$ and $w_{23\parallel}$ (bottom-right panel) is always close
to $-1/2$ as expected from 
drawing independent los velocities from $\mathcal{P}^{(1)}_{v_\parallel}$ at every vertex of $\triangle_{123}$
(see also figure~\ref{fig:equi} and the detailed discussion in section~\ref{sec:discuss}).

\section{\label{sec:3ptGSM}The 3-point Gaussian streaming model}

\subsection{Definitions}
\label{sec:defin}

The streaming model for the 3PCF
given in equation~(\ref{eq:3ptstreaming_w}) is exact within the distant-observer approximation. However, it requires knowledge of
the function ${\mathcal P}^{(3)}_{\boldsymbol{w}_{\parallel}}(w_{12\parallel}, w_{23\parallel} |\triangle_{123})$
which is challenging to derive from 
first principles. In analogy to the literature on the 2-point
correlation function, we propose the use of a scale-dependent bivariate Gaussian distribution to model 
${\mathcal P}^{(3)}_{\boldsymbol{w}_{\parallel}}(w_{12\parallel}, w_{23\parallel} |\triangle_{123})$.
This choice is motivated by a number of considerations: 
i) For large inter-particle separations, the function 
${\mathcal P}^{(3)}_{\boldsymbol{w}_{\parallel}}(w_{12\parallel}, w_{23\parallel} |\triangle_{123})$
extracted from our simulation appears to be approximately Gaussian close to its peak (e.g. see the bottom right panel in figure~\ref{fig:equi});
ii) Simplicity, as the Gaussian is the only probability density function that only requires two cumulants to be fully specified; 
iii) As shown in section~\ref{sec:linear-theory}, on large scales, we can accurately model the scale dependence of these cumulants by using perturbation theory at LO.

In the resulting phenomenological model, which we dub
the `3-point Gaussian streaming model' (3ptGSM in short), the joint probability density function of $w_{12\parallel}$ and $w_{23\parallel}$ is given by a bivariate Gaussian distribution with mean values
$m_1=\langle w_{12\parallel}|\triangle_{123} \rangle_\mathrm{t}$,
$m_2=\langle w_{23\parallel}|\triangle_{123} \rangle_\mathrm{t}$
and covariance matrix with elements
$C_{11}=\langle w_{12\parallel}^2|\triangle_{123}\rangle_\mathrm{t}-\langle w_{12\parallel}|\triangle_{123} \rangle_\mathrm{t}^2$,
$C_{12}=C_{21}=\langle w_{12\parallel}\,w_{23\parallel}|\triangle_{123}\rangle_\mathrm{t}-\langle w_{12\parallel}|\triangle_{123} \rangle_\mathrm{t}\langle w_{23\parallel}|\triangle_{123} \rangle_\mathrm{t}$,
$C_{22}=\langle w_{23\parallel}^2|\triangle_{123}\rangle_\mathrm{t}-\langle w_{23\parallel}|\triangle_{123} \rangle_\mathrm{t}^2$.

\begin{figure}
    \centering
    \includegraphics[scale=0.8]{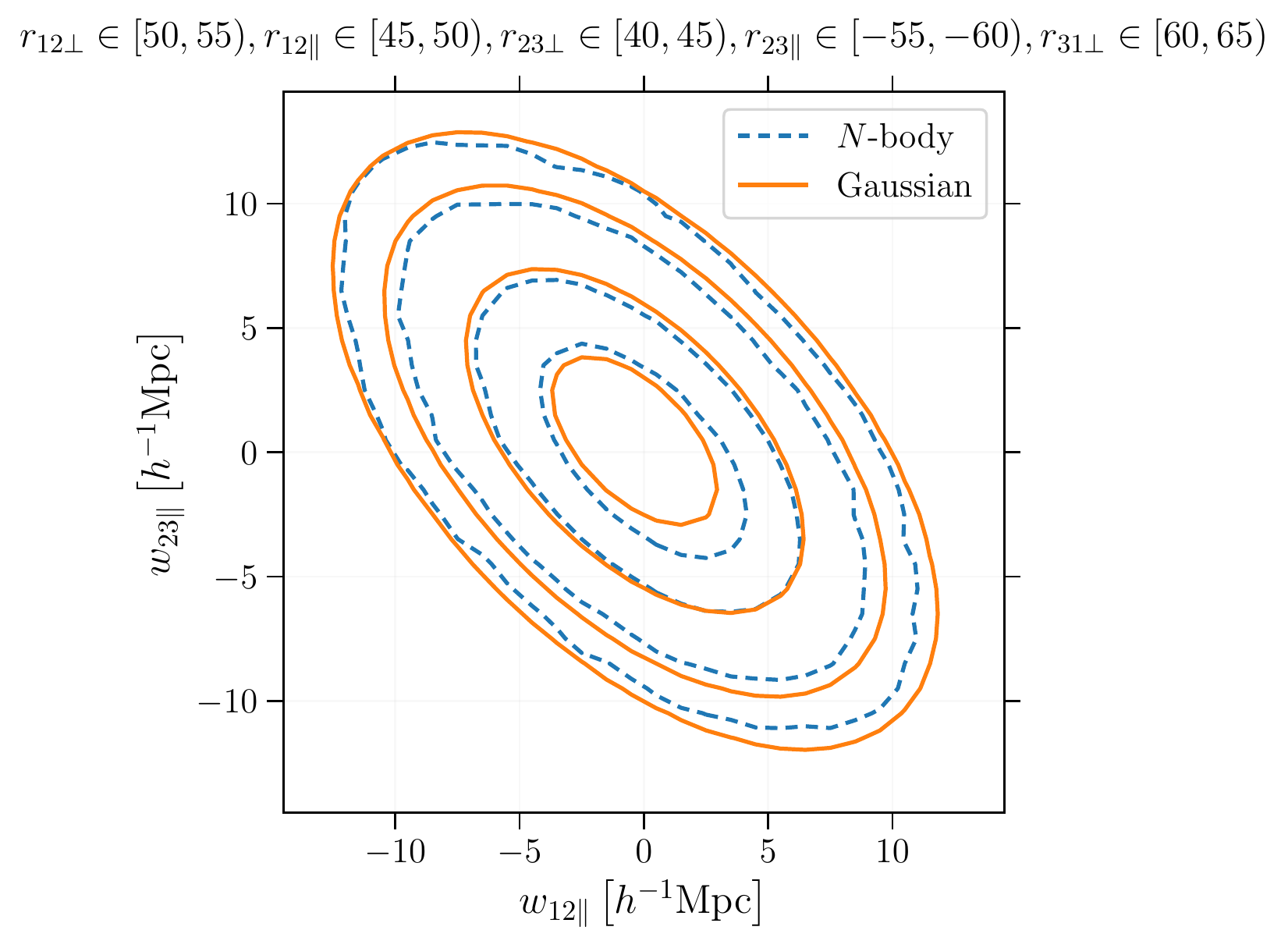}
    \caption{Contour levels of the joint PDF ${\mathcal P}^{(3)}_{\boldsymbol{w}_{\parallel}}(w_{12\parallel}, w_{23\parallel} |\triangle_{123})$
    extracted from our $N$-body simulation (dashed) are compared with those of the Gaussian model (solid) with cumulants predicted from perturbation theory at LO.
    The triangular configuration we consider is specified in the label on top of the figure in units of $h^{-1} \mathrm{Mpc}$. Contours correspond to the levels $\{6, 3, 1, 0.4\} \times 10^{-3}$ with the values decreasing from inside to outside.}
    \label{fig:bivariate-gaussian}
\end{figure}

In what follows, we investigate the simplest possible
implementation
of the 3ptGSM based on the perturbative predictions at LO
given in 
equations~(\ref{eq:meanw12final}),
(\ref{eq:meanw23final}), (\ref{eq:dispw12final}), and (\ref{eq:w12w23final}).
In figure~\ref{fig:bivariate-gaussian}, we compare the resulting PDF with that
extracted from our simulation for a particular triangular configuration which is specified on top of the figure.
To first approximation, the Gaussian model provides a very good description of the PDF. Looking into more details reveals that
it slightly underestimates the probability density around the peak.
The Kullback-Leibler (KL) and the Jensen-Shannon (JS) divergences\footnote{The KL divergence is the expectation of the logarithmic difference between the actual distribution $\mathcal{P}$ and the approximating Gaussian $\mathcal{G}$:
$D_{\mathrm{KL}}(\mathcal{P}\;\|\;\mathcal{G}) = \int_{\mathbb{R}^2} \mathcal{P}(\bm{x})\,\ln\!\left[\mathcal{P}(\bm{x})/{\mathcal{G}(\bm{x})}\right] \, \mathrm{d}\bm{x}$.
Since this statistic is not symmetric and is unbounded, it cannot be used to define the distance between two PDFs. 
However, starting from the KL divergence, a similarity measure between two PDFs which is symmetric was introduced in \cite{Rao87} and generalised in \cite{Lin91}. This is known as the JS divergence (JSD) which is given as
$
    \mathrm{JS}(\mathcal{P}\;\|\;\mathcal{G}) =\left[  D_{\mathrm{KL}}(\mathcal{P}\;\|\;\mathcal{G}) +  D_{\mathrm{KL}}(\mathcal{G}\;\|\;\mathcal{M})  \right]/2
$
where $\mathcal{M} =(\mathcal{P}+\mathcal{G})/2$. The JS divergence is bounded, $0 \leq \mathrm{JS} \leq \ln 2$, which makes the interpretation of its values easier. Additionally, the square root of the JS divergence is a pairwise distance metric.
} 
are 0.028 and 0.005 nats, respectively, indicating that the information loss 
associated with using the Gaussian approximation in place of the actual PDF is minimal. Similar values are obtained for different triangular configurations on large scales.
However, the approximation clearly fails at smaller separations
as is evident visually from figure~\ref{fig:equi}. In this case, for the triangular configuration considered in the top-left panel, we find KL and JS divergences of 1.31 nats and 0.39 nats (the upper bound being $\ln(2) \simeq 0.69$ nats), respectively.

\subsection{3-point correlations in the $N$-body simulation}
Our plan is to test the predictions of the 3ptGSM against our $N$-body simulation.
In order to measure $\mathcal{G}_3$, we first generate catalogs of `random' particles with uniform density within the simulation box and then use the `natural' estimator $DDD/RRR$ where the symbols
$DDD$ and $RRR$ denote the normalised data-data-data and random-random-random triplet counts in a bin of triangular configurations, respectively \cite{PeeblesGroth75, Peebles80}.
We characterize the shape and orientation of each triplet using the five-dimensional space
$(s_{12\perp}, s_{12\parallel}, s_{23\perp}, s_{23\parallel}, s_{31\perp})$ and, for all separations, we use bins that are $5 \,h^{-1}$ Mpc wide.
To speed the calculation up, we analyse ten subsamples of $100^3$ particles each randomly selected from the simulation. 
For each subsample, we employ five random catalogs containing $1.5\times 100^3$ objects each to measure $RRR$.\footnote{At fixed computing time,
errors are minimised by using a factor of 1.5-2 more random particles than simulation particles \cite{SlepianEisenstein15}.
In order to further reduce the uncertainties, we combine
the estimates of RRR obtained using different random sets. This is less computationally expensive than employing a single larger random catalog.}
Our final estimates for $\mathcal{G}_3$ are obtained by averaging
the partial results from the ten subsamples.
Error bars are computed by resampling the measurements from the different subsamples with the bootstrap method.
Since measuring the 3PCFs is very time consuming
and perturbation theory is only expected to be accurate
on large-enough scales,
we consider a limited number of triangular configurations with fixed $s_{12\perp}\in [50, 55)\,h^{-1}$ Mpc  and $s_{23\perp}\in [40, 45)\,h^{-1}$ Mpc. We vary $s_{12\parallel}$, $s_{23\parallel}$ in the range $[15, 65)\,h^{-1}$ Mpc
and $s_{31\perp}$ between 25 and $80\,h^{-1}$ Mpc.
We present some examples of our results in figure~\ref{fig:streaming-sperp31} and discuss them in detail in section~\ref{sec:zetatest}.

We also measure the connected 3PCF using the Szapudi-Szalay estimator that we schematically write as
$(D-R)(D-R)(D-R)/RRR$ \cite{SzapudiSzalay98,SlepianEisenstein15}.
We implement three versions of the estimator for the 3PCF obtained by binning the triplet counts in different ways.
\begin{enumerate}
    \item 
To begin with, we consider the same binning scheme in five dimensions we have used to measure ${\mathcal G}_3$.
This accounts for all the degrees of freedom in $\zeta_\mathrm{s}$ but also provides relatively noisy
estimates as the triplet counts are partitioned between many bins.
We use the same data subsamples, random catalogs
and separation ranges 
that have been described above for the full 3PCF.
Results with
their bootstrap standard errors are presented in
figure~\ref{fig:streaming-sperp31} and discussed in section~\ref{sec:zetatest}.
We anticipate here that the final uncertainty of the individual estimates is comparable with the signal. 
\item
In order to measure the connected 3PCF in redshift space with a much higher signal-to-noise ratio, we average $\zeta_\mathrm{s}$ over the orientation
of $\triangle_{123}$ with respect to the line of sight (and the winding order) while keeping the shape of the triangle fixed.
The resulting correlation function, $\bar{\zeta}_\mathrm{s}(s_{12},s_{23},s_{31})$,
only depends on three variables. While the averaging procedure does not lead to any information loss in real space (as $\zeta$ is isotropic and $\bar{\zeta}=\zeta$), it obviously gives a lossy compression in redshift space.
We use the Szapudi-Szalay method to measure 
$\bar{\zeta}_\mathrm{s}$ and $\zeta$ in our simulation
after binning the triplet counts in terms of the leg lengths of $\triangle_{123}$ (once again we use bins that are $5 \,h^{-1}$ Mpc wide).\footnote{We acknowledge that, in this case, all triplet counts involving random particles could be computed analytically \cite{PearsonSamushia19}.
However, for consistency with our study of the anisotropic 3PCF, we adopt the traditional approach based on random catalogs. Note that this choice does not influence our measurements that have relatively small errors (see figure~\ref{fig:streaming-isotropic}).}
We apply the estimator to five of the subsamples
introduced above.
We eventually average the resulting 3PCF
over the subsamples and compute bootstrap standard errors.
These results are shown in figure~\ref{fig:streaming-isotropic} and discussed in section~\ref{sec:isotropic}.
\item Finally, as an intermediate step between those discussed above, we combine narrow  ($5 \,h^{-1}$ Mpc wide) bins in $s_{12}$, $s_{23}$, and $s_{31}$ with a few  
broad ($0.5$ wide) bins in $\mu_{12}$ and $\mu_{23}$. 
This is similar to the `clustering wedges' that have been used to characterize the 2PCF in redshift space
\cite{Kazin+12, Sanchez+13}.
Note that changing sign to both $\mu_{12}$ and $\mu_{23}$ at the same time does not affect the 3PCF
as it is equivalent to reversing the sign of all the separation vectors that form the triangle $\triangle_{123}$ (see also section 3.2.2 in \cite{YankelevichPorciani18}). 
After 
summing up the triplet counts from pairs of corresponding bins under the transformation $(\mu_{12},\mu_{23})\to (-\mu_{12},-\mu_{23})$,
we end up considering eight wedges for each triangular shape.
We denote the resulting correlation function with the symbol
${\zeta}_\mathrm{s}^{(ij)}(s_{12},s_{23},s_{31})$
where the index $i\in\{1,2\}$ refers to the bins in
$\mu_{12}\geq 0$ 
and the index $j\in\{1,2,3,4\}$ maps to the bins in
 $-1\leq \mu_{23}\leq 1$.
We apply the estimator to five of the data subsamples
described above.
Examples of our results are shown in figure~\ref{fig:zeta_wdge} and discussed in section~\ref{sec:wedges}.
\end{enumerate}

\subsection{Results for the 3-point correlation function}
\label{sec:zetatest}

\subsubsection{Full correlation function}
We now solve equation~(\ref{eq:3ptstreaming_r}) for the 3ptGSM.
As input, we first use the real-space $\mathcal{F}_3$ evaluated at LO in perturbation theory. This means that we Fourier transform the linear matter spectrum to get $\xi$ and neglect $\zeta$ (case A). 
In order to estimate the influence of higher-order terms, we repeat the calculation by also considering the LO expression for $\zeta$ as in \cite{JingBorner97} (case B) although this is not fully consistent with the approximation we use for $\xi$ as we do not consider one-loop corrections.\footnote{Since $\zeta$ is given by the ensemble average of the product of two mass overdensities evaluated at linear order and one at second order, we make sure that the final expression for $\zeta$ is properly symmetrized in the coordinates of the three points.}  
Finally, we account for non-linear evolution in the 2PCF by Fourier transforming the matter power spectrum given by the halo model \cite{Mead+16} and also calculate $\zeta$ at LO (case C).
As output, we obtain the redshift-space $\mathcal{G}_3$. 
In the left panels of figure~\ref{fig:streaming-sperp31}, we compare the outcome of the 3ptGSM for different triangular configurations against measurements from our numerical simulation.
Results are plotted as a function of $s_{31\perp}$ by keeping the remaining four variables that define a triangular configuration fixed.
In the top panel, we consider a nearly
isosceles triangle with $s_{12}\simeq s_{23}\simeq 71\,h^{-1}$ Mpc, $s_{12\parallel}\simeq 47.5\,h^{-1}$ Mpc (i.e. $\mu_1\simeq 0.67$) and
$s_{23\parallel}\simeq -57.5\,h^{-1}$ Mpc (i.e. $\mu_2\simeq -0.80$) which corresponds to
$s_{31\parallel}\simeq 10 \,h^{-1}$ Mpc.
By increasing $s_{31\perp}$, we change the shape of the triangle (i.e. increase $s_{31}$ from 29 to $78\,h^{-1}$ Mpc or, equivalently,
$\cos \chi$ from $-0.92$ to $-0.4$) and simultaneously reduce $\mu_3$ from 0.34 to 0.13.
For $s_{31\perp}\simeq 70\,h^{-1}$ Mpc, we obtain an equilateral configuration. 
On the other hand, in the bottom panel, we consider
a scalene triangle with $s_{12}\simeq 55\,h^{-1}$ Mpc
($\mu_1 \simeq 0.32$) and $s_{23\parallel}\simeq 67.5\,h^{-1}$ Mpc ($\mu_2 \simeq -0.78$).
In this case, we vary $s_{31}$ from 44.5 to $85\,h^{-1}$ Mpc 
and $\mu_3$ from 0.62 to 0.4. For $s_{31\perp}\simeq 42$ and $58 \,h^{-1}$ Mpc,
we obtain isosceles triangles.
Overall, the model and the measurements show the same trends: the general agreement is rather good. 
The three different implementations of the model give very similar results and it is impossible to prefer one
over the others based on our measurements.

\subsubsection{Connected correlation function}
\begin{figure}[t]
 \begin{subfigure}[b]{0.49\textwidth}
 \centering
  	\includegraphics[scale=0.49]{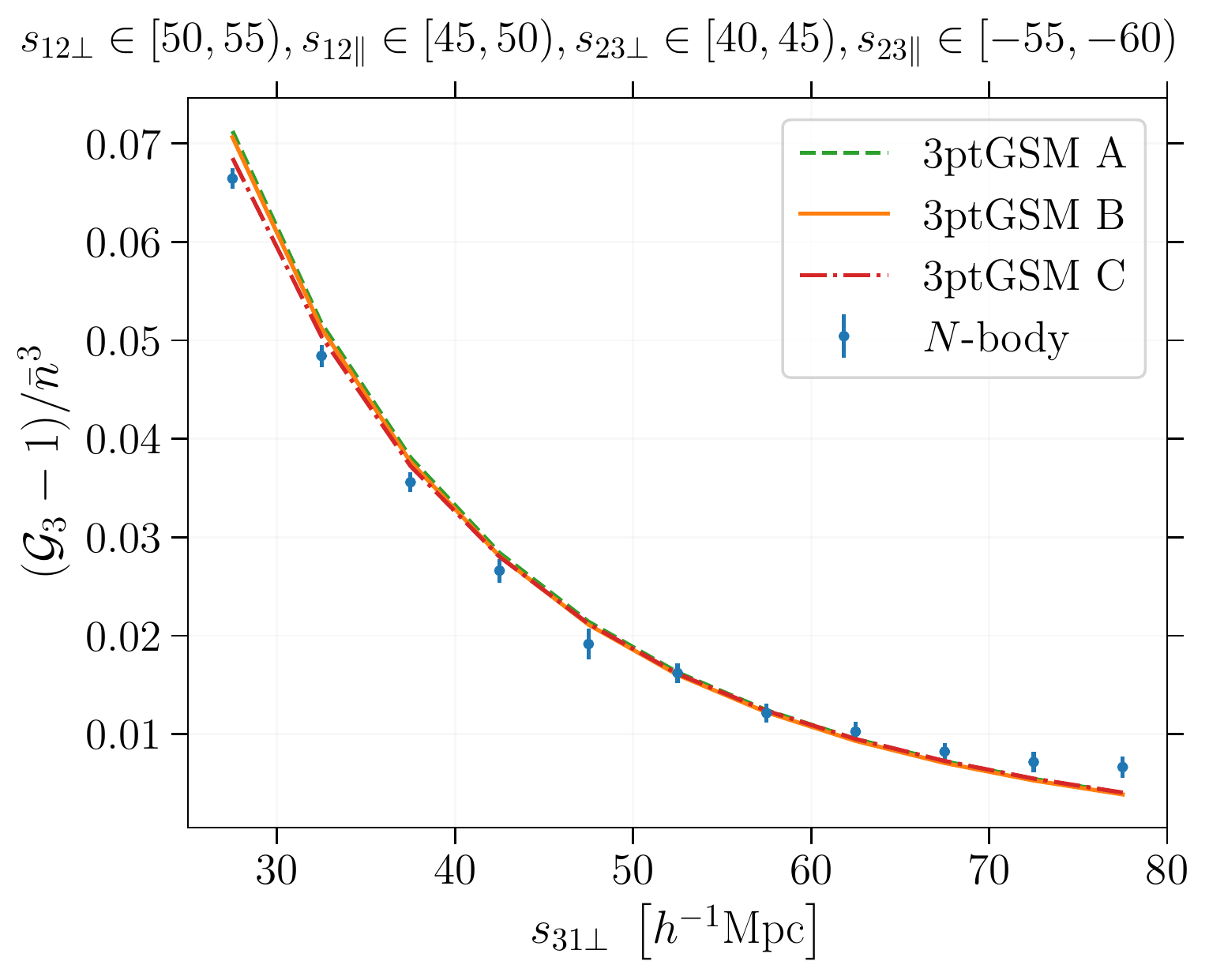}
    \end{subfigure}
 \begin{subfigure}[b]{0.49\textwidth}
 \centering
  	\includegraphics[scale=0.49]{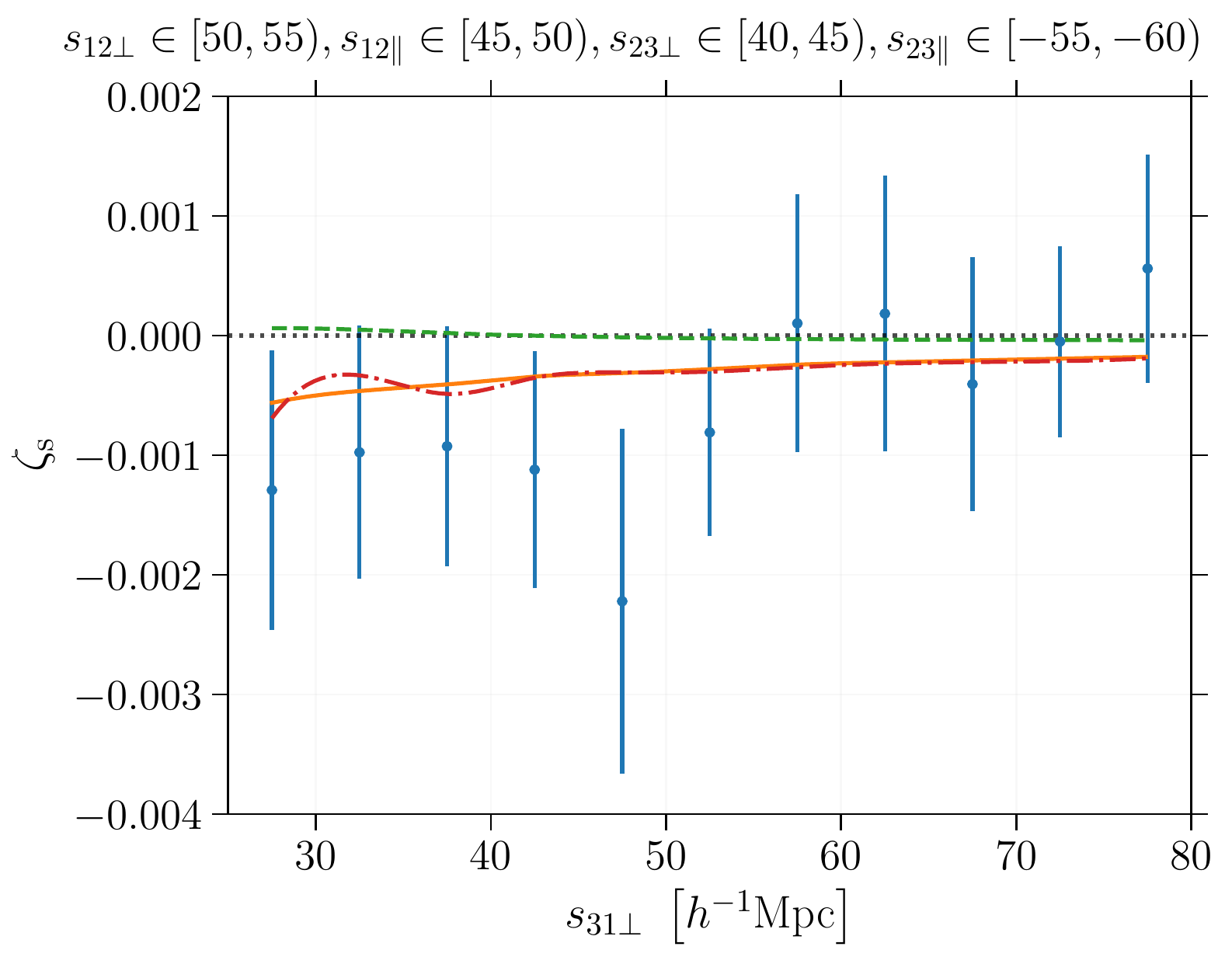}
    \end{subfigure}\\
 \begin{subfigure}[b]{0.49\textwidth}
 \centering
  	\includegraphics[scale=0.49]{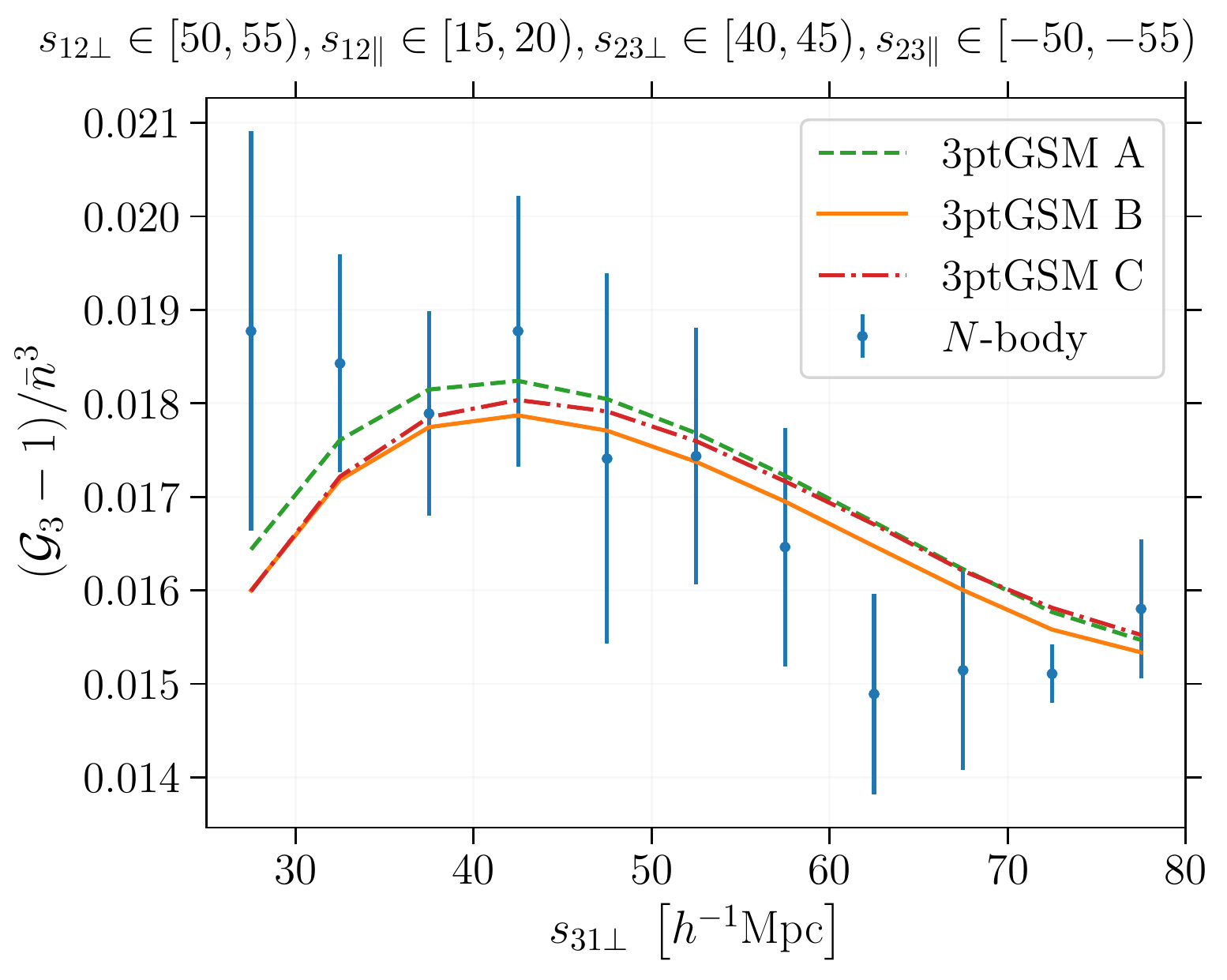}
    \end{subfigure}
 \begin{subfigure}[b]{0.49\textwidth}
 \centering
  	\includegraphics[scale=0.49]{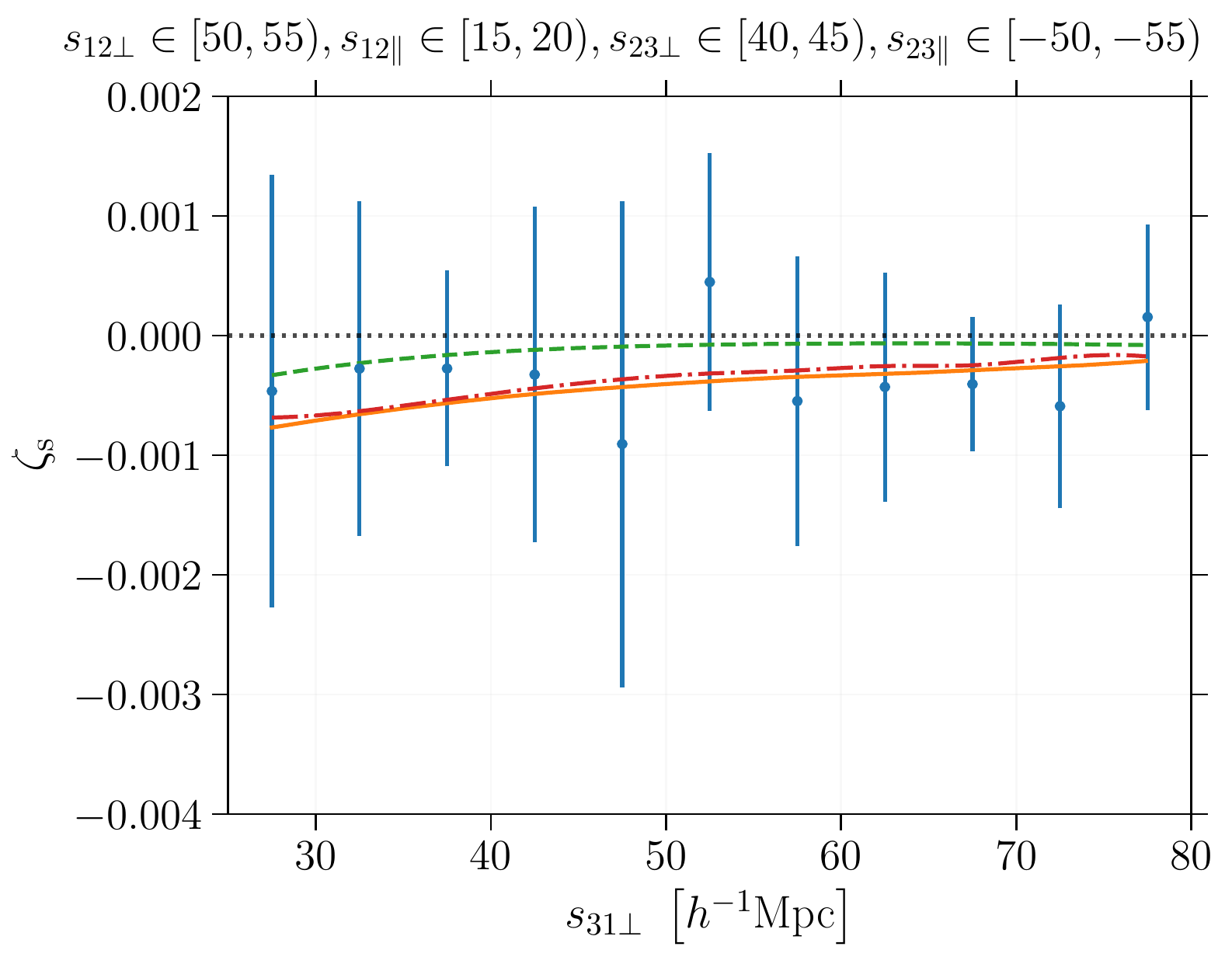}
    \end{subfigure}
     \caption{Left: Predictions from the 3ptGSM (lines) for the full redshift-space 3PCF $\mathcal{G}_3$ are compared with measurements from our $N$-body simulation (symbols with error bars). Three versions of the model are considered: case A uses as input the full 3PCF in real space $\mathcal{F}_3$ evaluated at LO in perturbation theory (dashed), case B also includes LO terms for $\zeta$ (solid), and case C combines the halo model for $\xi$ with the perturbative model for $\zeta$ at LO. The redshift-space separations listed on top of the figures are given in units of $h^{-1} \mathrm{Mpc}$.
     Right: As in the left panel, but for the connected 3PCF in redshift space.}
     \label{fig:streaming-sperp31}
\end{figure}

We obtain predictions for the connected 3PCF $\zeta_\mathrm{s}$ using equation~(\ref{eq:3ptstreamrversion}).
We model the PDF of the pairwise velocities, ${\mathcal P}^{(2)}_{w_\parallel}(w_\parallel| \boldsymbol{r})$ with a Gaussian distribution whose moments are derived from  
equations~(\ref{eq:mean-radial-velocity}) and (\ref{eq:pairwisetransvdisp}) as well as
(\ref{eq:mu12}), (\ref{eq:mu23}) and (\ref{eq:mu31}) for the los
projections:
$\langle w_{ij\parallel}|\boldsymbol{r}_{ij}\rangle_\mathrm{p}=\bar{w}(r_{ij})\,\mu_{ij}$ and 
$\langle w_{ij\parallel}^2|\boldsymbol{r}_{ij}\rangle_\mathrm{p}=2[\sigma_v^2-\psi_\parallel(r_{ij})]$.
In the right panel of figure~\ref{fig:streaming-sperp31}, we compare the results for $\zeta_{\mathrm{s}}$ 
with the data extracted from our simulation. Note that $\zeta_{\mathrm{s}}$ is very small on the scales we consider (remember that, in perturbation theory, $\zeta\sim \xi^2$) and our measurements are rather noisy due to the fact that estimating $\zeta_{\mathrm{s}}$ requires binning
the triplet counts in five dimensions. 
Anyway, the $N$-body results are in very good agreement 
with the predictions of the 3ptGSM for cases B and C
and show the same behaviour as a function of $s_{31\perp}$.

The fact that
the signal-to-noise ratio of our measurements is rather low (for many configurations $\zeta_{\mathrm s}$ is compatible with zero) may cast some doubts on the usefulness of extracting cosmological information from the galaxy 3PCF on large scales. It is thus important to mention here several reassuring indications that this conclusion is unfounded.
First, our simulation only covers a volume of ($1.2\,h^{-1} $ Gpc)$^3$ which is relatively small with respect 
to the redshift shells that will be used in the forthcoming generation of galaxy redshift surveys.
Second, galaxy biasing can substantially boost the amplitude of the 3PCF. Third, compressing the information
stored in $\zeta_{\mathrm s}$ by making use of summary statistics that depend on simpler (read lower-dimensional) configurations 
(e.g. multipoles or wedges, see e.g. figure~\ref{fig:zeta_wdge}) greatly increases the signal-to-noise ratio of the measurements. 
Fourth, although the individual measurements might be noisy, there are many triangular configurations to consider. Recent forecasts based on the joint analysis of two- and three-point statistics in Fourier space indicate that
adding the fully anisotropic bispectrum or its multipoles
help breaking degeneracies that are present in power-spectrum studies and thus set tighter constraints on several parameters.
\cite{YankelevichPorciani18,GualdiVerde20}.

\subsubsection{Connected correlation function in real space}
\label{sec:real}

\begin{figure}[t]
\begin{subfigure}[b]{0.49\textwidth}
 \centering
  	\includegraphics[scale=0.52]{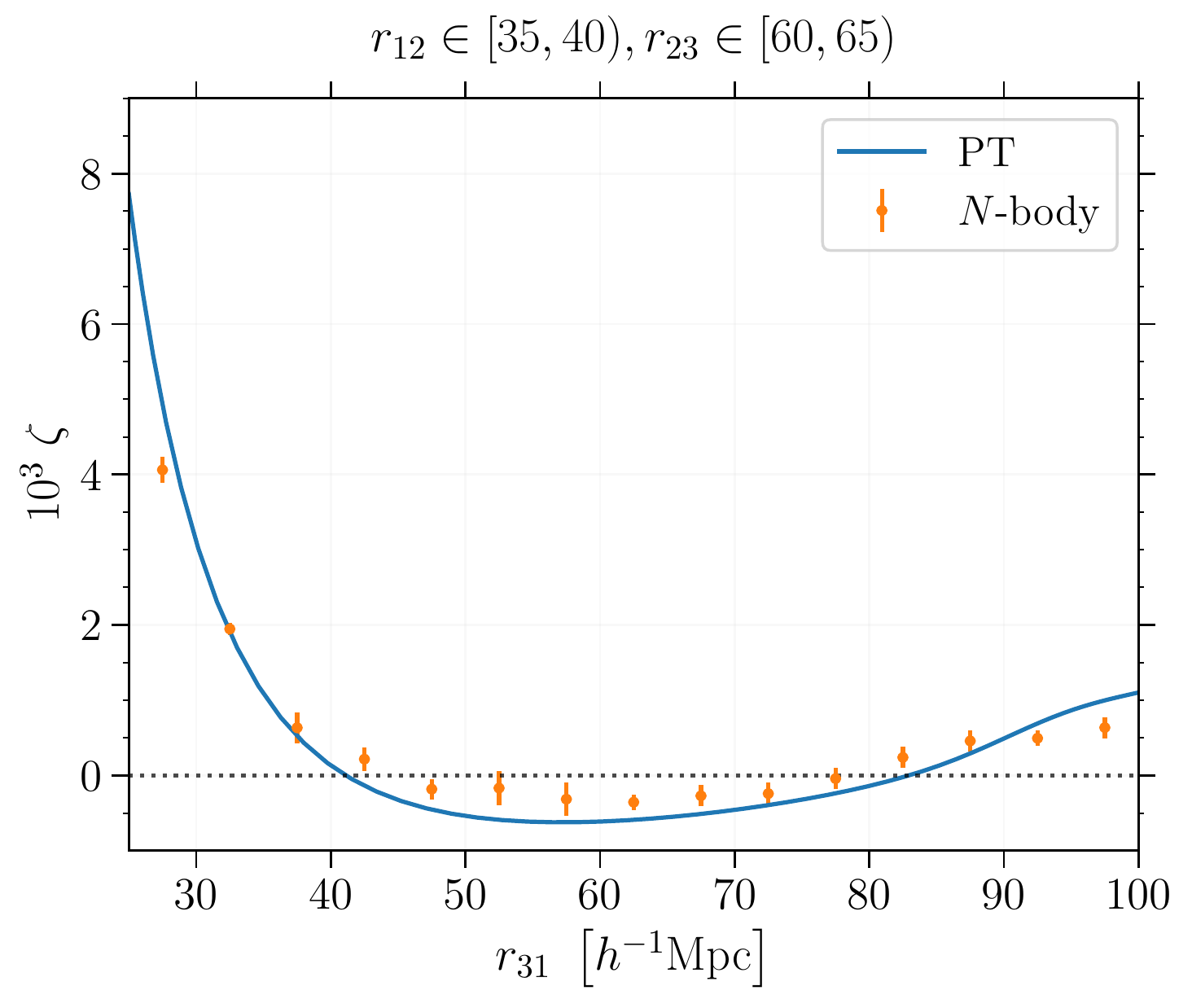}
 \end{subfigure}
  \begin{subfigure}[b]{0.49\textwidth}
 \centering
  	\includegraphics[scale=0.52]{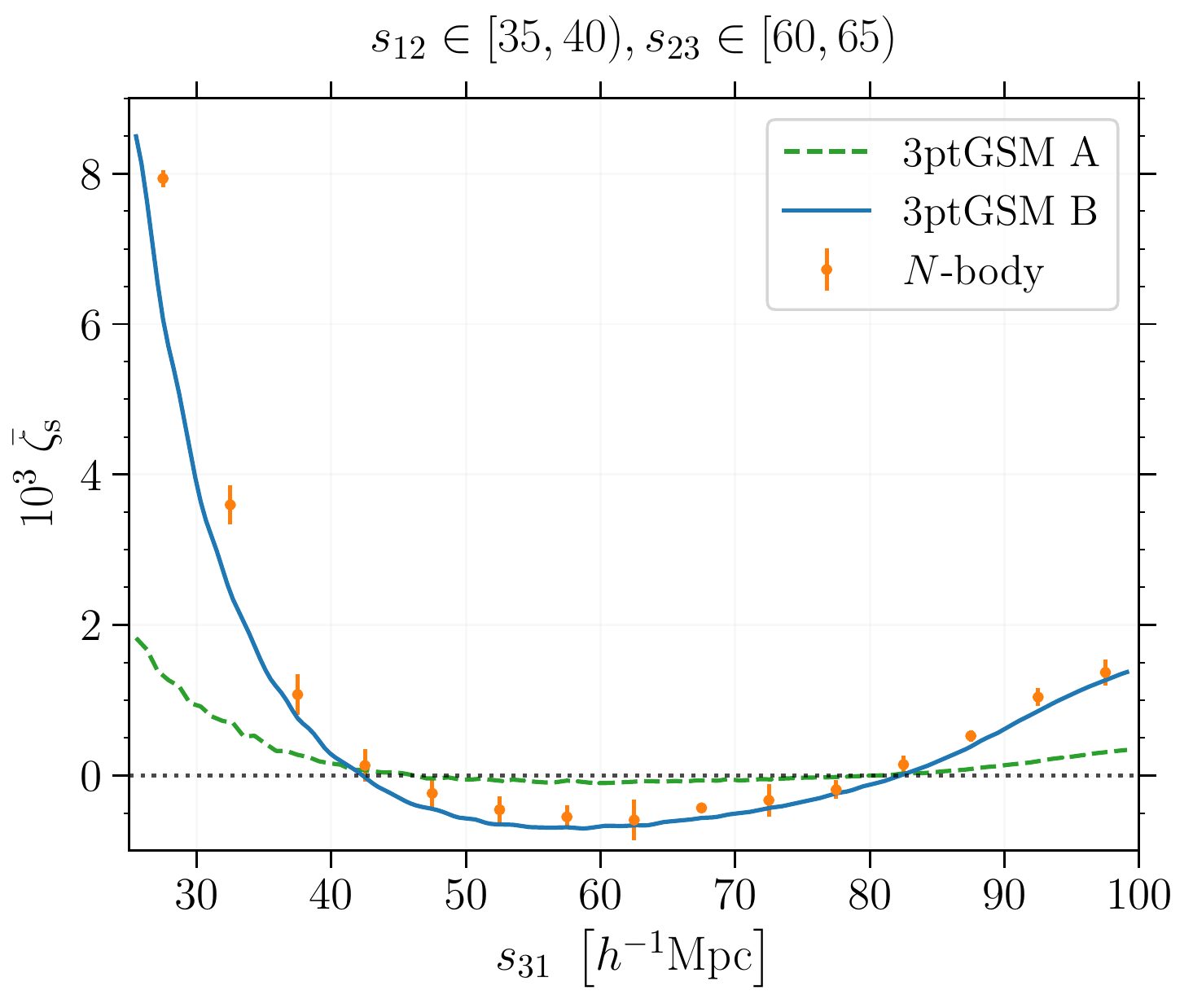}
 \end{subfigure}
     \caption{Left: The 3PCF in real space measured in the $N$-body simulation (symbols with error bars) is compared with the predictions from PT at LO (solid line) for a set of triangular configurations obtained by varying $r_{31}$ while keeping $r_{12}$ and $r_{23}$ fixed (as indicated by the top labels that give separations
     in units of $h^{-1} \mathrm{Mpc}$).
     Right: As in the left panel but for the
     spherically-averaged 3PCF in redshift space, $\bar{\zeta}_\mathrm{s}$. In this case, the solid line indicates the predictions of the 3ptGSM.}
     \label{fig:streaming-isotropic}
\end{figure}

Feeding the 3ptGSM with an accurate input for $\zeta$ 
is a necessary prerequisite in order
to properly test its capacity to model RSD.
Therefore, in the left-panel of figure~\ref{fig:streaming-isotropic}, we compare
the real-space 3PCF obtained from 
the perturbative model at LO against
the measurements in the simulation.
We consider two narrow bins centred around $r_{12}=37.5\,h^{-1}$ Mpc and $r_{23}=62.5\,h^{-1}$ Mpc
and vary $r_{31}$ within the full range.
Although the agreement is not perfect,
we find that the 
model at LO is in the same ballpark as the simulation results.
Overall, the model shows the same shape dependence of the data but relative deviations range typically between 20 and 50\% and, obviously, become larger around the zero-crossing points. 
For larger triangles with sides $r_{12}\simeq 50\,h^{-1}$ Mpc and $r_{23}\simeq 100\,h^{-1}$ Mpc, the model appears to work better (see e.g. figure 11 in \cite{MICE-I})
but $\zeta$ becomes very small and requires large
simulated volumes for an accurate measurement.
All these findings are consistent with other studies on
the matter 3PCF \cite{JingBorner97,BarrigaGaztanaga02} and
bispectrum \cite{ScoccimarroCouchman01, GilMarin+12, McCullagh+16, Hoffmann+18, Takahashi+19}.

\subsubsection{Connected correlation function averaged over all orientations}
\label{sec:isotropic}
In order to compute a theoretical prediction for 
the spherically-averaged $\bar{\zeta}_\mathrm{s}$,
we use equation~(\ref{eq:3ptstreamrversion}) to evaluate $\zeta_\mathrm{s}$ with the 3ptGSM and 
calculate
\begin{align}
\label{eq:sphaverage}
    \bar{\zeta}_\mathrm{s}(s_{12},s_{23},s_{31})&=
    \frac{\displaystyle
    \int \zeta_\mathrm{s}(\bm{s}_{12},\bm{s}_{23})\,\delta_\mathrm{D}^{(1)}\left(s_{31} - \sqrt{s_{12}^2+s_{23}^2+2s_{12}s_{23}\,
    \hat{\bm{s}}_{12}\cdot \hat{\bm{s}}_{23}}\right)\,
    \rd\hat{\bm{s}}_{12}\,\rd\hat{\bm{s}}_{23}}
    {\displaystyle \int \delta_\mathrm{D}^{(1)}\left(s_{31} -
    \sqrt{s_{12}^2+s_{23}^2+2s_{12}s_{23}\,
    \hat{\bm{s}}_{12}\cdot \hat{\bm{s}}_{23}}\right)\,
    \rd\hat{\bm{s}}_{12}\,\rd\hat{\bm{s}}_{23}}\;,
    \end{align}
    where the integrals are performed by independently varying
$\hat{\bm{s}}_{12}$ and $\hat{\bm{s}}_{23}$ over
the unit sphere.
Note that\footnote{The scalar product $\hat{\bm{s}}_{12}\cdot \hat{\bm{s}}_{23}$ gives the component of $\hat{\bm{s}}_{23}$ along the direction of $\hat{\bm{s}}_{12}$. Since the two vectors are independent and uniformly distributed on the unit sphere, $\hat{\bm{s}}_{12}\cdot \hat{\bm{s}}_{23}$ is distributed as any projection along the coordinate axes, i.e. uniformly between $-1$ and 1.}
\begin{align}
\bar{\zeta}_\mathrm{s}(s_{12},s_{23},s_{31})&=\frac{\displaystyle \int \zeta_\mathrm{s}(\bm{s}_{12},\bm{s}_{23})\,\delta_\mathrm{D}^{(1)}(s_{31} - \sqrt{s_{12}^2+s_{23}^2+2s_{12}s_{23}\,
    \hat{\bm{s}}_{12}\cdot \hat{\bm{s}}_{23}})\,
    \rd\hat{\bm{s}}_{12}\,\rd\hat{\bm{s}}_{23}}
    {\displaystyle 8\pi^2\,\int_{-1}^{+1} \delta_\mathrm{D}^{(1)}(s_{31} - \sqrt{s_{12}^2+s_{23}^2+2s_{12}s_{23}\,
    \cos\chi)}\,
    \rd\cos\chi}\nonumber\\
    &=\frac{\displaystyle \int \zeta_\mathrm{s}(\bm{s}_{12},\bm{s}_{23})\,\delta_\mathrm{D}^{(1)}(s_{31} - \sqrt{s_{12}^2+s_{23}^2+2s_{12}s_{23}\,
    \hat{\bm{s}}_{12}\cdot \hat{\bm{s}}_{23}})\,
    \rd\hat{\bm{s}}_{12}\,\rd\hat{\bm{s}}_{23}}
    {\displaystyle 8\pi^2\,\displaystyle{
    \frac{s_{31}}{s_{12}\,s_{23}}} \left[\Theta(s_{31}-|s_{12}-s_{23}|)-
    \Theta(s_{31}-s_{12}-s_{23})\right]}\;,
\end{align}
where $\Theta(x)$ denotes the Heaviside step function. 
In practice, we use the Monte Carlo method to integrate the numerator and the denominator of  equation~(\ref{eq:sphaverage}) and average over $1\,h^{-1}$ Mpc wide bins for $s_{31}$ at fixed $s_{12}$ and $s_{23}$.

In the right panel of figure~\ref{fig:streaming-isotropic}, we plot  $\bar{\zeta}_\mathrm{s}$ as a function of $s_{31}$
for the same triangular configurations we 
considered in real space.
Shown are both the measurements from the simulation and the model predictions (excluding case C as it practically coincides with case B).
Comparing the left and right panels reveals that
RSD markedly enhance the clustering signal in the simulation, particularly for small $r_{31}$.
The 3ptGSM nicely captures this trend.
The agreement of our case B implementation with the simulation is rather good: the model nicely reproduces
the dependence of $\bar{\zeta}_\mathrm{s}$ on $s_{31}$
with typical systematic deviations at the 20\% level.

 \begin{figure}
    \centering
    \includegraphics[scale=0.73]{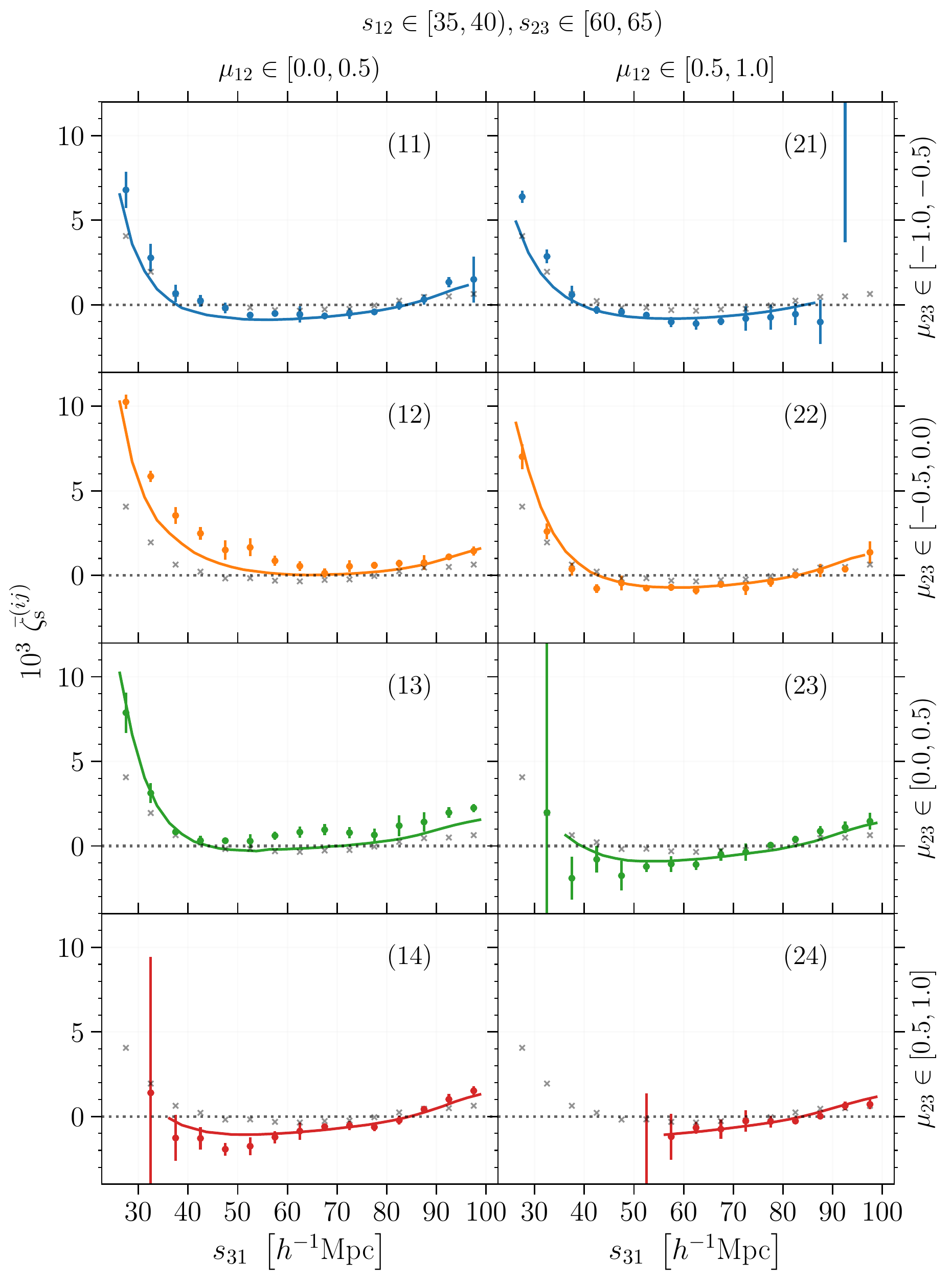}
    \caption{The wedge-averaged 3PCF measured in the simulation (symbols with error bars) is compared with the predictions of the 3ptGSM (solid lines). As a reference to help comparing the different panels, we also plot the real-space 3PCF extracted from the simulation (light $\times$ marks) and already shown in the left panel of figure~\ref{fig:streaming-isotropic}.
     The side lengths of $s_{12}$ and $s_{23}$ are listed on top of the figure in units of $h^{-1}\mathrm{Mpc}$.}
    \label{fig:zeta_wdge}
\end{figure}

\subsubsection{Connected correlation function averaged over wedges}
\label{sec:wedges}

Model predictions for the wedge-averaged 3PCF
are obtained using
equation~(\ref{eq:3ptstreamrversion}) in combination
with
\begin{align}
\label{eq:wedgeaverage}
    \bar{\zeta}^{(ij)}_\mathrm{s}(s_{12},s_{23},s_{31})&=
    \frac{\displaystyle
    \int \zeta_\mathrm{s}(\bm{s}_{12},\bm{s}_{23})\,
    W^{(ij)}(\hat{\bm{s}}_{12}\cdot \hat{\bm{s}},\hat{\bm{s}}_{23}\cdot \hat{\bm{s}})\,
    \rd\hat{\bm{s}}_{12}\,\rd\hat{\bm{s}}_{23}}
    {\displaystyle \int W^{(ij)}(\hat{\bm{s}}_{12}\cdot \hat{\bm{s}},\hat{\bm{s}}_{23}\cdot \hat{\bm{s}})\,
    \rd\hat{\bm{s}}_{12}\,\rd\hat{\bm{s}}_{23}}\;,
    \end{align}
    where
    \begin{align}
    W^{(ij)}(\mu_{12},\mu_{23})=\displaystyle \Pi_{\frac{i-1}{2},\frac{i}{2}}(\mu_{12})\,
    \Pi_{\frac{j-3}{2},\frac{j-2}{2}}(\mu_{23})\;,
    \end{align}    
    and $\Pi_{a,b}(x)=\Theta(x-a)-\Theta(x-b)$  denotes the boxcar function. Once again we perform the integrals with the Monte Carlo method.

In figure~\ref{fig:zeta_wdge}, 
we compare the wedge-averaged correlation function $\zeta_\mathrm{s}^{(ij)}$
 obtained from the 3ptGSM (case B) and from the simulation
for the same triangular configurations considered in figure~\ref{fig:streaming-isotropic}.
The eight panels are organised as follows.
The left and right columns correspond to $i=1$ (i.e.
$0\leq \mu_{12}<0.5$) and $i=2$ (i.e. $0.5\leq \mu_{12}\leq1$), respectively. Rows, from top to bottom, refer to $j=1$ ($-1\leq\mu_{23}<-0.5)$,
$j=2$ ($-0.5\leq\mu_{23}<0)$,
$j=3$ ($0\leq\mu_{23}<0.5)$ and
$j=4$ ($0.5\leq\mu_{23}\leq 1)$.
As a reference,
in each panel we also show the real-space 3PCF measured in the simulation.
The figure shows that RSD can enhance the 3PCF
by a factor of a few
(see, for instance, $\zeta_\mathrm{s}^{(14)}$, 
$\zeta_\mathrm{s}^{(21)}$, and $\zeta_\mathrm{s}^{(23)}$) as well as 
change its sign 
(as in $\zeta_\mathrm{s}^{(12)}$ and
$\zeta_\mathrm{s}^{(13)}$).
Independently of $s_{31}$,
the 3ptGSM provides an excellent description of the numerical results
for  $\zeta_\mathrm{s}^{(24)}$, 
$\zeta_\mathrm{s}^{(23)}$ and $\zeta_\mathrm{s}^{(22)}$.
In other cases, it works well only for large values of $s_{31}$ 
(see $\zeta_\mathrm{s}^{(21)}$,
$\zeta_\mathrm{s}^{(11)}$, 
$\zeta_\mathrm{s}^{(12)}$ and $\zeta_\mathrm{s}^{(14)}$).
On the other hand, the model tends to underestimate the effect of RSD for $\zeta_\mathrm{s}^{(13)}$ even for
large opening angles.

The main conclusion
emerging from the analysis of
figures~\ref{fig:streaming-sperp31}, \ref{fig:streaming-isotropic}, and \ref{fig:zeta_wdge} 
is that our implementation of the 3ptGSM, although
very simple, is already able to reproduce many features measured in the simulations.
This very encouraging result motivates further work
into building novel tools based on the 3ptGSM
for modelling $\zeta_{\mathrm s}$ on large scales and
analyse data from galaxy redshift surveys.
As a first step in this direction,
in the remainder of this paper, we analyse some key
aspects of the 3ptGSM
and
discuss how the current implementation could be improved. 

\subsection{Discussion}
\label{sec:discuss}

\subsubsection{Dissecting the 3ptGSM}
\begin{figure}
    \centering
    \includegraphics[scale=0.6]{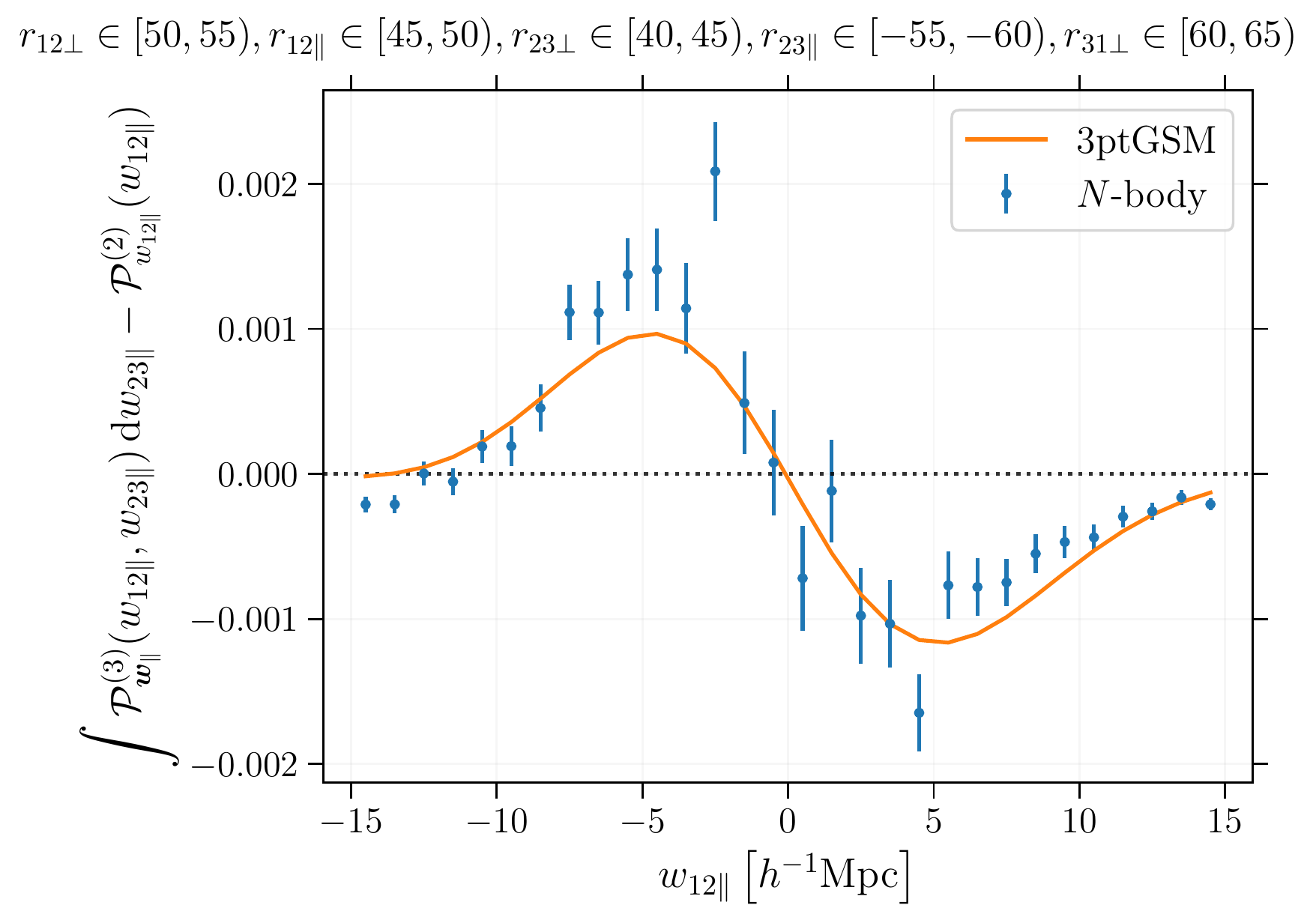}
    \caption{The difference of probability densities appearing in the rhs of equation~(\ref{eq:3ptdfullexpanded}). Symbols with error bars represent measurements from the $N$-body simulation while the dashed line shows the predictions of the 3ptGSM.}
    \label{fig:pdf-difference}
\end{figure}

Based on equations~(\ref{eq:connectedzetas}) and (\ref{eq:3ptdfullexpanded}), whenever
$\int {\mathcal P}^{(3)}_{\bm{w}_\parallel}(w_\parallel, q_\parallel | \boldsymbol{r}_{12},\boldsymbol{r}_{23})\,\mathrm{d}q_\parallel\neq
{\mathcal P}^{(2)}_{w_\parallel}(w_\parallel| \boldsymbol{r}_{12})$, RSD generate
a non-vanishing connected 3PCF $\zeta_\mathrm{s}$ even when $\zeta=0$. 
In the 3ptGSM, the marginalised distribution
gives a Gaussian PDF with mean $m_1$ and variance $C_{11}$. 
On the other hand, ${\mathcal P}^{(2)}_{w_\parallel}$ is a Gaussian with mean $\langle w_{ij\parallel}|\boldsymbol{r}_{ij}\rangle_\mathrm{p}$ and variance $\langle w_{ij\parallel}^2|\boldsymbol{r}_{ij}\rangle_\mathrm{p}-\langle w_{ij\parallel}|\boldsymbol{r}_{ij}\rangle_\mathrm{p}^2$. Note that the mean values are slightly shifted and so are also the variances (although by an even smaller amount). 
It follows that the difference between the two PDFs does not identically vanish. In practice, however, the effect is very small. By considering, for example, the triangular configuration analysed in figure~\ref{fig:bivariate-gaussian}, we find that the mean $w_{12\parallel}$ is $-0.36$ and $-0.25$ $h^{-1} \mathrm{Mpc}$ for the marginalised ${\mathcal P}_{\boldsymbol{w}_\parallel}^{(3)}$ and for ${\mathcal P}_{w_\parallel}^{(2)}$, respectively, while the standard deviation in $\simeq 4.67 \,h^{-1} \mathrm{Mpc}$ for both.
It follows that the term that multiplies $1+\xi(r_{12})$
in equation~(\ref{eq:3ptdfullexpanded}) is at best of 
the order of $10^{-3}$ and switches sign as $w_{12}$ grows past the mean value.
This is shown in figure~\ref{fig:pdf-difference} 
where we
also plot the difference between the PDFs estimated from the simulation.
The 3ptGSM provides a reasonable approximation to the numerical results. 
The total contribution of terms like this one to $\zeta_\mathrm{s}$
is shown in the right panels of figures~\ref{fig:streaming-sperp31} and \ref{fig:streaming-isotropic} as the result of our case A model. Note that it is always subdominant with respect to the contribution generated by $\zeta$, at least for the configurations considered here. 
There is some evidence that
the terms proportional to $1+\xi$ in
equation~(\ref{eq:3ptdfullexpanded})
might become more relevant at small scales
where the mean relative velocities are not so small compared to the dispersion. 
For instance, they appear to give a $\sim25\%$ contribution to $\bar{\zeta}_\mathrm{s}$ for the smallest values of $s_{31}$ shown in figure~\ref{fig:streaming-isotropic}.
However, it is unclear whether such small scales can be
robustly analysed with the 3ptGSM.

\begin{figure}
    \centering
   \includegraphics[scale=0.65]{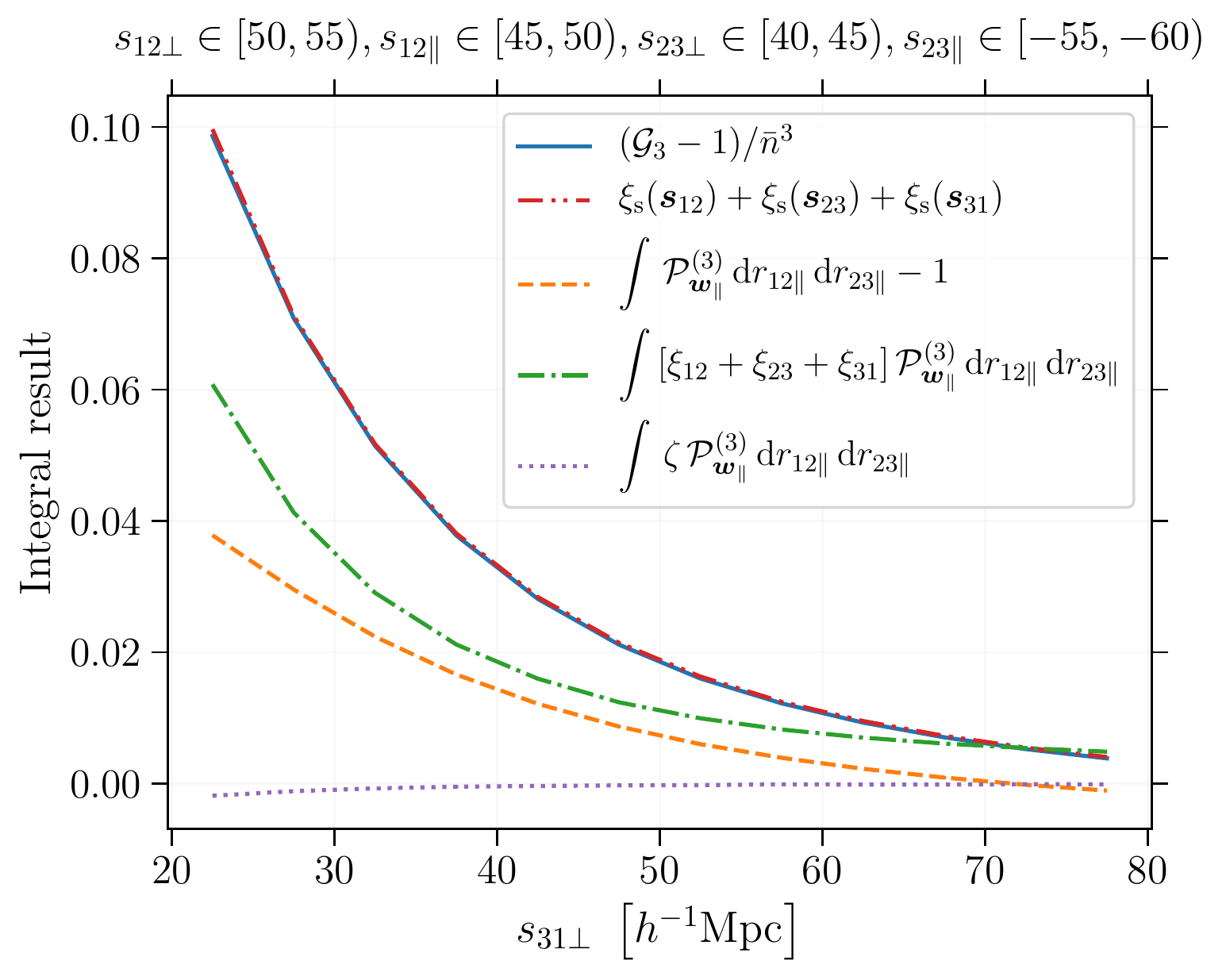}
    \caption{Partial contributions to the rhs of equation~(\ref{eq:3ptstreamrversion}) in the 3ptGSM for the same triangular configurations displayed in the top row of figure~\ref{fig:streaming-sperp31} (particle separations are given on top of the figure in units of $h^{-1} \mathrm{Mpc}$). The solid curve represents the integral containing the full 3PCF in real space. The dash double dotted line (hardly distinguishable from the solid one) displays the sum of the three integrals containing the two-point correlation function.
    The connected 3PCF in redshift space is derived by subtracting the second contribution from the first. Note that the value of $\zeta_\mathrm{s}$ is a small number obtained by subtracting two much larger numbers.
    The dashed, dash-dotted and dotted lines isolate the three sub-components of the solid curve. Namely, they show the part proportional to 1, $\xi$ and $\zeta$, respectively. 
 }
    \label{fig:integrand-disentangle}
\end{figure}

In figure~\ref{fig:integrand-disentangle}, we plot the different terms
that appear in the rhs of equation~(\ref{eq:connectedzetas})
using the same triangular configurations as in figure~\ref{fig:streaming-sperp31}.
The first thing to notice is that $\zeta_\mathrm{s}$ is obtained by subtracting two much larger numbers. This evidences the need for 
modelling $\mathcal{P}_{w_\parallel}^{(2)}$ and $\mathcal{P}_{\boldsymbol{w}_\parallel}^{(3)}$ 
in a consistent way.
Also note that the integral $\int \mathcal{P}_{\boldsymbol{w}_\parallel}^{(3)}\, \mathrm{d}w_\parallel\,\mathrm{d}q_\parallel$ appearing
in the last row of equation~(\ref{eq:connectedzetas}) is not
identically equal to one as the conditional PDF needs to be evaluated considering  
different triangular configurations that reflect the running of the real-space parallel separations in the integral.

\begin{figure}
    \centering
    \includegraphics[scale=0.8]{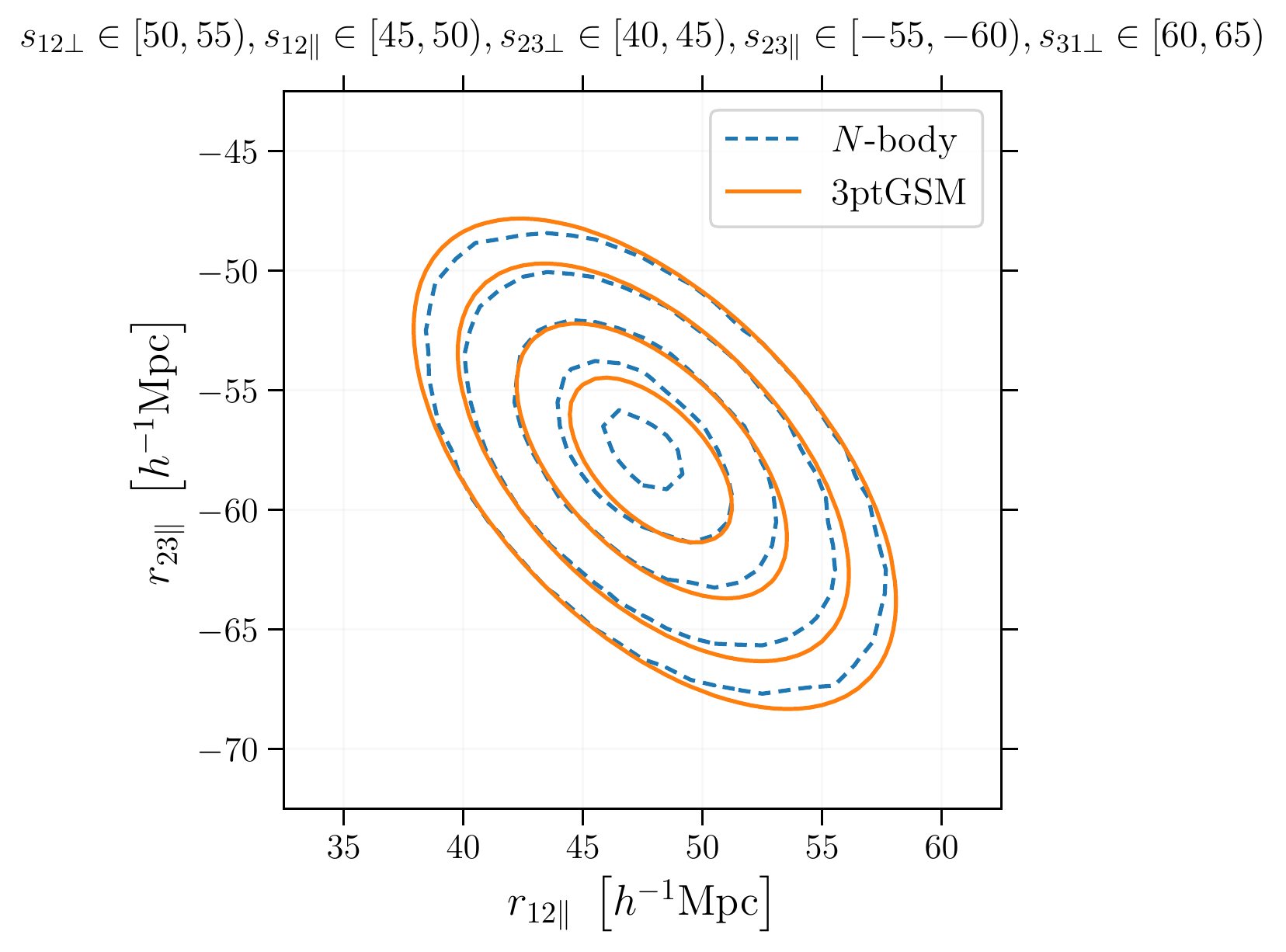}
    \caption{Contour levels of the integrand appearing in the rhs of equation~(\ref{eq:3ptstreaming_r}) for one of the triangular configuration shown in figures~\ref{fig:streaming-sperp31} and \ref{fig:integrand-disentangle} (particle separations are given on top of the figure in units of $h^{-1} \mathrm{Mpc}$).
Solid and dashed lines correspond to the 3ptGSM and the $N$-body simulation, respectively.
Contours correspond to the levels $\{8, 6, 4, 2, 1 \}\times 10^{-3}$ with the values decreasing from inside to outside.
Note that the predictions of the GSM do not reach the value $8\times 10^{-3}$.
}
    \label{fig:integrand-check}
\end{figure}

Although the Gaussian approximation for ${\mathcal P}^{(2)}_{w_\parallel}$ is not perfect, the Gaussian streaming model provides a very good description of $\xi_{\mathrm s}$ on large scales \citep[e.g.][]{ReidWhite11}.
This success originates from fortuitous
cancellations between the contributions of the peak and the
wings in the integrand of equation~(\ref{eq:2ptstreamfinal})
(see figure~4 in \cite{KuruvillaPorciani18}).
In figure~\ref{fig:integrand-check}, we show that the same phenomenon takes place in the 3ptGSM. Shown with solid lines are contour levels of the integrand appearing in the rhs of equation~(\ref{eq:3ptstreamrversion}) for a configuration in which
the 3ptGSM accurately reproduces the full 3PCF measured in the simulation. 
We extract the same quantity from the simulation by 
creating a bivariate histogram of $s_{ij\parallel} - w_{ij\parallel}$ for the particle triplets that form the same triangular configuration and making sure
that its integral gives ${\mathcal G}_3$.
The corresponding contour levels are plotted with dashed lines. From the figure, it is evident that
the peak of the integrand in the 3ptGSM is underestimated and the tails are overestimated when compared to the numerical results.
This provides motivation for improving the modelling of
${\mathcal P}^{(3)}_{\bm{w}_\parallel}$
along the lines that have been already used for the 2PCF
\cite[e.g.][]{Bianchi+15,Uhlemann+15,Bianchi+16,KuruvillaPorciani18,Cuesta+20}.

An even more direct consistency test of the 3ptGSM can be performed
by measuring the real-space correlations  and  velocity  moments from the simulation and inserting them into the key equations of the model
to isolate the impact of the Gaussian assumption,
as it has been done for 2-point statstics \cite{KuruvillaPorciani18}.
However, this investigation would be very time consuming as it requires accurate measurements 
for all possible triangular configurations up to (at least) $r \sim 150\ h^{-1}\ \mathrm{Mpc}$ (for the range
of separations considered in this work).
For this reason, we postpone this study to future work.

\subsubsection{Directions for future improvements}
\label{sec:improvements}
Overall, the simple version of the 3ptGSM we have implemented captures the main trends that can be observed in the simulation. However, there are some discrepancies.  
We identify a number of reasons for this partial
agreement.
First of all, the model we use for $\zeta$ needs to be substantially improved. As discussed above (left panel of figure~\ref{fig:streaming-isotropic}) perturbation theory at LO only provides a sketchy description of the simulation data for the corresponding triangular configuration
in real space
(i.e. using $r_{12}=s_{12}$ and $r_{23}=s_{23}$).
However, the situation is worse than that.
In fact, the integral that gives $\zeta_\mathrm{s}(\triangle_{123})$ in the streaming model receives contributions from 
triangles with pairwise separations $r_{ij\parallel}$ and $r_{ij\perp}$ that differ by up to 40-50 $h^{-1}$ Mpc from those that define $\triangle_{123}$.
For some of them, the model for $\zeta$ at LO does not perform very well.
Moreover, the second moments of the pairwise velocities predicted with standard perturbation theory at LO become progressively less accurate for squeezed triangles. One can notice this trend in some capacity already in the rightmost panels of figure~\ref{fig:grid_radial_std} and in the
top-right panel of figure~\ref{fig:cov-los}: the model increasingly departs from the simulation results as
$r_{31}$ and $r_{31\perp}$ decrease.
Since the double integral in equation~(\ref{eq:3ptstreamrversion}) runs over all sorts of triangular configurations including some squeezed ones, this generates inaccuracies.
As we  have seen
in section~\ref{sec:discuss}, 
the 3ptGSM prediction for $\zeta_\mathrm{s}$, which is of the order of $\xi^2$, is obtained from the subtraction of two much larger numbers of order $\xi$
(this can also be noticed
by comparing the left and right panels in figure~\ref{fig:streaming-sperp31}). Therefore, relatively small errors in the terms that need to be subtracted can shift $\zeta_\mathrm{s}$ substantially. 
We thus expect that the 3ptGSM will considerably benefit from more sophisticated input models for $\xi$, $\zeta$ and the moments of the pairwise velocities as it has already happened at the 2-point level \cite{Carlson+13,Wang+14,Vlah+16}.
Implementing these improvements, however,
clearly goes beyond the scope of this paper.

\subsubsection{Connection with dispersion models for the bispectrum}
Fourier transforming equation~(\ref{eq:2ptstreamfinal}) provides an expression for the anisotropic power spectrum in redshift space,
$P_\mathrm{s}(k_\parallel,k_\perp)$.
If one is ready to assume, for simplicity, that ${\mathcal P}^{(2)}_{w_\parallel}(w_\parallel|\boldsymbol{r})$
does not depend on $\bm{r}$, the convolution theorem then gives
$P_\mathrm{s}(k_\parallel,k_\perp)=S^{(2)}(k_\parallel)\,P(k_\parallel, k_\perp)$ with $S^{(2)}(k_\parallel)$ the Fourier transform of ${\mathcal P}^{(2)}_{w_\parallel}$.
This situation occurs if
${\mathcal P}^{(2)}_{w_\parallel}(w_\parallel|\boldsymbol{r})$ is replaced
by the scale-independent function ${\mathcal R}_{w_\parallel}^{(2)}(w_\parallel)$ we have introduced in equation~(\ref{eq:R2def1d}). This defines the so-called `dispersion model'.
The basic underlying idea 
(originally proposed in \cite{Peacock92})
is to imagine that, due to highly non-linear physics taking place on small scales, the los velocity at each spatial location
has a random component which is independently drawn from a distribution with variance $\sigma^2_v$
and the los relative velocities between two locations
have thus a variance of
$\sigma^2_{\rm p}=2\,\sigma^2_v$.
Assuming that ${\mathcal P}^{(1)}_{v_\parallel}$ is well approximated by a zero-mean
Gaussian with variance $\sigma^2_{{v}}$ gives $S^{(2)}(k_\parallel)=\exp(-k_\parallel^2\, \sigma_{\mathrm{p}}^2/2)$ which reduces to
$S^{(2)}(k_\parallel)\simeq 1-k_\parallel^2\,\sigma_{\mathrm{p}}^2/2$ on large scales.\footnote{Note that, at quadratic order in the wavenumbers, Gaussian and Lorentzian damping functions coincide.}
This expression 
is commonly used to analyse survey and simulation data
\citep[e.g.][]{PeacockWest92,Park+94,Peacock-Dodds94,2dFHawkins+03,Guzzo+08,6dFBeutler+12,Chuang13}
and $\sigma^2_{\mathrm p}$ is treated as a free parameter.\footnote{For dark matter,
the LO perturbative contribution to $\sigma_v^2$
is given in equation~(\ref{eq:sigmav}) but, as we have shown at the end of section~\ref{sec:disptrip}, 
this does not accurately describe  $N$-body data.}
The `damping factor' $S^{(2)}(k_\parallel)$ thus
accounts for the suppression of the clustering amplitude in redshift space due 
to incoherent relative motions along the los generated within
collapsed structures (e.g. the `finger-of-god' effect \cite{Jackson72,SargentTurner77}).

We now use equation~(\ref{eq:3ptstreamrversion}) to generalise the dispersion model to 3-point statistics.
The 3PCF and the bispectrum
$B(\boldsymbol{p},\boldsymbol{q},\boldsymbol{k})$ form a Fourier pair, i.e.
\begin{align}
\zeta(\triangle_{123})&= \langle \delta(\bm{x}_2)\,\delta(\bm{x}_2+\bm{r}_{21})\,\delta(\bm{x}_2+\bm{r}_{23})\rangle \nonumber \\ &= 
\int B(\boldsymbol{p},\boldsymbol{q},-\boldsymbol{p}-\boldsymbol{q})\,e^{-i\left(\boldsymbol{p}\cdot\boldsymbol{r}_{21}+\boldsymbol{q}\cdot\boldsymbol{r}_{23}\right)}\,
\frac{\mathrm{d}^3p\,\mathrm{d}^3q}{(2\pi)^6}\;,
\end{align}
(note that the correlation function is defined in terms of the `star ray' separation $\bm{r}_{21}$  introduced in footnote~\ref{footone} while we have always used $\bm{r}_{12}$ so far).
Let us now consider the simplest possible case in which:
(i) $\mathcal{P}_{\bm{w}_\parallel}^{(3)}$ does not depend on $\bm{r}_{21}$ and $\bm{r}_{23}$, (ii) the PDF of the pairwise velocities can be approximated by a Gaussian distribution with covariance matrix ${\boldsymbol{\mathsf {\Sigma}}}$,
and (iii) the contribution from the two-point terms in the rhs of equation~(\ref{eq:connectedzetas})
is subdominant (as discussed above).
In this case, the convolution theorem gives
\begin{equation}
    B_\mathrm{s}(\boldsymbol{p},\boldsymbol{q},-\boldsymbol{p}-\boldsymbol{q})= S^{(3)}(p_\parallel, q_\parallel)\,
    B(\boldsymbol{p},\boldsymbol{q},-\boldsymbol{p}-\boldsymbol{q})\;,
\end{equation}
with
\begin{align}
    S^{(3)}(p_\parallel,q_\parallel)&=\int {\mathcal P}^{(3)}_{\bm{w}_{\parallel}}(w_{21\parallel},w_{23\parallel}|\triangle_{123})\,e^{i (p_\parallel w_{21\parallel}+q_\parallel w_{23\parallel})}\,
    \mathrm{d}w_{21\parallel} \,\mathrm{d}w_{23\parallel}\nonumber\\
    &=\exp\left[-\frac{1}{2}
    (\Sigma_{11}\,p_\parallel^2+2\,\Sigma_{12}\,p_\parallel\, q_\parallel+\Sigma_{22}\, q_\parallel^2)
    \right]\;.
    \label{eq:Ffinal}
\end{align}
However, the result must be invariant with respect to changing the pair of wavevectors we use to evaluate the damping factor,
i.e. $S^{(3)}(p_\parallel,q_\parallel)=S^{(3)}(p_\parallel, -p_\parallel-q_\parallel)=S^{(3)}(-p_\parallel-q_\parallel,q_\parallel)$. It follows that the covariance matrix must have the form
\begin{equation}
{\boldsymbol{\mathsf {\Sigma}}}= \sigma^2
\begin{pmatrix}
1 & 1/2\\
1/2& 1
\end{pmatrix} \;,
\label{eq:sigmamatrix}
\end{equation}
with $\sigma^2$ a free parameter. 
For convenience,
in this calculation we have used the variable
$w_{21\parallel}$ while in the remainder of the paper
we always dealt with $w_{12\parallel}=-w_{21\parallel}$. Therefore, equation~(\ref{eq:sigmamatrix}) can be re-written 
in terms of the covariance matrix ${\boldsymbol{\mathsf {C}}}$
we have introduced in section~\ref{sec:defin} as
\begin{equation}
{\boldsymbol{\mathsf {C}}}= \sigma^2
\begin{pmatrix}
1 & -1/2\\
-1/2& 1
\end{pmatrix} \;.
\label{eq:Cmatrix}
\end{equation}
\begin{figure}
    \centering
    \includegraphics[scale=0.8]{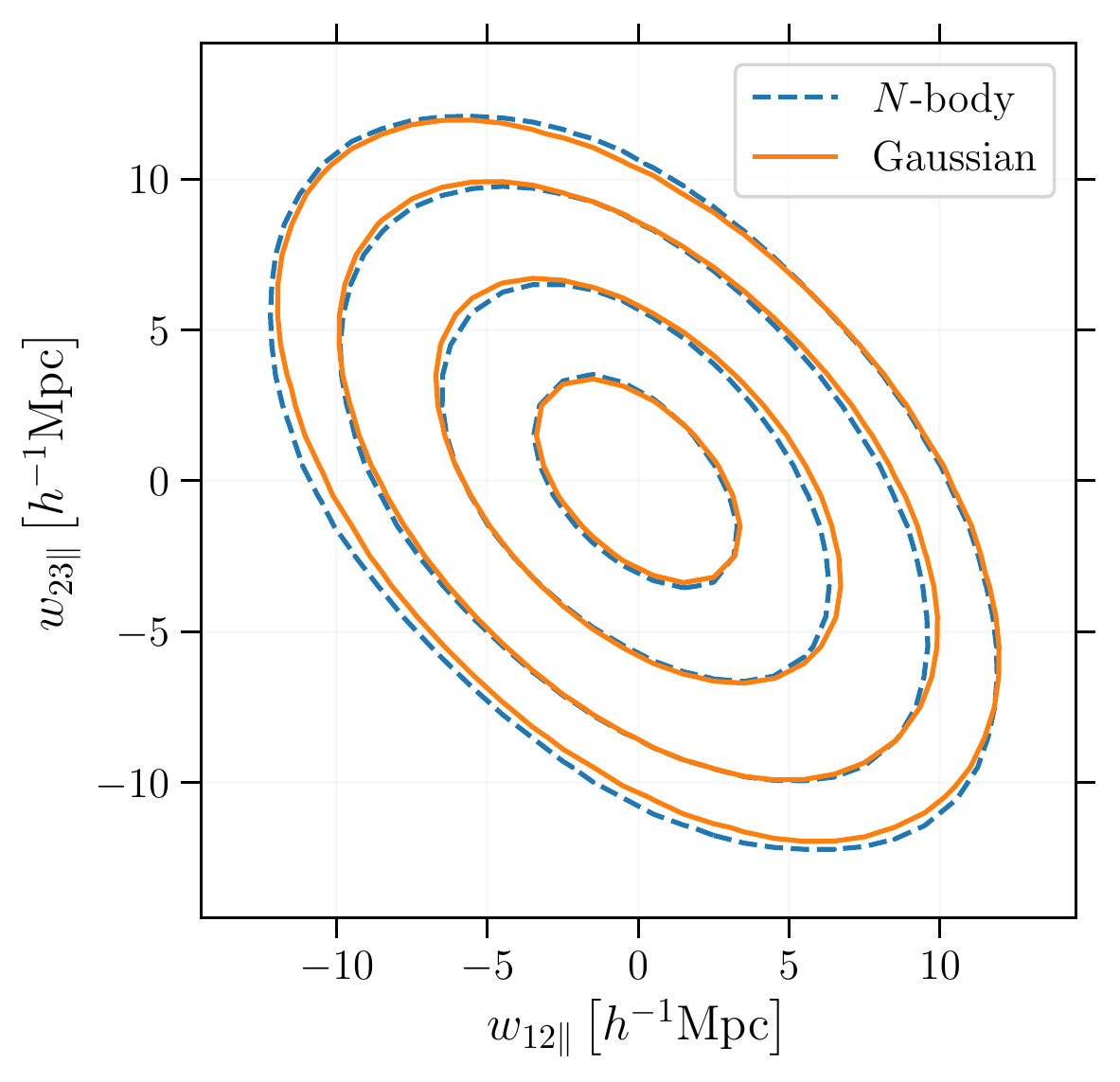}
    \caption{Contour levels of the scale-independent PDF $\mathcal{R}^{(3)}_{\boldsymbol{w}_\parallel}$, as given in equation~(\ref{eq:r3_pdf}).  The dashed lines correspond to the direct measurement from the simulation while the solid lines represent a zero mean bivariate Gaussian with a covariance matrix of the same form as in equation~(\ref{eq:Cmatrix}) and $\sigma^2=24.2\ h^{-2}\mathrm{Mpc}^2$.
    Contours correspond to levels $\{6, 3, 1, 0.4\} \times 10^{-3}$  with the values decreasing from inside to outside.}
    \label{fig:r123}
\end{figure}
It is reassuring to see that this result provides a zeroth-order approximation to the
velocity statistics we measure in the $N$-body simulation as shown in the
bottom-right panel of figure~\ref{fig:equi} and in figure~\ref{fig:cov-los}.
On  large scales, the mean pairwise velocities are much smaller than their dispersions which are nearly scale independent. Moreover, the linear correlation coefficient between $w_{12\parallel}$ and $w_{23\parallel}$ is always close to $-1/2$.

Equation~(\ref{eq:Cmatrix}) has a simple and straightforward interpretation within the context of the dispersion model: if the los velocity at each
location is independently drawn from a distribution with variance $\sigma^2_v$, then ${\boldsymbol{\mathsf {C}}}$ is the covariance matrix
of the velocity differences $w_{12\parallel}$ and $w_{23\parallel}$. The non-vanishing off-diagonal term comes
from the fact that location number 2 appears in both  pairs as evidenced in equation~(\ref{eq:crossexpanded}).
Therefore, we can write that $\sigma^2=\sigma^2_{\mathrm p}=2\,\sigma^2_v$. In other words, in full analogy with the 2-point case, the dispersion model is obtained by 
replacing ${\mathcal P}^{(3)}_{\boldsymbol{w}_\parallel}$ with
the function
${\mathcal R}^{(3)}_{\boldsymbol{w}_\parallel}$
introduced in equation~(\ref{eq:r3_pdf}).
While completing this work, we became aware
that this line of reasoning was first pursued in 
reference \cite{Matsubara94} to model the galaxy 3PCF
on small scales.
This publication also introduces a very rudimentary form of our equation~(\ref{eq:general-streaming-model-npoint}) in which ${\mathcal R}^{(3)}_{\boldsymbol{w}_\parallel}$ appears instead
of ${\mathcal P}^{(3)}_{\boldsymbol{w}_\parallel}$.
In figure~\ref{fig:r123},
we show that a Gaussian PDF provides an excellent approximation to ${\mathcal R}^{(3)}_{\boldsymbol{w}_\parallel}$.

In the literature on the bispectrum, the damping factor is generally written as a symmetric function of three wavenumbers,
${\mathcal F}(p_\parallel,q_\parallel,k_\parallel)$
with the condition $p_\parallel+q_\parallel+k_\parallel=0$. 
Equations~(\ref{eq:Ffinal}) and (\ref{eq:Cmatrix}) say
that ${\mathcal F}(p_\parallel,q_\parallel, -p_\parallel-q_\parallel)=S^{(3)}(p_\parallel, q_\parallel)$. 
There are multiple functional forms for ${\mathcal F}$
that satisfy this condition. For instance, we could obtain a valid ${\mathcal F}$ by applying a symmetrization method either to the function $S^{(3)}$ (i.e. ${\mathcal F}(p_\parallel,q_\parallel,k_\parallel)=[S^{(3)}(p_\parallel,q_\parallel)+S^{(3)}(p_\parallel,k_\parallel)+S^{(3)}(k_\parallel,q_\parallel)]/3$)
or to the argument of the exponential function that appears in $S^{(3)}$ (i.e. ${\mathcal F}(p_\parallel,q_\parallel,k_\parallel)=\exp[-(2p_\parallel^2+2q_\parallel^2+2k_\parallel^2+
pq+kp+kq)\sigma^2/6] $). 
A simpler solution is found by further requiring that 
$\mathcal{F}$ only depends on the square of the wavenumbers
which gives\footnote{This is the
most commonly used ansatz 
and provides a reasonable fit to numerical simulations \citep[e.g.][]{SCF99,Hashimoto+17}.} 
${\mathcal F}(p_\parallel,q_\parallel,k_\parallel)=\exp[-(p_\parallel^2+q_\parallel^2+k_\parallel^2)\,\sigma^2_{\mathrm p}/4]\simeq 1-(p_\parallel^2+q_\parallel^2+k_\parallel^2)\,\sigma^2_{\mathrm p}/4$.
In brief, providing an expression for ${\mathcal F}$ is somewhat arbitrary. All what matters in practice is the function $S^{(3)}$.

\subsubsection{Dependence on the growth rate of structure}
\label{sec:fofz}
\begin{figure}[t]
 \centering
  	\includegraphics[scale=0.9]{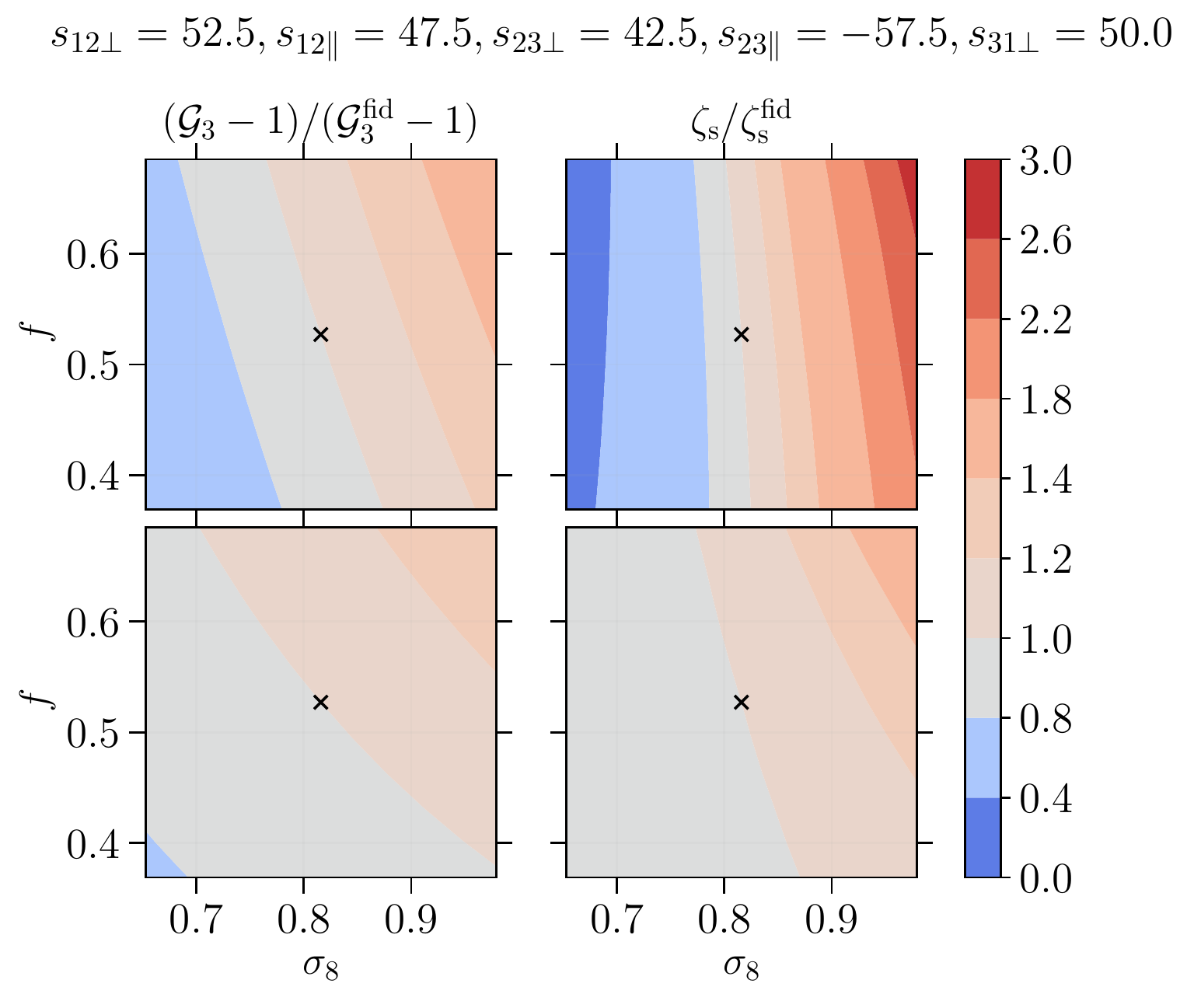}
     \caption{Dependence of ${\mathcal G}_3$ and $\zeta_{\mathrm s}$ on $f$ and $\sigma_8$ in the 3ptGSM. Shown are the contour levels of the ratios $({\mathcal G}_3-1)/({\mathcal G}^{\mathrm{fid}}_3-1)$ (left) and $\zeta_{\mathrm s}/\zeta_{\mathrm s}^{\mathrm {fid}}$ (right) where
     the fiducial values for $f$ and $\sigma_8$ coincide with those used in our simulation and are highlighted by the crosses. 
     The top panels represent the full dependence of the 3PCFs while
     the bottom ones do not consider the overall normalisation of the real-space functions $\xi$ and
     $\zeta$ (which would be degenerate with the linear bias factor in actual survey data).
     The separations that define the triangular configuration are indicated on top of the figure in units of $h^{-1} \mathrm{Mpc}$.}
     \label{fig:fsigma8}
\end{figure}

Measurements of the linear growth rate of structure $f(z)$ through RSDs
are used to probe gravity and the nature of dark energy \cite{PercivalWhite09,SongPercival09}.
This is one of the main drivers for developing the next generation of galaxy redshift surveys. 
Although the implementation of the 3ptGSM discussed in this paper applies to the clustering of matter and extensions to biased tracers will be needed for direct applications to survey science, it is anyway interesting to provide a few illustrative examples of how the 3ptGSM responds to variations in the growth rate of structure.

\begin{figure}[t]
  \begin{subfigure}[b]{0.49\textwidth}
 \centering
  	\includegraphics[scale=0.50]{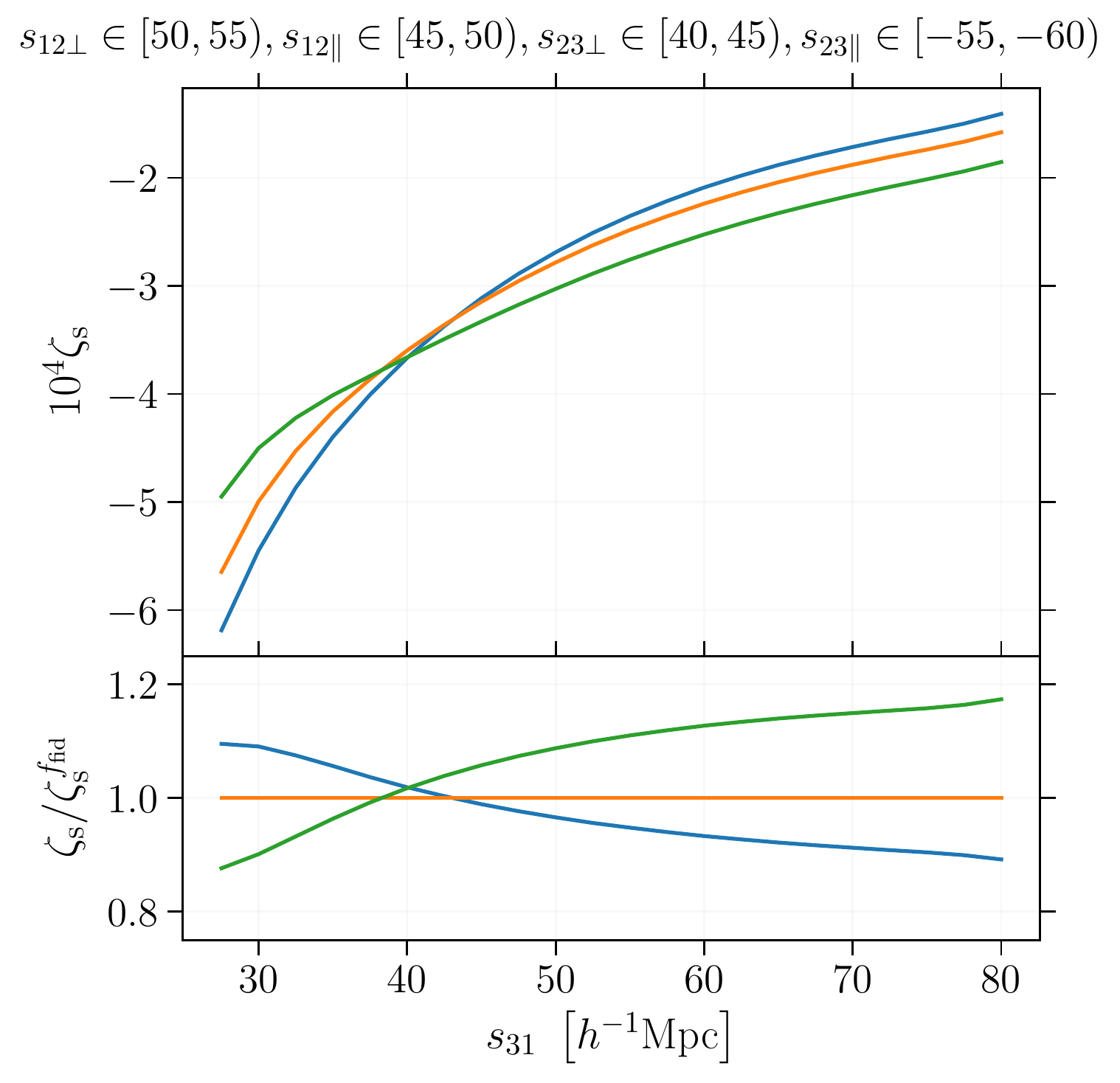}
  \end{subfigure}
  \begin{subfigure}[b]{0.49\textwidth}
 \centering
  	\includegraphics[scale=0.50]{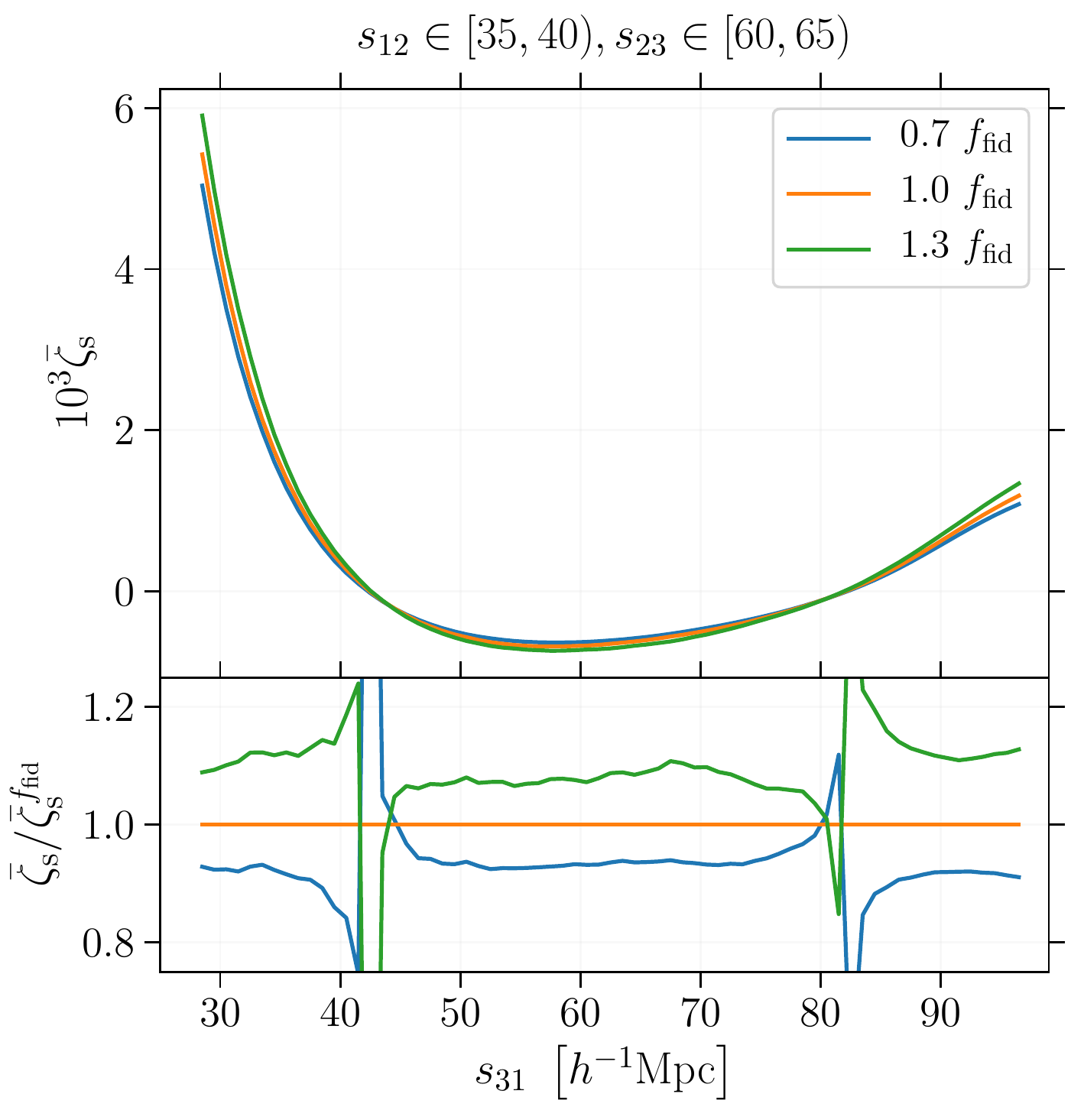}
 \end{subfigure}
 \caption{Variations of $\zeta_{\mathrm s}$ (left) and $\bar{\zeta}_{\mathrm s}$ (right) with $f$ in the 3ptGSM
 for some of the triangular configurations shown in figures~\ref{fig:streaming-sperp31} and \ref{fig:streaming-isotropic}.}
     \label{fig:streaming-varying-f}
\end{figure}

In the top panels of figure~\ref{fig:fsigma8}, we show the dependence of
${\mathcal G}_3$ and $\zeta_\mathrm{s}$ on $f$ and $\sigma_8$ for a particular triangle configuration. 
All the other parameters that determine the linear power
spectrum are kept unchanged. The shape of the contours 
is determined by the dependencies
of the different ingredients of the streaming model:
the mean pairwise
velocity in equation~(\ref{eq:mean-radial-velocity}) scales as $f\sigma_8^2$, the velocity dispersion in equations~(\ref{eq:pairwise-radial-dispersion}) and (\ref{eq:pairwisetransvdisp})
as $(f\sigma_8)^2$, and the leading-order perturbative terms for $\xi$ and $\zeta$ as $\sigma_8^2$ and $\sigma_8^4$, respectively.
In the bottom panels, we scale out the normalisation of the real-space statistics in order to focus on the effects of the RSD.
It is worth noticing that $\zeta_{\mathrm s}$ and ${\mathcal G}_3$ (which is dominated by 2-point statistics) display different degeneracies in the $f$-$\sigma_8$ plane. 
Since RSD in the galaxy power spectrum on large scales are only sensitive to the degenerate combination $f\sigma_8$
and $b\sigma_8$ (where $b$ denotes the linear bias parameter), the results above suggest that measuring $\zeta_{\mathrm s}$ with sufficient accuracy should be able to break the $f$-$\sigma_8$ degeneracy (see also \cite{GilMarin+14} for a related conclusion based on Fourier-space statistics). 
In figure~\ref{fig:streaming-varying-f}, we show the dependence of $\zeta_{\mathrm s}$ and $\bar{\zeta}_{\mathrm s}$ on $f$ while keeping
$\sigma_8$ fixed at its fiducial value (for some of the configurations displayed in figures~\ref{fig:streaming-sperp31} and \ref{fig:streaming-isotropic}).
Thirty per cent variations in $f$ induce
scale-dependent changes
in $\zeta_{\mathrm s}$ at the 10-20\% level and modulate the amplitude of $\bar{\zeta}_{\mathrm s}$ by  7-12\%.
On large scales (and for bins that are similar to ours),
the Euclid mission is expected to measure $\bar{\zeta}_{\mathrm s}$ with an accuracy of $\sim$10\%
on the individual data points
(A. Veropalumbo, private communication). 
Slightly larger uncertainties should be expected for
the wedge-averaged 3PCF while measurements of 
$\zeta_{\mathrm s}$ should suffer from a
substantially lower signal-to-noise ratio.
Considering
the large number of possible triangular configurations,
our results suggest
that the 3PCF should be able to provide a competitive measurement of the growth rate of structure.
Performing 
an accurate forecast, however, requires an estimation of the covariance matrix of the measurements and goes clearly beyond the scope of this paper.

\section{\label{sec:conclusions}Summary}
\label{sec:summary}
We have derived, from first principles, 
the equations that relate the 
$n$-point correlation functions in real and redshift space.
We have followed a particle-based approach using statistical-mechanics techniques based on the $n$-particle phase-space densities.\footnote{In section~\ref{sec:collisionless}, we have provided a
dictionary to translate our formalism into the language used by many previous papers that discuss collisionless systems.}
Our results are exact (within the distant-observer approximation)
and completely independent of the nature of the tracers
we consider and of their interactions.
They generalise the so-called streaming model
to $n$-point statistics.
The theory is formulated more naturally in terms of the
full $n$-point correlations.
In this case, the redshift-space correlation function is  obtained
as an integral of its real-space counterpart
times the joint PDF of $n-1$ relative los peculiar velocities.
Equation~(\ref{eq:general-streaming-model-npoint})
expresses this relation succinctly and the velocity PDF is defined in equation~(\ref{eq:multipdfdef}).

We have shown that 
it is possible to re-formulate the theory entirely
in terms of connected correlation functions although the price to pay is a velocity term that is not a PDF (and can be negative) as well as a higher degree of abstractness.
This result is expressed by equations
(\ref{eq:general-streaming-model-connected})
and (\ref{eq:cndef}).

In the second part of the paper, 
we have focused on 3-point statistics. First of all, by combining the
streaming model for the 2PCF and the 3PCF, we have derived equation~(\ref{eq:3ptstreamrversion})
which provides a computationally-friendly framework
to calculate connected 3-point correlations in redshift space. 
A key ingredient appearing in this equation is
the bivariate PDF for the los relative velocities between particles pairs in a triplet, ${\mathcal P}^{(3)}_{\bm{w}_{\parallel}}(w_{12\parallel}, w_{23\parallel} | \triangle_{123})$.
Making use of a large $N$-body simulation, we have characterised
the properties of this function for unbiased tracers of the matter-density field.
Figures~\ref{fig:equi} and \ref{fig:bivariate-gaussian}
show that the PDF is unimodal and, for large triangles, has a quasi-Gaussian peak.
The dispersion of $w_{12\parallel}$ and $w_{23\parallel}$
is always much larger than the mean.
Moreover, $w_{12\parallel}$ and $w_{23\parallel}$ tend to be anti-correlated, especially on large scales.

In section~\ref{sec:linear-theory},
we have derived theoretical predictions for the first two moments of $\bm{w}_{12}$ and $\bm{w}_{23}$
using standard perturbation theory at LO.
Equation~(\ref{eq:mean-radial-velocity-three}) shows that
the mean relative velocity between a particle pair in a triplet is not purely radial but has also a transverse component in the plane of the triangle defined by the particles. Individual expressions for the different components are given in 
equations (\ref{eq:decomposition}), (\ref{eq:R(triangle)})
and (\ref{eq:Ttriangle}). Figures~\ref{fig:grid_radial} and \ref{fig:grid_transverse} show that the LO predictions
accurately match the simulation results from quasi-linear scales onward ($r_{ij} \gtrsim 20\,h^{-1} \mathrm{Mpc}$).
Perturbative expressions for the second moments are given in equations~(\ref{eq:matrix1212}) and (\ref{eq:matrix1223}).
In this case, a constant offset
needs to be added to the theoretical results (that neglect small-scale physics) in order to reproduce the simulations
on large scales. Figure~\ref{fig:grid_radial_std} shows that, after applying the correction, the model is accurate to better than a few per cent for separations $r_{ij} \gtrsim 50\,h^{-1} \mathrm{Mpc}$. 
In section~\ref{sec:losproj}, we have discussed the projection of the relative velocities along the los.
Figure~\ref{fig:cov-los} shows that the perturbative predictions agree well with the simulation for triangles with legs  $r_{ij} \gtrsim 50\,h^{-1} \mathrm{Mpc}$.
Our results lay the groundwork for investigating 3-point statistics of the los pairwise velocities with future experiments based on the kinetic Sunyaev-Zel'dovich effect like the Simons Observatory \cite[SO,][]{SO}, CMB-S4 \cite{CMBS4}, CMB-HD \cite{CMBHD-I, CMBHD-II};
 as well as with other peculiar-velocity surveys 
like the Taipan galaxy survey \cite[Taipan,][]{Taipan} and the Widefield ASKAP  L-band  Legacy  All-sky  Blind  Survey \cite[WALLABY,][]{Wallaby-20}.

In section~\ref{sec:3ptGSM}, we have introduced the 3ptGSM that brings together several elements of our study.
This model is based on the exact equation~(\ref{eq:3ptstreamrversion})
but phenomenologically approximates ${\mathcal P}^{(3)}_{\bm{w}_{\parallel}}(w_{12\parallel}, w_{23\parallel} | \triangle_{123})$
with a bivariate Gaussian distribution whose moments are computed using 
perturbative techniques (and offsetting the velocity dispersion with a constant so that to match its direct measurement in the simulation). We have then presented a simple practical implementation of the 3ptGSM by deriving  
all its ingredients 
(real-space clustering amplitudes and velocity statistics) from standard perturbation theory at LO.
The comparison of the model predictions against the correlation function from the simulation performed in 
figures~\ref{fig:streaming-sperp31}, \ref{fig:streaming-isotropic} and \ref{fig:zeta_wdge} is very encouraging, in particular considering
that the model has no free parameters.

The forthcoming generation of galaxy surveys will cover
large-enough volumes to permit accurate measurements of the 3PCF on large scales.
This achievement will inform us about galaxy formation, cosmology, neutrino masses, the nature of primordial perturbations, dark energy, and the gravity law.
It is thus timely to create new theoretical tools 
that facilitate this endeavour.
In this paper,
we have developed a general framework for the analysis of RSDs in the $n$-point correlation functions.
This pilot work sets the foundation for future 
developments including:
(i) considering biased tracers of the matter-density field, 
(ii) extending our calculations
to different flavours of PT \cite[e.g.][]{ReidWhite11,Carlson+13,Vlah+16}
for both real-space clustering and velocity statistics, and
(iii) going beyond the Gaussian approximation for the PDF of the relative los velocities 
by introducing multivariate distributions with non-zero skewness and that are leptokurtic \cite[e.g.][]{Bianchi+15,Uhlemann+15,Bianchi+16,KuruvillaPorciani18,Cuesta+20}.

\begin{acknowledgments}
We thank the anonymous referee for suggesting to add section~\ref{sec:fofz}, Alfonso Veropalumbo for useful discussions, and Daniele Bertacca for exchanges about a parallel line of research. JK has been partially supported by the funding for the ByoPiC project from the European Research Council (ERC) under the European Union's Horizon 2020 research and innovation program grant agreement ERC-2015-AdG 695561.
We are thankful  to  the  community for developing  and  maintaining open-source software packages extensively used in our work, namely \textsc{Cython} \citep{cython}, \textsc{Matplotlib} \citep{matplotlib}, \textsc{Numpy} \citep{numpy}
and \textsc{Scipy} \citep{scipy}.
\end{acknowledgments}

\bibliographystyle{JHEP}
\bibliography{streaming}

\end{document}